\documentclass{emulateapj}
\usepackage{graphics,graphicx}
\usepackage{subfigure}
\usepackage{epsfig}
\usepackage{natbib}
\usepackage{longtable}
\usepackage{amssymb}
\usepackage{lscape}

\begin{document}

\title{Stepwise Filter Correlation Method and Evidence of Superposed Variability Components in GRB Prompt Emission Lightcurves}

\author{He Gao, Bin-Bin Zhang, Bing Zhang}
\affil{Department of Physics \& Astronomy, University of Navada Las
Vegas, Las Vegas, NV, 89154}

\begin{abstract}
Gamma-ray bursts (GRBs) have variable lightcurves. Although most
models attribute the observed variability to one physical origin
(e.g. central engine activity, clumpy circumburst medium,
relativistic turbulence), some models invoke two physically distinct
variability components. We develop a method, namely, the {\em
stepwise filter correlation} (SFC) method, to decompose the
variability components in a GRB lightcurve. Based on a low-pass
filter technique, we progressively filter the high frequency signals
from the lightcurve, and then perform a correlation analysis between
each adjunct pair of filtered lightcurves. Our simulations suggest
that if a mock lightcurve contains a ``slow'' variability component
superposed on a rapidly varying time sequence, the correlation
coefficient as a function of the filter frequency would display a
prominent ``dip'' feature around the frequency of the slow
component. Through simulations, we demonstrate that this method can
identify significant clustering structures of a lightcurve in the
frequency domain, and proved that it can catch superposed signals
that are otherwise not easy to retrieve based on other methods (e.g.
the power density spectrum analysis method). We apply this method to
266 BATSE bright GRBs. We find that the majority of the bursts have
clear evidence of such a superposition effect. We perform a
statistical analysis of the identified variability components, and
discuss the implications for GRB physics.

\end{abstract}


\section{Introduction}

The temporal structure of Gamma-Ray Bursts (GRBs) exhibits diverse
morphologies \citep{Fishman95}. They can vary from a single smooth
pulse to extremely complex lightcurves with many erratic pulses with
different durations, amplitudes, and fine structures. Based on
temporal information, it has been difficult to categorize GRBs.

Physically, several mechanisms have been invoked to interpret GRB
temporal variability. The leading scenario is to attribute the
lightcurve variability to the irregularity of the central
engine\footnote{The central engine in general sense refers to the
central compact object (e.g. an accreting black hole or a spinning
down neutron star) as well as the stellar envelope (if any) that
regulates the time history of the outflow.}. For the commonly
discussed internal shock scenario \citep{Rees94,Sari97,Daigne03,Maxham09},
the observed time sequence tracks that of the central engine very
well \citep{Kobayashi97,Maxham09}. Alternatively, if the emission is
from the central engine photosphere, then the observed lightcurve
time history tracks that of the central engine directly
\citep[e.g.][]{Lazzati09}. Within such a scenario, the observed
lightcurves can be directly connected to the behavior of the central
engine \cite[e.g.][]{Lei07,Lu08}. A second scenario takes the
opposite view: The observed variability originates in the emission
region, which is not directly related to the history of central
engine activity. Since the GRB outflow is relativistic, this
requires that the emission region is not uniform. Rather, it
contains locally Lorentz boosted emission regions, such as
mini-jets \citep{Lyutikov03} or relativistic turbulence
\citep{Narayan09,Kumar09}. A third scenario discussed in the
literature invoked a clumpy circumburst medium to interpret
variability within the external shock model of GRB prompt emission
\citep[e.g.][]{Dermer99}. Swift observations of early X-ray
afterglows of GRBs revealed a steep decay phase connected to the
prompt emission lightcurve \citep{Tagliaferri05,Barthelmy05}. This
suggests that the GRB prompt emission region is detached from the
afterglow region, and therefore prompt emission should be of an
``internal'' origin \citep{Zhang06}. This disfavors this third model
of GRB variability.

Recently, \cite{Zhang11} proposed a new model of GRB prompt emission
in the Poynting-flux-dominated regime, namely, the
Internal-Collisioninduced MAgnetic Reconnection and Turbulence
(ICMART) model. This model invokes a central engine powered,
magnetically dominated outflow, which self-interacts and triggers
fast magnetic turbulent reconnection to power the observed GRBs. An
important prediction of the ICMART model is that it has two
variability components: a broad (slow) component related to the
central engine activity, and a narrow (fast) component associated
with relativistic magnetic turbulence.  \cite{Zhang11} conjectured
that the visually apparent broad pulses in GRB lightcurves are
related to the time history of the central engine (with each broad
pulse corresponding to an ICMART event), while the much faster
variabilities superposed on the broad pulses are related to
relativistic, magnetic turbulence.

Alternatively, \cite{Morsony10} simulated jet propagation from a
massive star, and suggest that the broad pulses of several seconds
duration  are due to interaction of the jet with the progenitor,
while the shorter scale variability in the millisecond range is
related to that of the base of the inner engine (e.g. the black hole
or the millisecond pulsar).

It would be essential to use rigorous mathematical methods to study
GRB lightcurves to investigate whether the time sequence demands
superposition of multiple variability components. Power density
spectrum (PDS) is the most commonly used tool to study the temporal
behavior of astronomical objects. Beloborodov et al. (1998, 2000)
found that although the PDS of individual GRBs are diverse, the
average PDS of a stack of GRBs is in accord with a power law with
index -5/3 over 2 orders of magnitude in frequency. By locating the
PDS peaks, \cite{Shen03} revealed the typical variability time
scales of some GRBs. In general, this method is not powerful to
address whether a GRB lightcurve has superposed variability
components. This is because it is insensitive to the lower frequency
component (if it exists) since the GRB durations are typically not
much longer than the broad pulses themselves. In the time domain,
several methods have been developed to study temporal properties of
GRBs. For example, lightcurves were decomposed into individual
pulses using some parameterized empirical pulse functions
\citep{Norris96} or a peak-finding algorithm
\citep{Li96,McBreen01,Nakar02}, and the temporal properties of the
resulting pulses were analyzed. When performing the empirical pulse
modeling, \cite{Norris96} noted that some bursts are too complex to
fit, possibly indicating pulse superposition. However, their matrix
inversion algorithm failed to handle the problem. The peak finding
selection method can decompose a lightcurve into many individual
peaks. However, the method is not developed to reveal the superposed
variability components.

Nonetheless, the possibility of superposition was suggested from
other observational evidence. From a frequency-dependent analysis of
prompt X-ray lightcurves of a sample of BeppoSAX GRBs,
\cite{Vetere06} discovered that the lightcurve tends to become
smoother in softer energy bands. They then speculated that there
might be a slow component superposed on a fast component. It is
therefore of great interest to develop a rigorous mathematical
method to identify such a superposition effect, if any, in gamma-ray
lightcurves alone without the assistance of multi-wavelength data.
Hereafter, we define a {\em slow component} as an underlying broad pulse
component, while a {\em fast component} as the component of rapid
variability that overlaps on top of the slow component.

In this paper, we develop a mathematical method to process GRB
lightcurves as an effort of identifying the superpostion effect.
This method, known as the {\em stepwise filter correlation} (SFC)
method, is presented in detail in Section 2. We delineate its
mathematical basis, and justify its robustness in identifying the
superposed slow component through simulations. In Section 3, we
apply this method to a sample of bright BATSE GRBs, and indeed
identify the superposition effect in the majority of them. The
properties of the identified slow components are studied
statistically. The data analysis results and their physical
implications are presented in Section 4. We note that some
independent studies (e.g. R. Margutti et al. 2011, in preparation)
reached the similar conclusion as ours.

\section{Stepwise Filter Correlation Method}

\subsection{The method}

In signal processing, a filter is a device or process that removes
some unwanted component or feature from a time sequence signal. Our
method is based on a low-pass filter named Butterworth filter. For a
certain cutoff frequency, this filter passes low-frequency signals
below this frequency but attenuates signals above (see Appendix 1
for mathematical details). Our method is based on the following
concept. Suppose that one specific time series (lightcurve) can be
decomposed into the summation of $N$ pulses, either horizontally
(pulses are laid out side by side) or vertically (superposition).
One may denote the time scales (durations) of these pulses as
$t_{\rm j,~ j=1...N}$. The corresponding frequency for each pulse is
therefore $f_{\rm j}=1/t_{\rm j}$. If one applies a stepwise filter
in the frequency space, one would get a series of residual
lightcurves (RLCs). If there is no pulse falling into the range
between the cutoff frequencies $f_{\rm c,i}$ and $f_{\rm c,i+1}$,
then the two RLCs should be identical. If there are some pulses
whose frequencies fall into this range, the RLC with the lower
cutoff frequency (RLC$_{\rm i}$) should be smoother than the one
with the higher cutoff frequency (RLC$_{\rm i+1}$), since these
pulses are screened after performing the low-pass filter at $f_{\rm
c,i}$. The more pulses falling into this frequency range or the
highger the amplitudes of these pulses, the more
different the two RLCs look like. One can quantify the difference
between the two RLCs using a statistical correlation method. A
cluster of many pulses or a high amplitude of pulses
within a frequency range would result in a
correlation coefficient $R_{\rm i}$ (defined between RLC$_{\rm i}$
and RLC$_{\rm i+1}$) to be more less than unity. By plotting $R_{\rm
i}$ against $f_{\rm c,i}$, a ``dip'' in the curve would reveal such
a clustering, and hence, would lead to the identification of a
variability component around a particular frequency.

We realize such a concept based on the following procedure:

(1) For a time series, define a frequency range $(f_{\rm min},
f_{\rm max})$ to be searched from. We then divide this frequency
range into many discrete frequency bins uniformly in logarithmic
scale. For all the GRBs, our frequency step is uniformly chosen as
$\log(\Delta f)=0.05$. This gives a sequence of the cutoff
frequencies $f_{\rm c,i}(i=1,...M)$, where $M$ is the total number
of frequency bins. The low-pass filter to the original time series
is then performed with each cutoff frequency $f_{\rm c,i}$ in turn.
The RLC for each cutoff frequency, e.g. RLC$_{\rm i}$ corresponding
to $f_{\rm c,i}$, is recorded.

(2) Perform a correlation analysis between each pair of adjunct RLCs
(e.g. RLC$_{\rm i}$ vs. RLC$_{\rm i+1}$). Record the Pearson's
correlation coefficient $R_{\rm i}$ for each pair.

(3) Plot $R_{\rm i}$ against $f_{\rm c,i}$, and identify apparent
dips in the curve.

\subsection{Simulation tests}

To prove the validity of the method, we perform some simulations. We
start with a simple two-component lightcurve as shown in Fig.1 top
row left panel. We superpose two periodic signals, both with the
function form  $A| \sin (\pi t/T) |$, where the periods of the two
components are $T_{\rm s}=100 \pi$ s for the slow component and
$T_{\rm f}=10\pi$ s for the fast component, and the amplitude ratio
between the two components is $A_{\rm s}:A_{\rm f}=2:1$. The middle
panel of the top row shows the PDS of the lightcurve, which clearly
shows the two components. The right panel of top row is the $R_{\rm
i}-f_{\rm c,i}$ figure of our SFC method. Two dips that correspond
to the two frequencies ($f_{\rm s}=1/100 \pi~{\rm s^{-1}}$ and
$f_{\rm f} = 1/10 \pi ~{\rm s^{-1}}$ are clearly identified. We also
add some white noise to the mock lightcurve. We find that even when
the amplitude of the white noise is comparable to the signal, the
dip in the lower frequency still shows up. This suggests that this
method is powerful in identifying the low frequency component in the
superposition.

Next, we simulate a more realistic lightcurve (middle row of Fig.1).
The lightcurve (left panel of the middle row) is now a superposition
of a slow component with pulse widths randomly distributed in the
range of $T_{\rm s}= (10 - 20)$ s and a fast component with pulse
widths randomly distributed in the range of $T_{\rm f}= (1 - 3)$ s.
The amplitudes of the two components are randomly chosen in the
range of $A_{\rm s}=(0.5-3)$ and $A_{\rm f}=(0.5-1)$, respectively.
Since there is no strict periodicity and since the duration of the
time series is not much longer than the slow component time scale,
the PDS method (middle panel of the middle row) fails to identify
the two frequency components. On the other hand, our SFC method
(right panel of the middle row) clearly identifies a dip around 1/17
$s^{-1}$, which is right within the frequency range of the slow
component. This simulation suggests that the SFC method is much more
powerful in identifying the superposed components than the PDS
method.

We note that the absolute value of the correlation coefficient $R$
depends on the step length of the cutoff frequency. A smaller
frequency bin means smaller differences between consecutive RLCs, so
that $R$ would be closer to 1. In any case, the global shape of the
SFC $R_{\rm i}-f_{\rm c,i}$ curve (e.g. the location of the dips)
does not depend on the size of the frequency bin, as long as it is
small enough.

In order to further understand the SFC algorithm, we have performed
a series of additional simulation tests (see Appendix 2 for
details). These tests suggest that the SFC method is sensitive to
significant clustering structures of a lightcurve in the frequency
domain. A significant clustering structure is a cluster of
frequencies that is separated from other frequency clusters, and
that has a large enough amplitude. If a frequency cluster is too
wide, or is too close to another frequency cluster, the
corresponding dip is diminished. Similarly, if the amplitude of a
frequency cluster component is too small, the corresponding dip
would be too shallow or disappear completely. An interesting finding
is that the quiescent gaps that separate pulses in GRB lightcurves
\citep{RR01} would complicate the analysis. Only when the gaps are
removed manually, can one identify the corresponding frequencies of
the slow pulse components (see Appendix 3 for details). Another
finding is that if the slow component has only one pulse, the long
tail of pulse tends to extend the duration, so that the identified
duration can be much longer than the full width at half maximum
(Appendix 2).

\section{Application to GRB data}

We now apply the SFC method to real GRBs. In this paper, our aim is to
demonstrate the validity of the SFC method and to investigate
whether the superposition effect exists in GRBs. So we do not pursue
sample completeness. Rather, we only focus on some bright GRBs that
have clear temporal structures.

We select 266 bright GRBs detected by Burst and Transient Source
Experiment (BATSE) \citep{Kaneko06}, whose lightcurve data and
$T_{\rm 90}$ values are publically available from the online
database http://heasarc.gsfc.nasa.gov/docs/cgro/batse/. In our
analysis, we use light curves with 64 ms resolution obtained by
BATSE in the four Large Area Detector energy channels, $20 - 300$
keV. The background is subtracted in each channel using linear fits
to the 1024 ms data. The SFC method is then applied to these bursts.
For all the analyses, we adopt a fixed maximum frequency of $f_{\rm
max}=5$, which is based on the consideration of having at least 3
time bins for a 64 ms time resolution. The minimum frequency varies
from burst to burst, but is related to a duration at least
(sometimes larger than) $T_{90}$, in order to catch the slowest
variability component. After fixing the frequency range, the
frequency step is chosen evenly in the logarithmic space with $\log
(\Delta f)=0.05$.

In order to quantitatively delineate the significance and confidence
level of each dip, we define two parameters. The significance
parameter, $s$, delineates the deepness/shallowness of a dip in the
SFC $R_{\rm i}-f_{\rm c,i}$ curve. A dip is typically asymmetric, we
apply the shallower wing of the dip to define its shallowness. We
first identify the local minimum point at the bottom of the dip,
e.g. for
%can be identified if there is a minimum
%in the data sequence in the SFC $R_{\rm i}-f_{\rm c,i}$ curve with a
%certain frequency bin size. For each dip, e.g.
the $n$-th dip the
coordinate ($f_{\rm n}, R_{n}$) in the SFC curve.
Next, we find out the
inflection points (where the second derivatives change sign), or the
turning points (where the first derivatives change sign) if an
inflection point does not exist, in the left and right wings of the
dip, respectively. These two points,
% on both the left and right
%sides (or the endpoints of the curve for the first or last dip case),
%whose coordinates are
i.e. ($f_{\rm n,left}, R_{n,left}$) and ($f_{\rm
n,right}, R_{n,right}$), could be defined as ``boundaries'' of the
dip. One can then define
\begin{equation}
S_n={\rm Min}\{\frac{R_{n,left}-R_{n}}{log(f_{\rm n})-log(f_{\rm
n,left})},\frac{R_{n,right}-R_{n}}{log(f_{\rm n,right})-log(f_{\rm
n})}\}
\end{equation}
for each dip. A larger $S_n$ means a more significant dip.
To reduce the bin-size effect, we normalize $S_n$ to the most
significant one, $S_n^{\rm max}$ (which is usually the slowest one
$S_1$), i.e. we define $s_n= S_n /S_n^{\rm max}$ for each dip.
% (the one with the largest
% as the critical parameter and then normalize it for every
%individual burst to reveal the $\mathfrak{n}$-th dip's significant
%level.
This $s$ parameter (which is $s \leq 1$) is then the significance
parameter.

Next, we define a confidence level
parameter, $c$, based on Monte-Carlo simulations.
%we turn to use Monte Carlo simulation to give the
%confidence level of dips in SFC curve.
For each time bin with a particular observed count rate, we can
generate a  mock count rate based on the observed count rate
$\mathcal{C}$ by randomly generating the data based on a normal
distribution with ($\mathcal{C},\mathcal{\sqrt{C}}$). We then
generate 1000 mock lightcurves by collecting these randomly
generated count rates for each time bin. We apply the SFC method to
each mock lightcurve, and identify the frequencies of the dips in
each realization. For each dip in the original lightcurve, we define
$c$ as the fraction that the simulations reproduce. We regard a
component with high confidence level if $c \geq 0.9$, i.e., more
than 900 simulations have revealed the component. We note that even
though in general high-significance dips have a high confidence
level, the two parameters are not always correlated. Some high ``s''
dips turn out to have a low ``c''. We therefore evaluate both
parameters for every dip measured in the SFC $R_{\rm i}-f_{\rm c,i}$
curves.

We take GRB\,930331A as an example (bottom row of Fig.1). The left
panel shows the original lightcurve, and the middle and right panels
show the PDS and SFC analysis results, respectively. Although the
PDS does not show any interesting feature, the SFC curve indeed
shows a prominent dip around 1/38 ${\rm s^{-1}}$ (both significance
and confidence level parameters equal unity, i.e. $s=1$, $c=1$).
Checking back in the lightcurve, one indeed sees
one broad pulse with a time scale of 38 s and another with the pulse
width slightly shorter. Rapid spikes overlap on top of these two
broad pulses. To verify whether the 38s component truly exists, we
apply the SFC method to a portion of the lightcurve, from the
trigger time $T_{\rm 0}$ to a certain time $T$, with $T$ stepwisely
increasing in the range of $0 < T-T_0 < T_{90}$. We found that once
$T-T_0$ is longer than 38s, the dip at 1/38$s^{-1}$ persists in all
SFC curves, indicating that the 38s component is a real slow
component.

Applying this method to our entire GRB sample, we find the following
interesting facts.

(1) The total 266 bursts could be grouped into four categories based
on both their lightcurves and SFC curves: (I) Good sample: 117/266
($44.0\%$) of the bursts can be included in this sample. They
clearly show at least one dip in the SFC curve. Checking back the
lightcurves, one can usually find one or more pulses with the
identified characteristic frequencies. Superposed on the identified
slow component, there are always more rapid variability features.
This clearly suggests a superposition of at least two variability
components in the lightcurves. (II) Gap/long tail sample:
88/266($33.1\%$) of the bursts have quiescent periods in the
lightcurve whose durations are comparable to the broad pulses, or
have one FRED-like pulse with extended tail. For these cases, dips
in the SFC curve are affected by the quiescent periods (gaps) and
the tails. For the gap case, the real slow component can be revealed
by manually removing the gaps in the lightcurves (see examples in
Appendix 3). These bursts also clearly show the superposition effect
as seen in the sample (I). However, since the identified frequencies
do not well match the pulse durations, we have excluded these bursts
in the statistical study presented below; (III) Irregular (noisy)
sample: 24/266 ($9\%$) of the bursts show dips in the SFC curve.
However, their lightcurves are too noisy to identify the
corresponding components (see examples in Appendix 3). To be
cautious, we do not include these bursts in the statistical
analysis. (IV) Short/low temporal resolution sample: Finally, 37/266
($13.9\%$) of the bursts are short bursts or long bursts whose
lightcurves have a poor temporal resolution (see example in Appendix
3). These bursts have too narrow a frequency range to perform the
SFC analysis. All the lightcurves and their corresponding SFC curves
for the samples I-III are presented at the UNLV GRB group website
http://grb.physics.unlv.edu/sfc. An example to each of the groups
II, III and IV is presented in Appendix 3.

(2) Within Sample I (good sample), 30/117 ($25.6\%$) bursts show
just one dip in the low frequency regime. Due to space limitation,
we only present some examples in Fig.2. All the other cases are
disseminated at the group website. The identified slow component
time scales, as well as the $s$ and $c$ parameters for each dip are
presented in Table 1. Since there is no strict periodicity in the
lightcurves, the time scales of all the components we have
identified are rough values, and we have rounded them to the nearest
0.5.

(3) The rest 87/117 bursts ($74.4\%$) in Sample I show more than one
dips. For each dip we try to identify the corresponding component in
the lightcurve. Some examples are presented in Fig.3. Others are
presented in the group website. We present the time scales of all
the identified components $T_i$ and their relevant $s$ and $c$
parameters also in Table 1, with increasing frequencies for
ascending number $i$. Only dips with $c \geq 0.9$ are selected.

Inspecting the lightcurves with multiple dips in the SFC curve, we
find that it is not always straightforward to relate dips with pulses
in the lightcurves. In some cases (e.g. GRB 910430 and
GRB 940414B), the low-frequency dip corresponds to a broad pulse
with overlapping fast variability whose frequency corresponds to the
high-frequency dips. In these cases, it is straightforward to
identify the underlying broad pulses as the slow component, while to
identify the overlapping narrow pulses as the fast component.
%In some
%cases (e.g. GRB 921209), although the fast component is found
%overlapping on the slow component, one can also identify another
%pulse with a shorter duration corresponding to the high frequency
%(3s) that lies side-by-side with the broader 9.5s pulse. So one may
%not readily attribute the 3s component to either a fast or a slow
%component.
In more complicated cases (e.g. GRB 940228A), besides identifying
some slow components (e.g. 8s and 4s) that correspond to
individual broad pulses, one also identifies a very slow (21s)
component. This is due to clustering of multiple pulses to make
a 21s ``cluster''.
%One may consider this 21s component as a ``slow''
%component, with the more rapid pulses (e.g. 2s) identified as the
%fast component. Alternatively, one may regard those 2s pulses as
%slow components on their own (since there are even finer structures
%on top of the pulses), while the 21s signal marks a time scale of
%clustering of those pulses.
In general, we caution that the SFC
method, although sensitive to identify variability components not
easy to unveil using the PDS method, may be over-sensitive to pick
up variability components. We therefore caution that one should
always go back to the lightcurves to clarify the physical nature of
the frequency components identified in the SFC curves.

To understand the results better, we perform a statistical analysis
on the time scales identified using the SFC technique. Figure 4
presents the identified variability timescales $T_i$ vs. the
duration $T_{90}$ of the bursts, histogram of $T_i$, and histogram
of $T_{90}/T_i$. First we focus on the one-dip only sample. The
characteristic time scales $T_1$ of this sample are marked in red in
Fig.4. The following trends can be observed: first, it seems that
there is a very rough positive correlation between $T_1$ and
$T_{90}$ (Fig.4a). This suggests that one tends to find longer slow
components in longer bursts. Since there is a wide distribution in
$T_{90}$, $T_1$ distributes from 2 s to 108 s (Fig.4b). On the other
hand, the scatter of correlation is large. The ratio $T_{90}/T_1$
spans in at least one decade for this one-dip only sample. In some
bursts, $T_1$ as small as $1/10$ of $T_{90}$ can be found (Fig.4c).
Next, we include all the identified components ($T_i$) in the
multi-dip sample (black circles in Fig.4a, and dashed histograms in
Figs.4b and 4c). It is found that all the distributions are much
wider. The overall histograms including all $T_i$ in the entire
sample (solid histogram in Fig.4b and Fig.4c) cover 2 orders of
magnitude in both $T_i$ and $T_{90}/T_i$. The spreading is mostly
caused by the fast components identified in the multi-dip sample,
but the long clusters (such as the 21s component identified in GRB
940228A) also contribute to the scatter. Since our frequency
interval $\log \Delta f = 0.05$ is uniform for all the bursts
regardless of their $T_{90}$, our result does not suffer from the
possible selection effect caused by different $T_{90}$. Inspecting
the one-dip only sample, one usually also see the overlaping fast
component, but with a lower amplitude than the multiple-dip ones. So
these bursts are intrinsically similar to the multiple dip sample.
The fast-component dips only show up when the high-frequency
component amplitudes are large enough.

Another way to look at the distribution is to isolate the slowest
component $T_1$ from other higher frequency ones. Figure 5 shows
such a separation: $T_1$ in red and $T_i$ ($i>1$) in black. It can
be seen that the $T_i-T_{90}$ correlation is more prominent for
$T_1$. This may be because the longer the burst, the more probable
that a long cluster (e.g. 21s cluster in GRB 940228A) would show up.
The correlation between $T_i$ ($i>1$) is much weaker. In particular,
variability time scales as short as seconds can appear in very long
GRBs (e.g. $T_{90} \sim 250$ s). This suggests that the fastest
variability component essentially does not depend on the duration of
the burst.

\section{Conclusions and Discussion}

We have developed a new method (Stepwise Filter Correlation) to
decompose variability components in a time series. Through Monte
Carlo simulations, we demonstrate that this method can identify
significant clustering structures of a lightcurve in the frequency
domain, and is more powerful than the traditional methods (e.g. PDS)
to identify superposed variability components, especially the slow
variability component with duration comparable to the duration of
the time series.
%A dip in the
%correlation curve would reflect a variability component
%in the lightcurve.}

We then apply this method to GRB lightcurves as an effort to
investigate whether the lightcurve is a superposition of multiple
variability components. Our findings can be summarized as follows:

(1) We have applied this method to 266 BATSE bright GRBs, which
may be grouped into 4 categories.
% and distribute 266 bright bursts into four groups.
In general, most bursts show a clear dip in the low frequency
range, suggesting a slow component. By checking back to the
lightcurves, we were able to identify the corresponding pulses
with the relevant dip frequency in most of bursts. We found that
such a slow component usually
has superposed rapid variability components. We therefore conclude
that GRB lightcurves are typically the superposition of multiple
variability components.

(2) We selected 117 bursts as the good sample, and carried out a
statistical statistical analysis sample. Among them, 30 show only
one dip in the correlation curve. The other 87 GRBs have more than
one dips.
%In all
%cases, one can identify the time scales corresponding to the
%frequency dips in the lightcurves. One can always identify some
%``slow'' component pulses whose durations correspond to the dips in
%the correlation curves and whose pulses are superposed with even
%faster variabilities. In a same burst, one may identify such slow
%components with different time scales. Our analysis therefore
%suggests that at least a fraction of bursts contain superposed
%variability components.
%(2) We have performed a statistical analysis of the identified
%variability components.
For the one-dip only sample in which the dip
corresponds to a slow component, the distribution of this time scale
$T_1$ spreads from several seconds to $\sim 100$s, with no typical
time scales, and $T_{90}/T_1$ spreads in one order of magnitude
(from $\sim 1$ to $\sim 10$). There is a rough trend of correlation
between $T_1$ and $T_{90}$. Including all the variability
components, the distributions of $T_i$ and $T_{90}/T_i$ spread in
two orders of magnitude, without a characteristic value. The fastest
time scale of order $\sim 1$s can be found in bursts with a wide
range of durations.

The identification of the variability superposition effect (i.e. the
existence of a slow component with overlapping faster variabilities)
suggests that the causes of GRB lightcurve variabilities may be
diverse. There might be more than one physical mechanisms that
define the observed variability. This is in align with the
prediction of the ICMART model \citep{Zhang11} and the jet
propagation model \citep{Morsony10}. The common aspect of these two
suggestions is that the slow component (duration of seconds to 10s
of seconds) is attributed to the engine that defines the jet
variability.
%This is because the central engine of \cite{Zhang11} is
%broadly defined as the central object and the stellar envelope above
%it \cite[e.g.][]{Zhang03} (see footnote 1).
The difference between
the two scenarios is the origin of the fast component. While the
envelope model \citep{Morsony10} attributes it to the intrinsic
variability at the base of the inner engine, i.e. the central black
hole or magnetar, the ICMART model \citep{Zhang11} attributes it to
relativistic magnetic turbulence in the emission region. These two
scenarios may be further differentiated through testing more
detailed predictions in both models. For example, in the jet-star
interaction model, the inner engine powered variability shows up
only if the input PDS is hard enough, e.g. $E(k) \propto k^0$, or
essentially the same power per decade. It is not known whether this
can be achieved in the inner engine, and such a hard PDS is not
observed in the high frequency regime of GRBs. On the other hand,
the fast variability in the ICMART model arises from locally
Lorentz-boosted mini-jets due to relativistic turbulent reconnection.
Simulations suggest that the it can reproduce the observed PDS
\citep{ZZ11}.
%emission from , which
%has a well predicted PDS (e.g. $E({k_\perp})
%\propto k_{\perp}^{-5/3}$ (Komolgorov-type) and $E(k_\parallel)
%\propto k_{\parallel}^{-2}$ \citep{Goldreich95,Cho02}. The former is
%in general agreement with the PDS analysis results.
In any case,
neither model predicts a characteristic time scale for the fast
component. It is therefore still a theoretical challenge to account
for the typical fast component time scales identified in some
bursts.

Finally, we'd like to justify the Butterworth low-pass filter we
have adopted. In principle, one can use low-pass, high-pass, or
band-pass filters. First, a band-pass filter only passes signals in
a certain frequency band, which is disfavored by the SFC method.
This is because a good correlation between two adjacent frequency
bins may simply reflect that the changes between the two frequency
bins are similar. One may not get a ``dip'' even though the changes
are significant.
Second, the purpose of this work is to find out whether superposition
exists in GRB lightcurves. We therefore care more about the underlying
slow component. We therefore choose a low-pass filter, which is
more sensitive to the slow component. A high-pass filter, on the other
hand, would be more sensitive to fast components.
As for specific digital low-pass filters, we have done simulation
tests for several types, including the Butterworth filter, the
Chebyshev filter, and the Gaussian filter. We find that different
choices of filter would not change the results significantly. Among
them, the Butterworth filter is designed to have as flat a frequency
response as possible in the unscreened bandpass, which is beneficial
to retain information of the slow component. We therefore adopt it
in this work. We have also tried the wavelet transform method. It is
a useful tool to unveil multiple variability components in the
lightcurves (see e.g. Vetere et al. 2006). However, we did not find
an easy way to quantify the results. We therefore do not apply the
wavelet transform in this work.

\acknowledgements We would like to thank an anonymous referee for
constructive comments and suggestions. We also thank
Zhe Chu and Nan Xing for
helpful discussion on the statistical methods, and Xue-Feng Wu and
Wei-Hua Lei for discussion on applications to GRBs.  This work is
supported by NSF through grant AST-0908362, and by NASA through
grants NNX10AD48G and NNX10AP53G.

{}

\clearpage
%%%%%%%%%%%%%%%%%%%%%%%% Tables %%%%%%%%%%%%%%%%%%%%%%%%%%%%%%

\begin{deluxetable*}{lcccccc}

\label{Table:sample} \tablecolumns{8} \tablewidth{0pc}
\tablecaption{Characteristic timescales identified in BATSE bright
Gamma-Ray Bursts}

\tablehead{ \colhead{GRB} & \colhead{$T_{\rm 90}$} &
\colhead{$T_{\rm 1}$\tablenotemark{a}(s/c)}  & \colhead{$T_{\rm
2}$\tablenotemark{b}(s/c)} & \colhead{$T_{\rm
3}$\tablenotemark{c}(s/c)} & \colhead{$T_{\rm
4}$\tablenotemark{d}(s/c)}
 }

\startdata

910627  &  15.2    &  5 (1/1)     &  0  &  0  &  0   \\
910807  &  59.6    &  12 (1/1)   &  0  &  0  &  0   \\
911031A &  90.0    &  23 (1/1)   &  0  &  0  &  0   \\
911118A &  19.2    &  22.5 (1/1)    &  0  &  0  &  0   \\
920218C &  122.5    &  55 (1/1)   &  0  &  0  &  0   \\
920511A &  48.5     &  3 (1/1)   &  0  &  0  &  0   \\
920524 &  66.1    &  7.5 (1/1)   &  0  &  0  &  0   \\
920622B &  36.0    &  7  (1/1)   &  0  &  0  &  0   \\
930331A &  119.1   &  38 (1/1)   &  0  &  0  &  0   \\
930425A &  29.2   &  30 (1/1)   &  0  &  0  &  0   \\
930916B &  74.3   &  4 (1/1)   &  0  &  0  &  0   \\
931106  &  152.1   &  108 (1/1)  &  0  &  0  &  0   \\
931221A &  57.9    &  29 (1/1)   &  0  &  0  &  0   \\
940306 &  42.6   &  47 (1/1)   &  0  &  0  &  0   \\
940520 &  32.8   &  2.5 (1/1)   &  0  &  0  &  0   \\
940529D &  37.6    &  42.5 (1/1)   &  0  &  0  &  0   \\
941020B &  56   &  28 (1/1)   &  0  &  0  &  0   \\
950111B &  46.3    &  48 (1/1)   &  0  &  0  &  0   \\
950403A &  14    &  12 (1/1)   &  0  &  0  &  0   \\
950425 &  59.1   &  42.5 (1/1)    &  0  &  0  &  0   \\
951202 &  28.5   &  34 (1/1)   &  0  &  0  &  0   \\
960114  &  36.5    &  32.5 (1/1)  &  0  &  0  &  0   \\
960807  &  12.7    &  6 (1/1)    &  0  &  0  &  0   \\
961102 &  71.4   &  85 (1/1)   &  0  &  0  &  0   \\
970202 &  26.7   &  34 (1/1)   &  0  &  0  &  0   \\
970223 &  16.3   &  16 (1/1)   &  0  &  0  &  0   \\
970807B &  37.6    &  7  (1/1)   &  0  &  0  &  0   \\
970912B &  65.6    &  37 (1/1)   &  0  &  0  &  0   \\
971029A &  89.9    &  23 (1/1)   &  0  &  0  &  0   \\
971220A &  13.6   &  15 (1/1)   &  0  &  0  &  0   \\
980124A &  45.1    &  32 (1/1)   &  0  &  0  &  0   \\
980225  &  127.7   &  81 (1/1)   &  0  &  0  &  0   \\
980329A &  18.5    &  20 (1/1)    &  0  &  0  &  0   \\
991121  &  112.2   &  71 (1/1)   &  0  &  0  &  0   \\
000302A  &  22.7   &  28.5 (1/1)   &  0  &  0  &  0   \\

\hline
%\newpage

910425  &  90.2   &  106 (1/1)   &  9.5 (0.65/1)&  4 (0.41/1) &  0   \\
910430  &  62.0    &  35 (1/1)   &  6 (0.11/1) &  0  &  0   \\
910601  &  28.5    &  20 (1/1)   &  4.5 (0.36/1)&  0  &  0   \\
910614  &  146.9   &  83  (0.65/1)  &  33 (1/1)&  2.5 (0.02/0.98)  &  0   \\
910619  &  106.1   &  67 (1/1)   &  11 (0.06/1)&  2 (0.05/1)  &  0   \\
910814A &  77.8    &  62 (1/1)   &  10 (0.3/1)&  0  &  0   \\
910905  &  81.5    &  58 (1/1)   &  18 (0.64/0.96) &  9 (0.77/1) &  0   \\
911127A  &  18.8   &  18 (1/1)   &  2 (0.046/1)&  0  &  0   \\
911202A  &  20.1   &  22 (1/1)   &  7 (0.2/0.99) &  0  &  0   \\
920110A  &  318.6   &  225.5 (1/1)   &  56.5 (0.33/1)&  11.5 (0.1/1) &  3 (0.027/1)  \\
920210B  &  51.8   &  60 (1/1)   &  15 (0.25/0.97)&  0  &  0   \\
920308A &  51.1    &  11 (1/1)   &  5 (0.53/1) &  1 (0.15/1) &  0.5 (0.03/0.99)  \\
920513  &  88.6    &  70 (1/1)   &  4 (0.74/1) &  2 (0.42/1) &  0   \\
920525B &  16.1    &  8  (1/1)   &  4 (0.67/1) &  0  &  0   \\
920617B &  67.7    &  34 (0.03/1)   &  27 (1/1) &  0  &  0   \\
920627B &  52.8    &  26.5 (1/1) &  4 (0.11/1) &  3 (0.1/1)  &  1 (0.17/1)  \\
921015  &  272.4   &  108 (0.6/1)  &  48 (1/1)&  0  &  0   \\
921118  &  174.7   &  78 (1/1)   &  10 (0.18/1)&  0  &  0   \\
921206B &  53.8    &  21 (1/1)   &  3 (0.46/1) &  0  &  0   \\
921209B &  38.1    &  9.5 (1/1)  &  3 (0.76/1) &  0  &  0   \\
921230A  &  18.8   &  16 (1/1)   &  2 (0.17/0.9)&  0  &  0   \\
930309A  &  90.1   &  72 (0.83/1)   &  7 (1/1)&  1 (0.8/1) &  0   \\
930506B  &  22.1   &  22 (1/1)   &  5.5 (0.03/1)&  2 (0.15/1) &  0   \\
930720A  &  45.9   &  26 (0.7/1)   &  5 (0.5/1)&  2.5 (1/1) &  0   \\
930910C &  83.1    &  52 (1/1)    &  4 (0.03/1) &  3 (0.07/1) &  1 (0.11/1)  \\
931026  &  134.7   &  142 (1/1)   &  6.5 (0.04/1) &  0  &  0   \\
940128B &  45.2    &  18 (1/1)   &  7 (0.2/1)&  0  &  0   \\
940210  &  30.7    &  12 (1/1)   &  1.5 (0.16/1)&  1 (0.28/1) &  0   \\
940228A &  33.3    &  21 (0.3/1)   &  8 (1/0.93) &  4 (0.65/1) &  2 (0.4/1)  \\
940301  &  42.5   &  42.5 (1/1)   &  2 (0.05/1) &  0  &  0   \\
940302  &  119.9   &  67 (0.82/1)   &  12 (0.59/1) &  2 (0.21/1) &  0   \\
940319  &  75.9    &  60 (1/1)   &  10 (0.3/1) &  2.5 (0.08/1)&  1 (0.11/1)   \\
910321  &  51.6   &  16 (1/1)    &  3 (0.2/1)& 0.5 (0.15/1)  &  0   \\
940323  &  60.7    &  6 (1/1)    &  2.5 (0.19/1) &  0  &  0   \\
940414B &  42.8    &  13.5 (1/1) &  4 (0.2/1) &  1 (0.16/1)  & 0   \\
940619  &  88.4   &  56 (0.34/1)   &  31.5 (0.67/1)&  20 (1/0.98)  &  5 (0.35/1)  \\
940703A  &  34.9   &  30 (1/1)   &  15 (0.4/1) &  3.5 (0.12/1) &  0.5 (0.03/1)   \\
940806D  &  10.2    &  3.5 (0.5/1)  &  0.5 (1/1) &  0  &  0   \\
940817  &  32.2   &  34 (0.62/1)   &  19 (1/1)  &  8.5 (0.5/1)  &  2.5 (0.7/1)   \\
\enddata

\end{deluxetable*}

\clearpage

\addtocounter{table}{-1}
\begin{deluxetable*}{lcccccc}

\tablecolumns{8} \tablewidth{0pc}

\tablecaption{Continued}

\tablehead{ \colhead{GRB} & \colhead{$T_{\rm 90}$} &
\colhead{$T_{\rm 1}$\tablenotemark{a}(s/c)}  & \colhead{$T_{\rm
2}$\tablenotemark{b}(s/c)} & \colhead{$T_{\rm
3}$\tablenotemark{c}(s/c)} & \colhead{$T_{\rm
4}$\tablenotemark{d}(s/c)}
 }

\startdata

941014A &  45.4    &  23 (1/1)   &  11 (0.65/1) &  6 (0.15/1)  &  4 (0.25/1)   \\
941017A &  77.1   &  85 (1/1)  &  7 (0.05/1)&  4 (0.03/1) &  1.5 (0.09/1)  \\
941023A &  34.9    &  22 (1/1)   &  11 (0.2/1)&  5.5 (0.34/1)&  0   \\
941119  &  33.4    &  24 (1/1)   &  5 (0.06/1)  &  2.5 (0.09/1)&  0   \\
941126E &  36.1    &  13 (1/1)   &  1 (0.03/0.99)  &  0  &  0   \\
950208  &  58.6    &  18.5 (1/1)  &  7 (0.24/1)  &  1 (0.21/1)  &  0   \\
950211B &  54.3    &  34 (0.43/1)   &  14 (1/1)&  3 (0.43/1)  &  0   \\
950608  &  142.0   &  101 (0.77/1)   &  45 (1/1) &  2.5 (0.15/0.97) &  0 \\
950701B &  10.6    &  7 (0.95/1)    &  4 (1/1)  &  0  &  0   \\
950706  &  68.9    &  27 (1/1)   &  14 (0.38/1) &  0  &  0   \\
950909  &  65.7    &  21 (1/1)   &  7 (0.2/1)  &  0  &  0   \\
951011  &  31.5    &  28 (1/1)   &  4 (0.55/1)  &  2.5 (0.75/0.98)&  0   \\
951219  &  58.8   &  21 (1/1)   &  4 (0.8/1)&  1.5 (0.15/1) &  0   \\
960322A  &  22.8   &  25 (1/1)   &  2.5 (0.14/1) &  0  &  0   \\
960524C  &  80.6   &  64 (0.84/1)   &  23 (1/1) &  3.5 (0.28/0.93) &  1 (0.3/1)   \\
960607B &  140.5    &  112 (0.11/1)   &  88.5 (1/1) &  14 (0.69/1)&  2 (0.11/0.91)   \\
960824  &  229.9   &  82 (1/1)   &  6.5 (0.05/1) &  2.5 (0.07/0.9) &  0   \\
961228C &  60.0    &  34 (0.54/1)   &  12 (0.4/1)&  4 (1/1) &  0   \\
970111  &  31.5    &  13 (0.07/1)    &  9 (1/1) &  0  &  0   \\
970306  &  122.5   &  34.5 (0.52/1)   &  14 (0.18/1) &  6 (1/1) &  0   \\
970315B  &  16.8   &  6.5 (0.98/1)    &  3.5 (0.1/1) &  2 (1/1) &  1 (0.76/1)   \\
970411  &  58.9   &  53.5 (1/1)   &  3.5 (0.06/1) &  0  &  0   \\
970420  &  10.5    &  8 (1/1)    &  2 (0.81/1)  &  1 (0.5/1) &  0   \\
970612B &  37.6    &  24  (0.07/1)  &  13 (1/1)&  2.5 (0.4/1)&  0   \\
970816  &  6.5     &  5 (1/1)    &  1.5 (0.36/1)&  0  &  0   \\
970831 &  114.5    &  126 (1/1)   &  28 (0.34/0.94) &  10 (0.11/1)  &  0   \\
971110  &  195.2   &  123 (1/1)  &  55 (0.21/1) &  28 (0.95/1) &  9 (0.53/1)  \\
971207C &  48.3    &  38 (0.74/1)   &  5.5 (1/1)&  0  &  0   \\
980105  &  36.8    &  10 (0.42/1)   &  6.5 (1/1)&  1 (0.24/1)  &  0   \\
980203B  &  23.0   &  6 (1/1)   &  2 (0.42/1) &  1.5 (0.9/1)  &  0.5 (0.42/1)   \\
980208B &  31.2    &  12 (1/1)   &  6.5 (0.42/0.96)&  1.5 (0.16/1) &  0   \\
980315B &  105.0   &  74 (0.93/1)   &  23.5 (1/1)& 2 (0.13/0.96) &  0   \\
980703B &  108.4   &  77 (1/1)   &  17 (0.25/1)&  0  &  0   \\
980803  &  19.8   &  21 (0.03/1)   &  7.5 (1/1) &  1.5 (0.1/1)  &  0.5 (0.29/1)   \\
980923  &  33.0    &  15 (1/1)   &  5 (0.44/1) &  2 (0.39/1) &  1 (0.15/1)   \\
990108  &  145.7   &  58 (0.04/0.98)   &  36 (1/1)&  3 (0.05/0.99)  &  0   \\
990111A  &  15.0   &  14 (1/1)   &  1.5 (0.12/1) &  0  &  0   \\
990123A &  63.4    &  56 (1/1)   &  16 (0.74/1) &  0  &  0   \\
990316B &  100.5   &  89 (1/1)   &  18 (0.14/1)&  2 (0.05/1)  &  1 (0.05/1)   \\
990323C  &  49.5   &  17.5 (1/1)    &  5 (0.18/1) &  2 (0.46/1) &  0   \\
990728  &  42.8    &  15 (1/1)   &  3 (0.39/1) &  0  &  0   \\
990803  &  19.4   &  1 (1/1)   &  0.5 (0.45/0.99) &  0  &  0   \\
991004D  &  77.4   &  39 (0.69/1)   &  9.5 (1/1) &  0.5 (0.22/1)  &  0   \\
991009  &  131.6   &  83 (1/1)   &  23.5 (0.29/1) &  6 (0.11/1)  &  0   \\
991113  &  61.4    &  13 (1/1)    &  3 (0.76/1)&  0  &  0   \\
991127  &  52.7    &  8 (1/1)    &  1.5 (0.39/1)&  0  &  0   \\
991216  &  15.2   &  16 (1/1)   &  3 (0.55/1)&  1 (0.11/1) &  0.5 (0.1/1)  \\
000101  &  51.8    &  33 (0.8/1)   &  7 (1/1)  &  0  &  0   \\
000103  &  67.4    &  21 (0.79/1)   &  10 (1/1) &  0  &  0   \\
000201A &  95.0    &  42 (1/1)   &  13.5 (0.19/1)& 6 (0.08/1)  &  0   \\
000221  &  26.2    &  12 (1/1)   &  0.5 (0.05/1)  &  0  &  0   \\
000511A &  115.0   &  73 (1/1)    &  14.5 (0.1/1)& 0  &  0   \\

\enddata

\tablenotetext{a} {Characteristic timescale corresponding to the
first dip}

\tablenotetext{b} {Characteristic timescale corresponding to the
second dip}

\tablenotetext{c} {Characteristic timescale corresponding to the
third dip}

\tablenotetext{d} {Characteristic timescale corresponding to the
fourth dip}

\end{deluxetable*}

%%%%%%%%%%%%%%%%%%%%%%%% Figure %%%%%%%%%%%%%%%%%%%%%%%%%%%%%%
\clearpage
\begin{figure}[h!!!]
\begin{minipage}[b]{0.35\textwidth}
\centering
\includegraphics[width=2in]{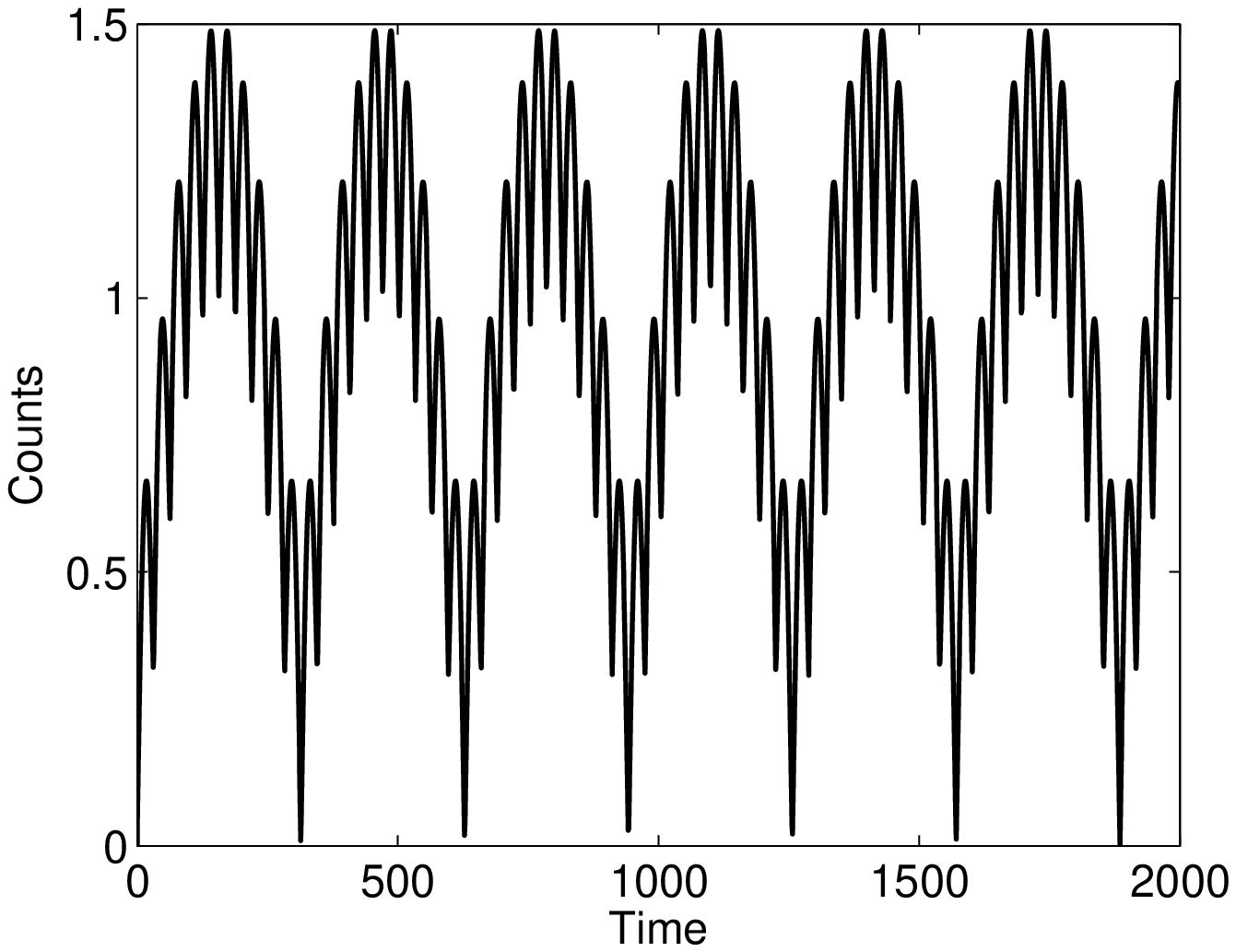}
\end{minipage}%
\begin{minipage}[b]{0.35\textwidth}
\centering
\includegraphics[width=2in]{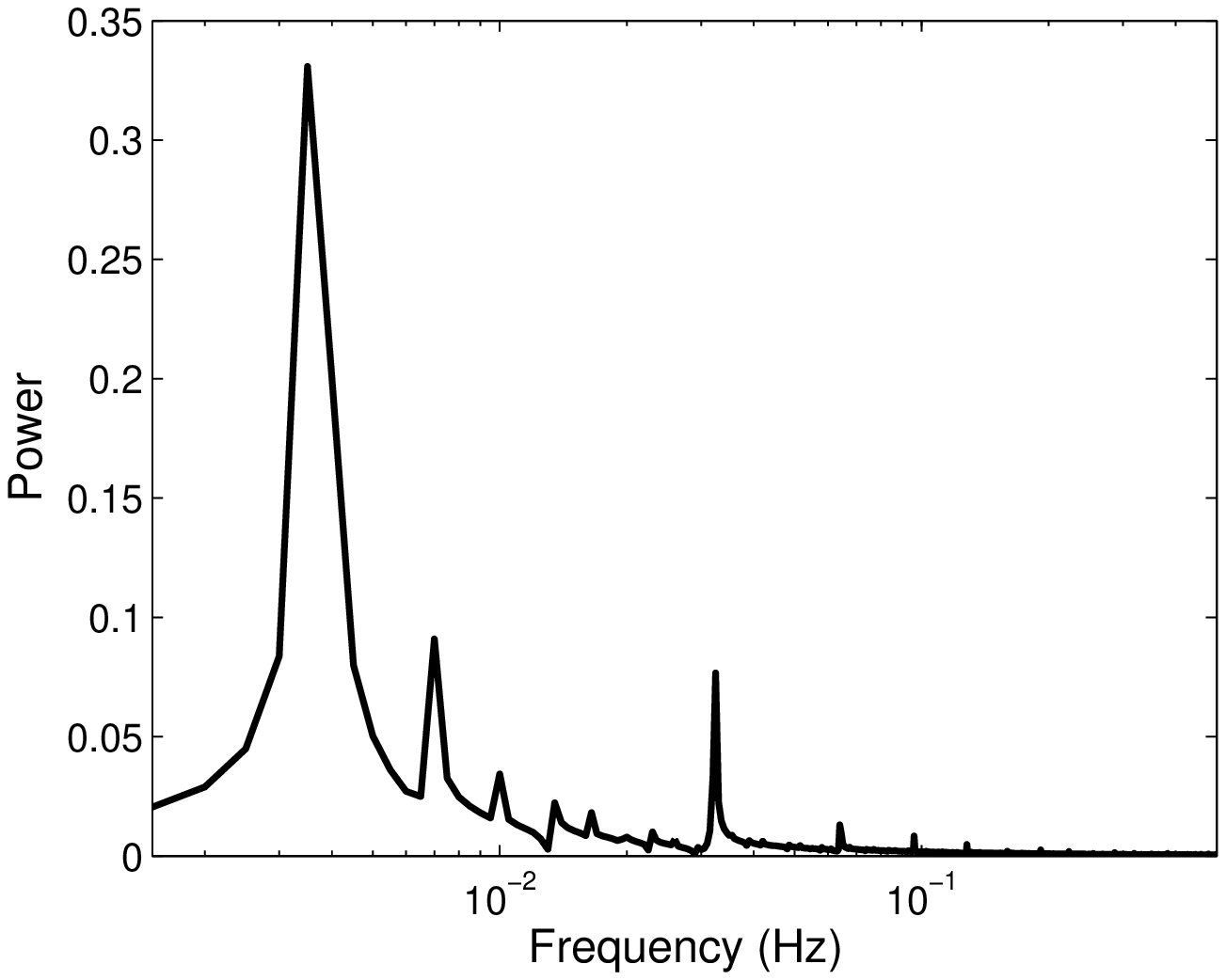}
\end{minipage}%
\begin{minipage}[b]{0.35\textwidth}
\centering
\includegraphics[width=2in]{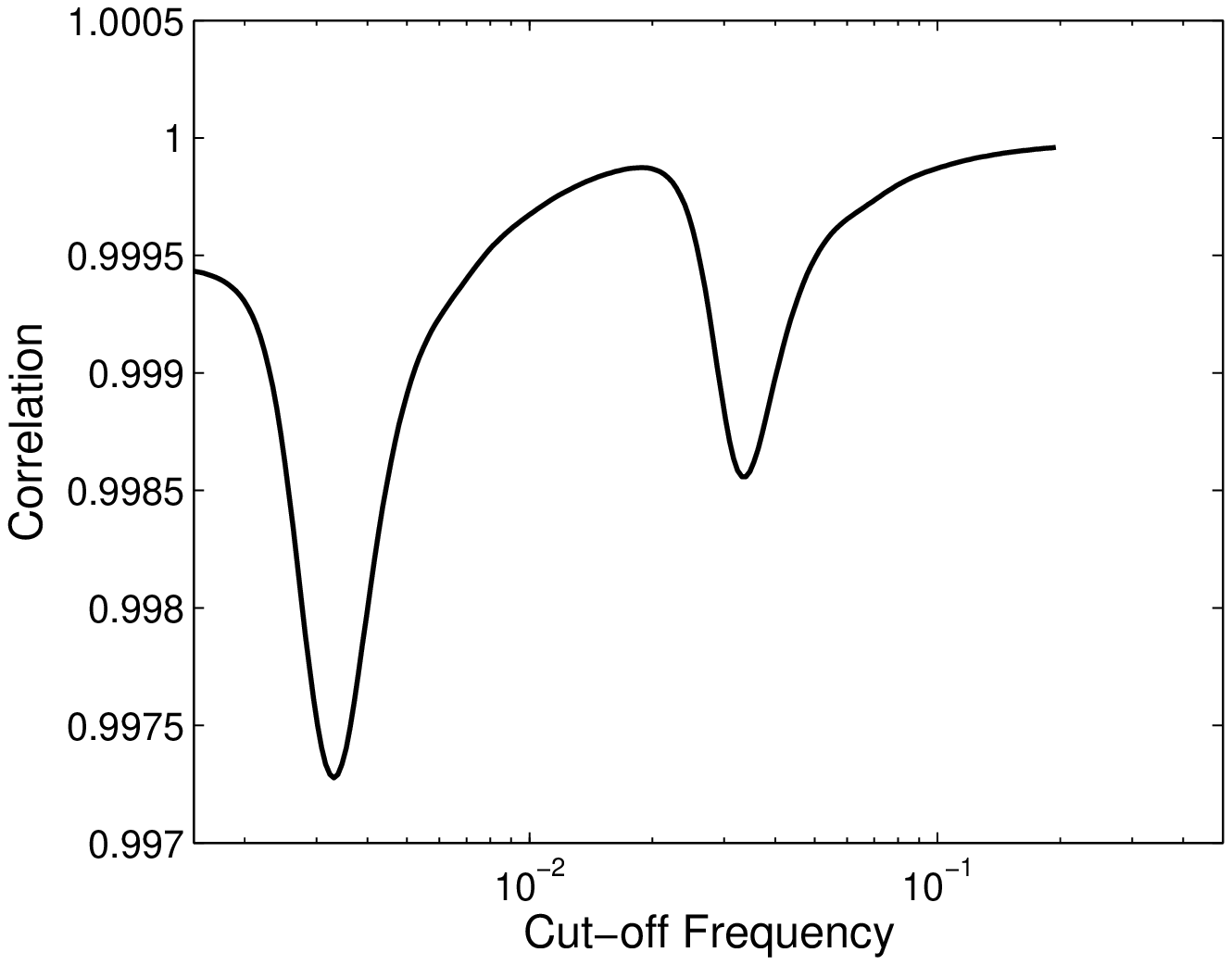}
\end{minipage}\\
\begin{minipage}[b]{0.35\textwidth}
\centering
\includegraphics[width=2in]{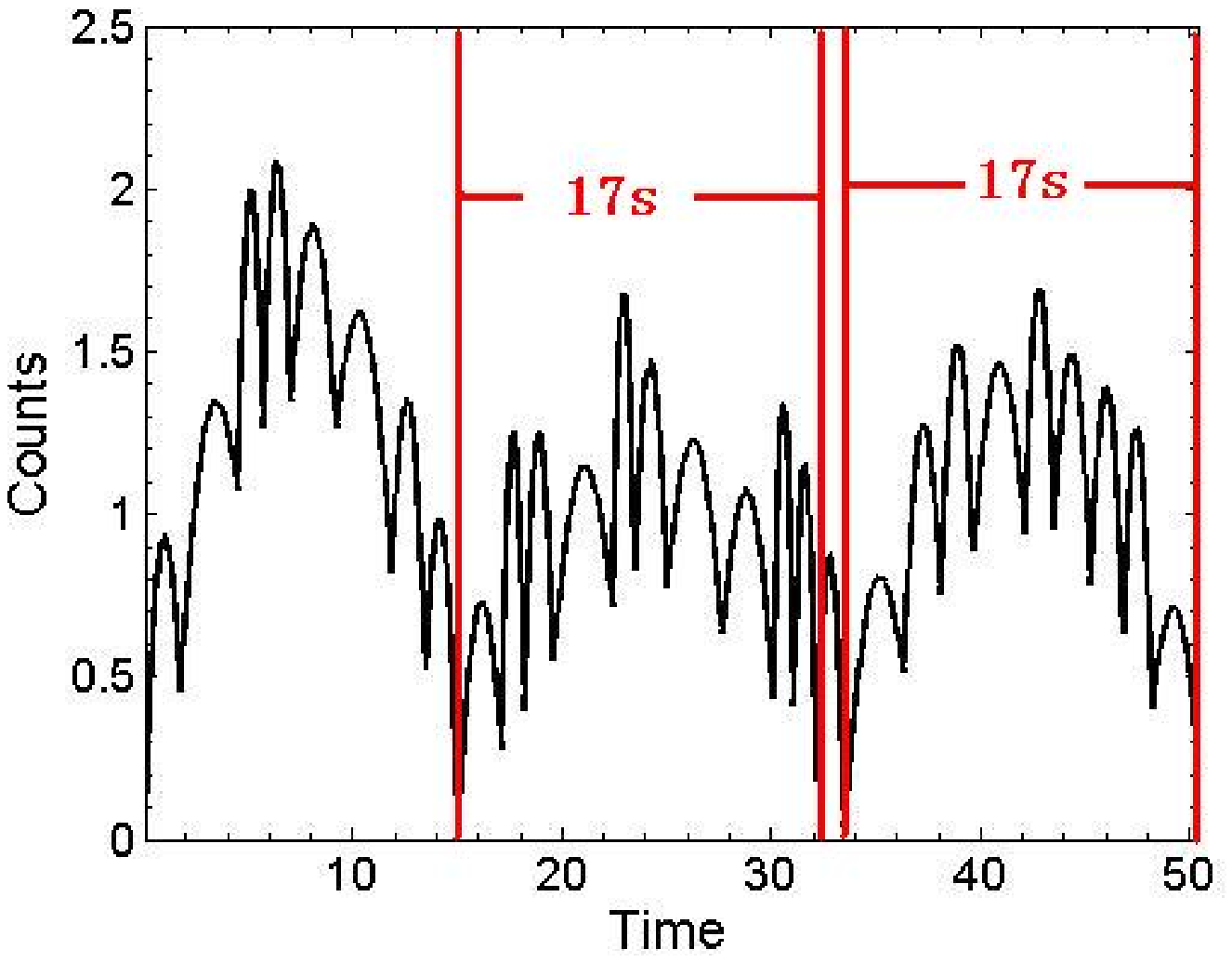}
\end{minipage}%
\begin{minipage}[b]{0.35\textwidth}
\centering
\includegraphics[width=2in]{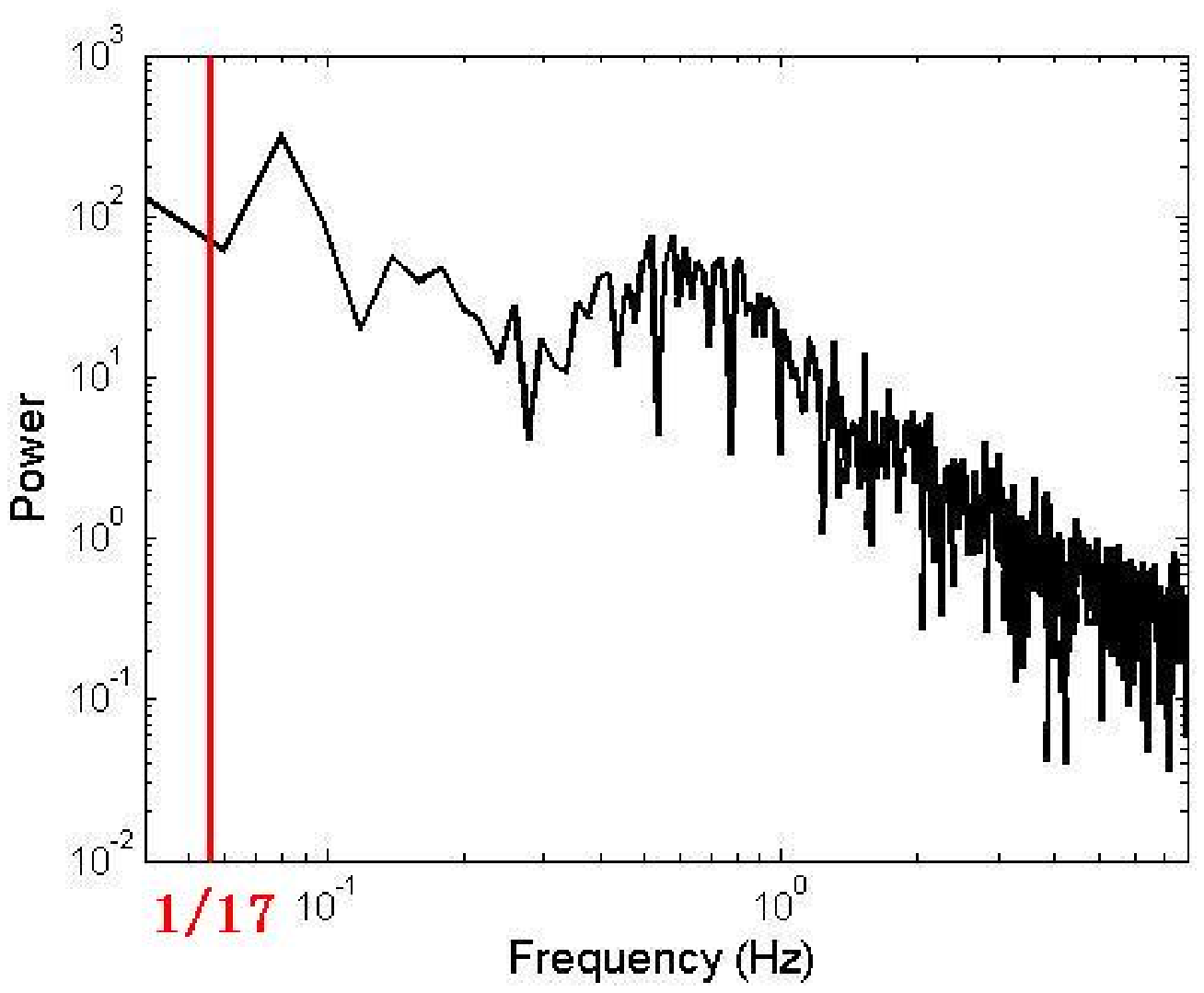}
\end{minipage}%
\begin{minipage}[b]{0.35\textwidth}
\centering
\includegraphics[width=2in]{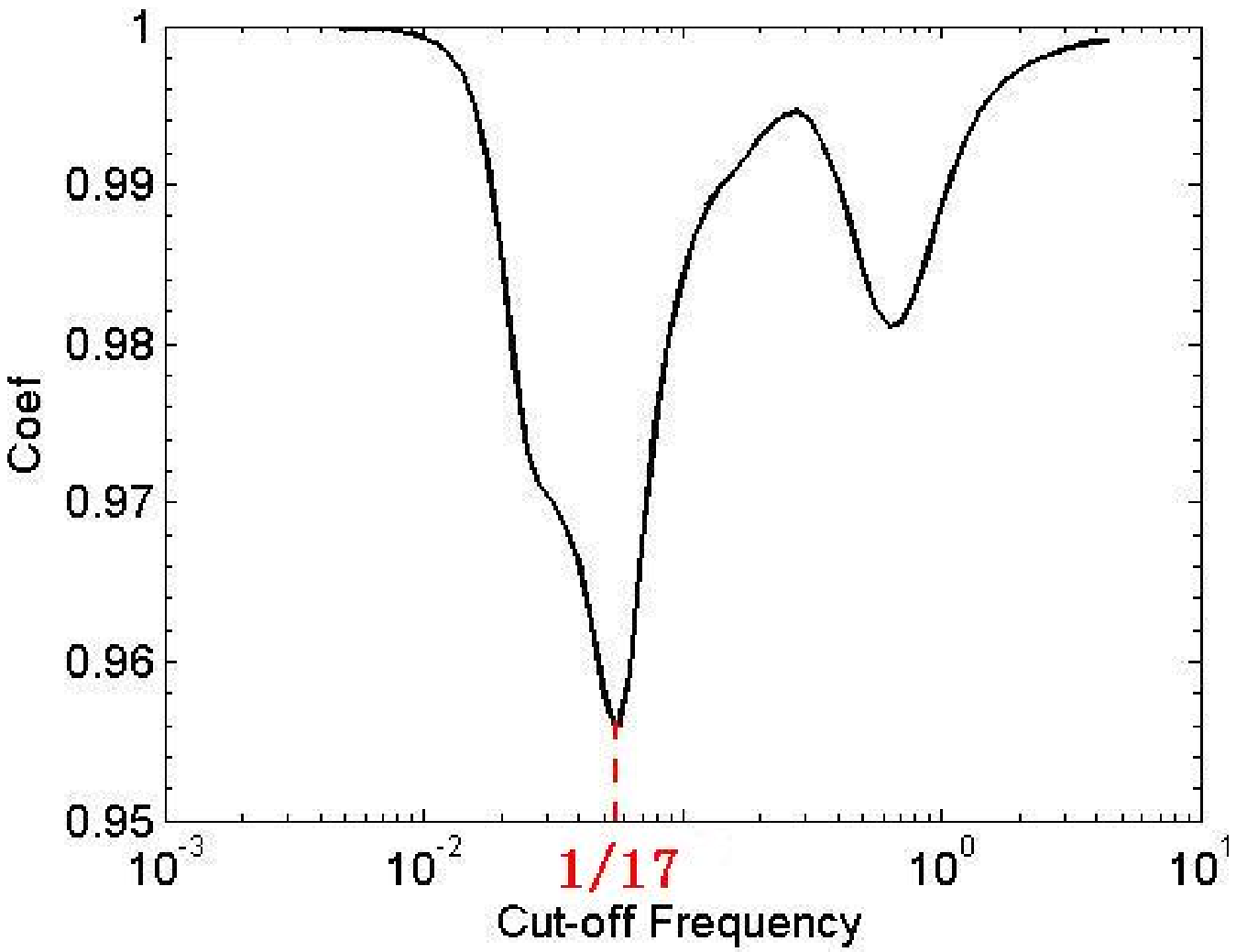}
\end{minipage}\\
\begin{minipage}[b]{0.35\textwidth}
\centering
\includegraphics[width=2in]{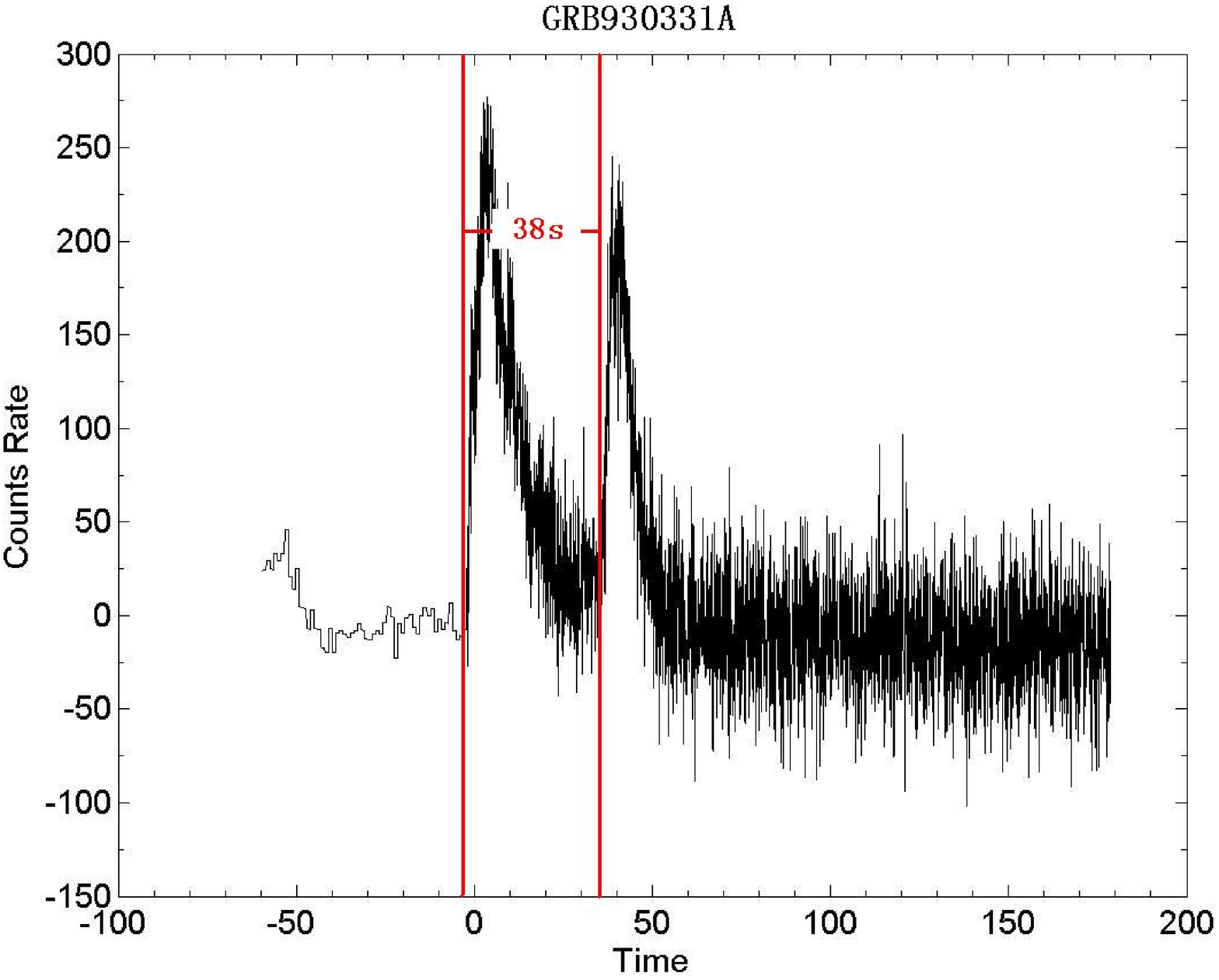}
\end{minipage}%
\begin{minipage}[b]{0.35\textwidth}
\centering
\includegraphics[width=2in]{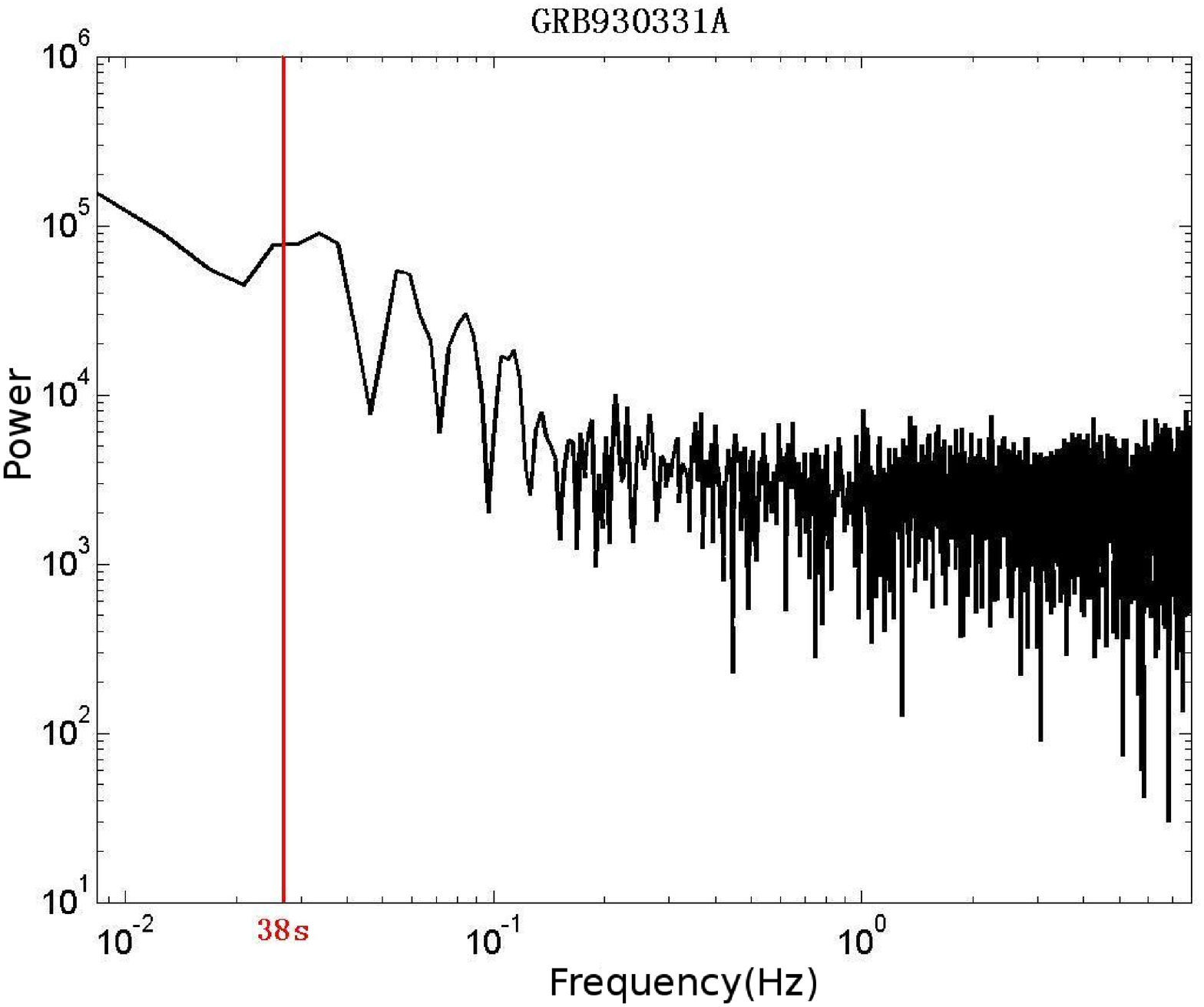}
\end{minipage}%
\begin{minipage}[b]{0.35\textwidth}
\centering
\includegraphics[width=2in]{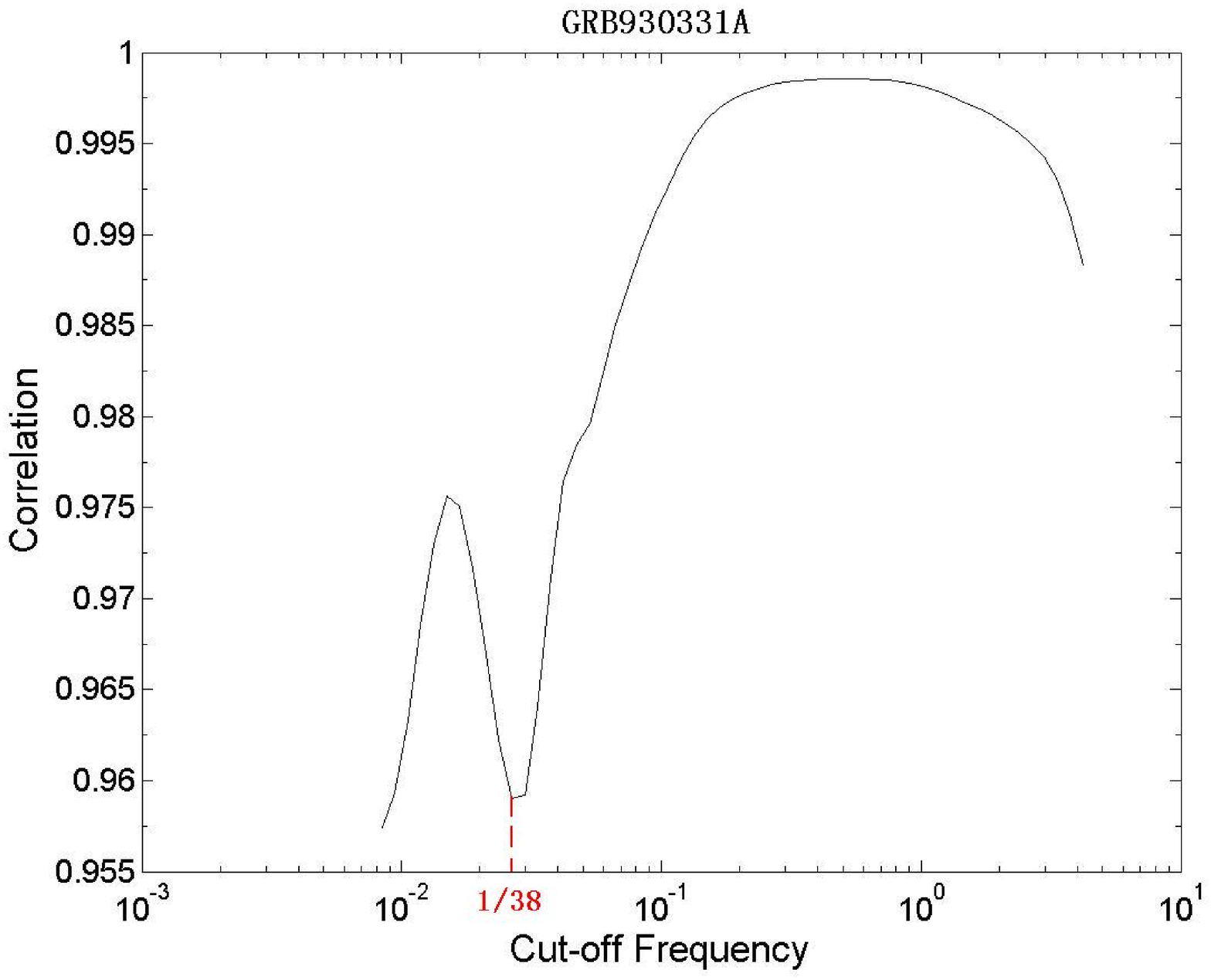}
\end{minipage}%
       \caption{Examples that prove the validity of the SFC method.
{\em Top panel:} The first simulation test; {\em Middle panel}: the
second simulation test; {\em Lower panel:} a real GRB 930331A). In
all three panels, the left figure is the simulated or real
lightcurve, the middle figure is power density spectrum, and the
right figure is the correlation curve, i.e. the correlation
coefficient $R_i$ versus the cutoff frequency $f_{\rm c,i}$.}
% $Coef_{\rm i}$ versus $Cutf_{\rm i}$ curve. The middle panel is for the second simulation
%      test, the order of figures is the same as above. The bottom panel is for a real GRB case, GRB\,930331A.}
           \label{Fig1}
            \end{figure}

\clearpage
\begin{figure}[h!!!]
\begin{minipage}[b]{0.5\textwidth}
\centering
\includegraphics[width=2.5in]{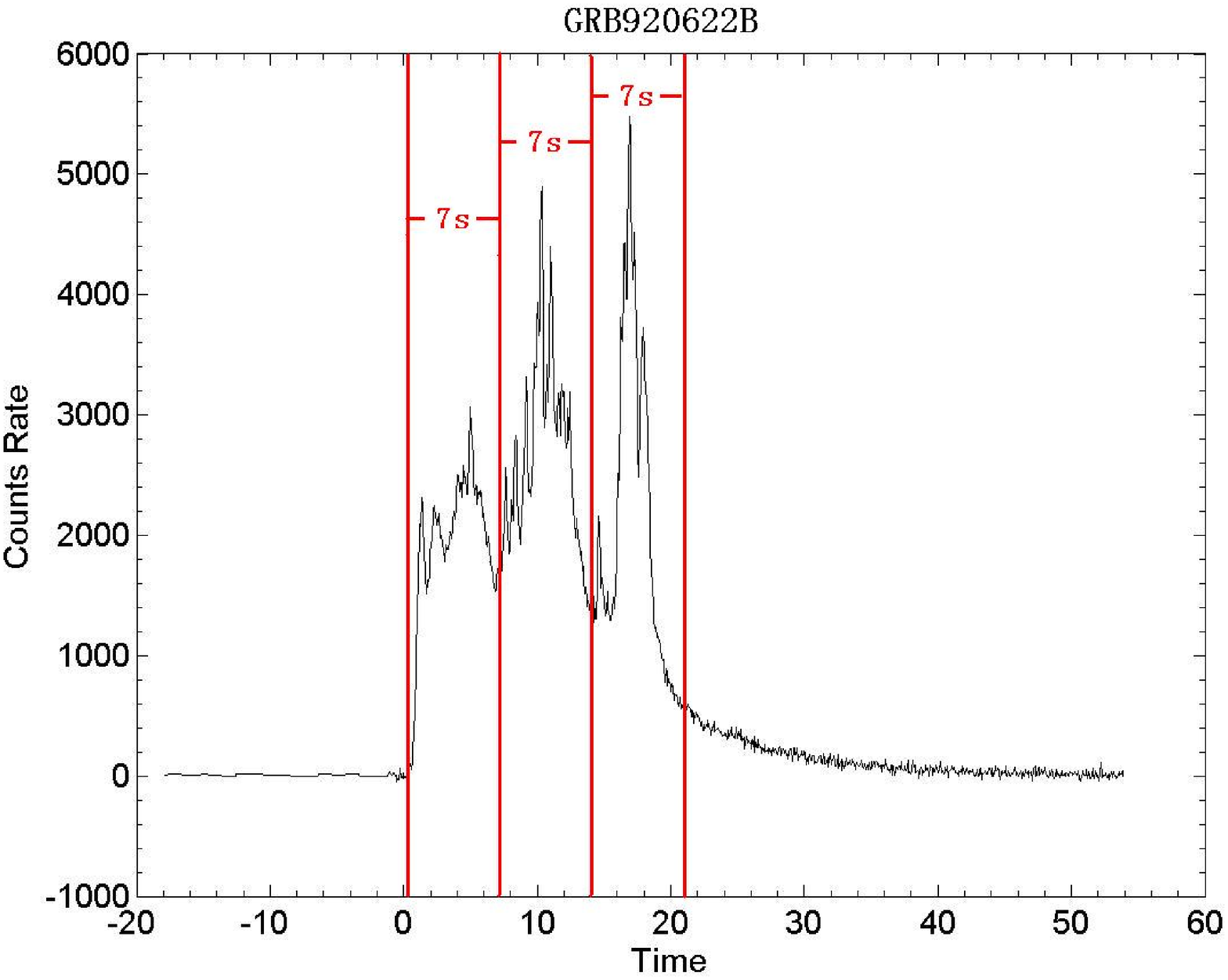}
\end{minipage}%
\begin{minipage}[b]{0.5\textwidth}
\centering
\includegraphics[width=2.5in]{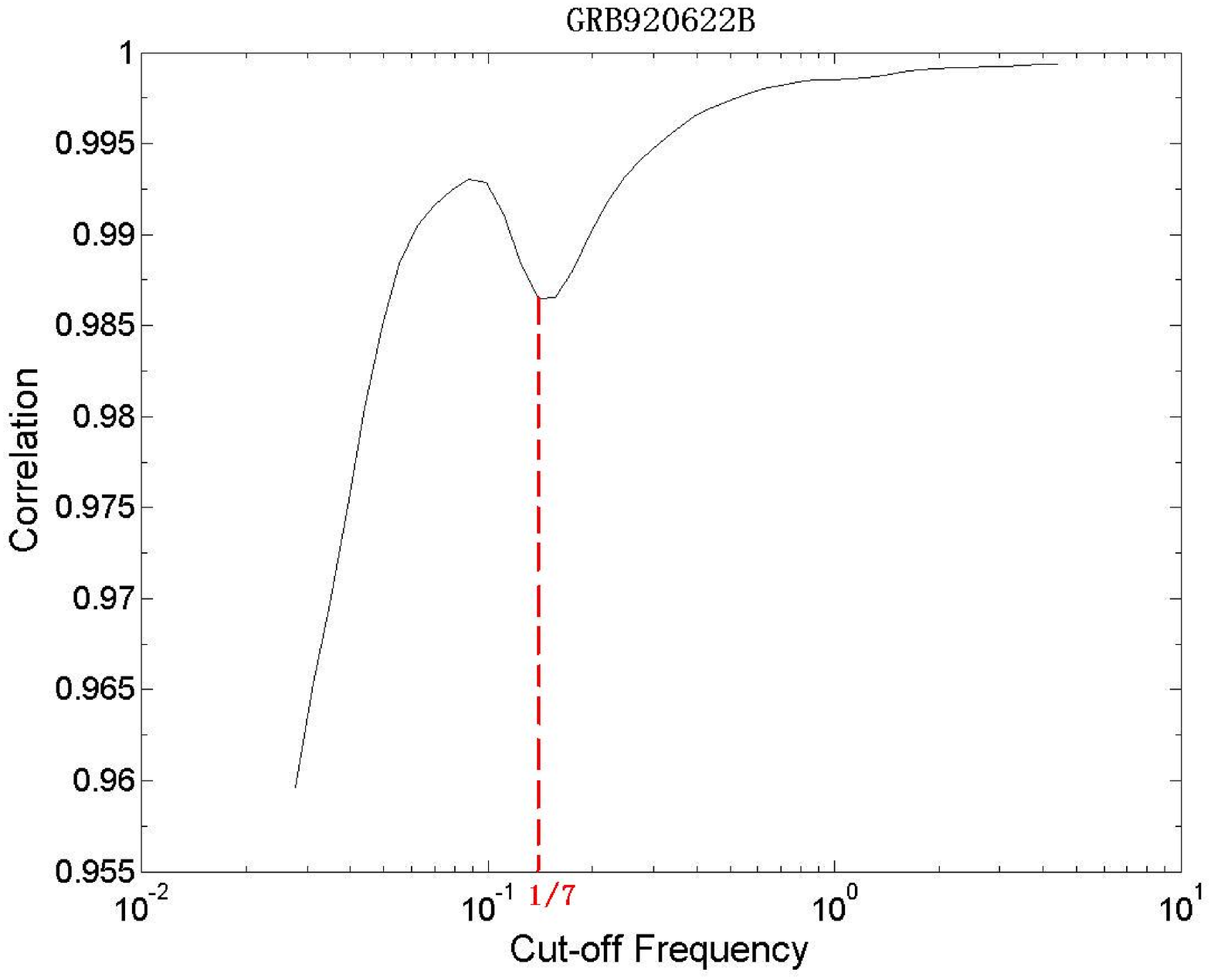}
\end{minipage}\\
\begin{minipage}[b]{0.5\textwidth}
\centering
\includegraphics[width=2.5in]{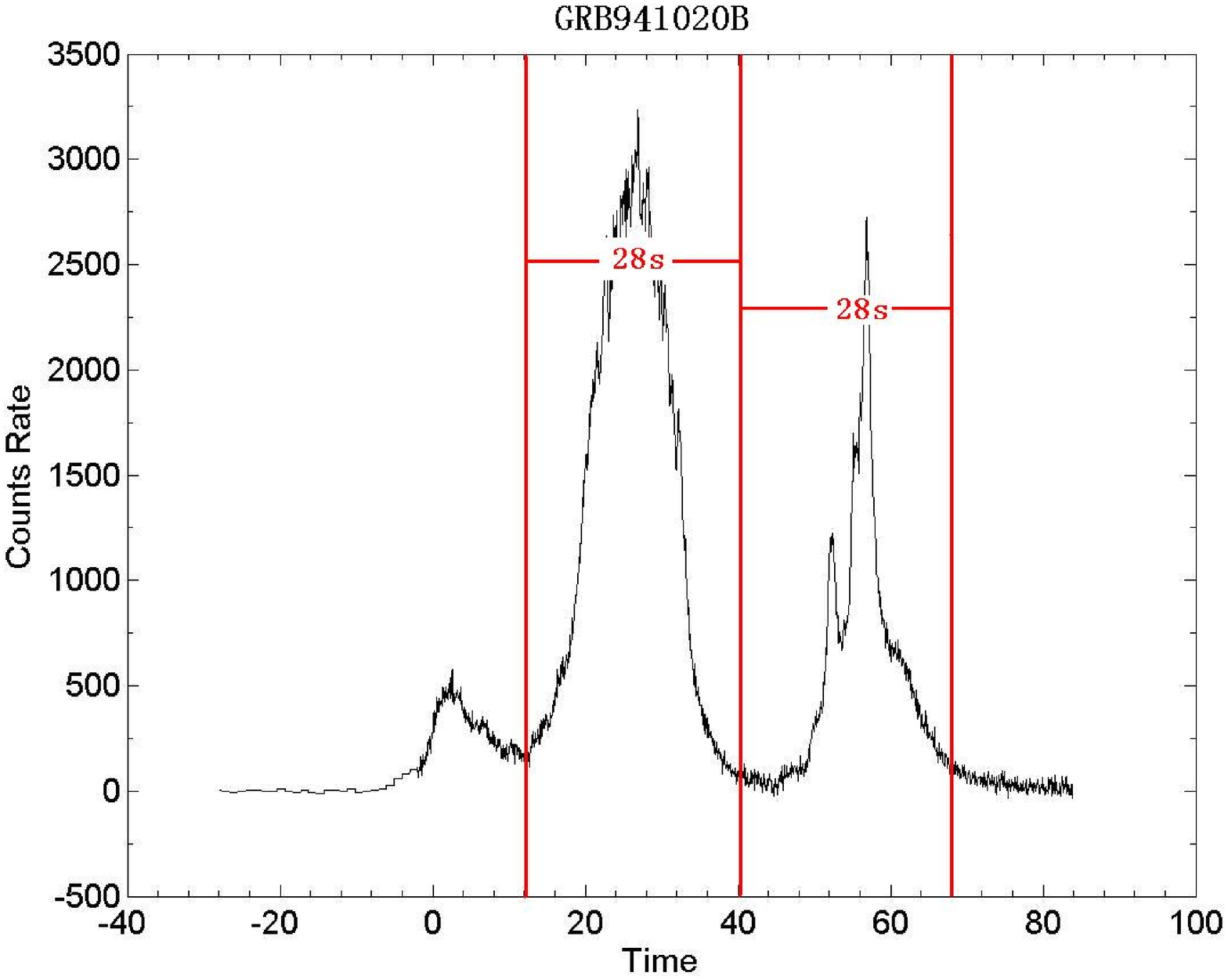}
\end{minipage}%
\begin{minipage}[b]{0.5\textwidth}
\centering
\includegraphics[width=2.5in]{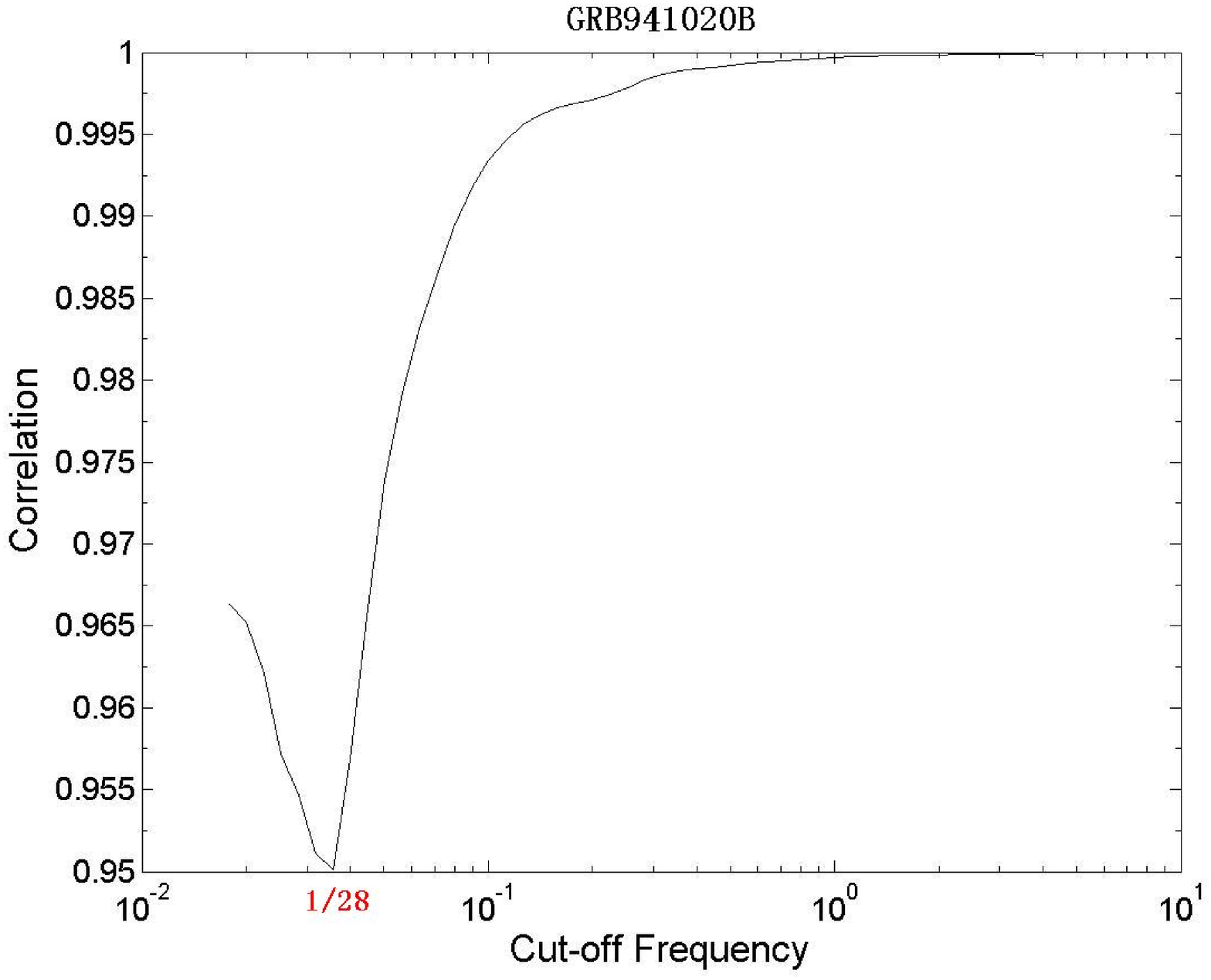}
\end{minipage}\\
\begin{minipage}[b]{0.5\textwidth}
\centering
\includegraphics[width=2.5in]{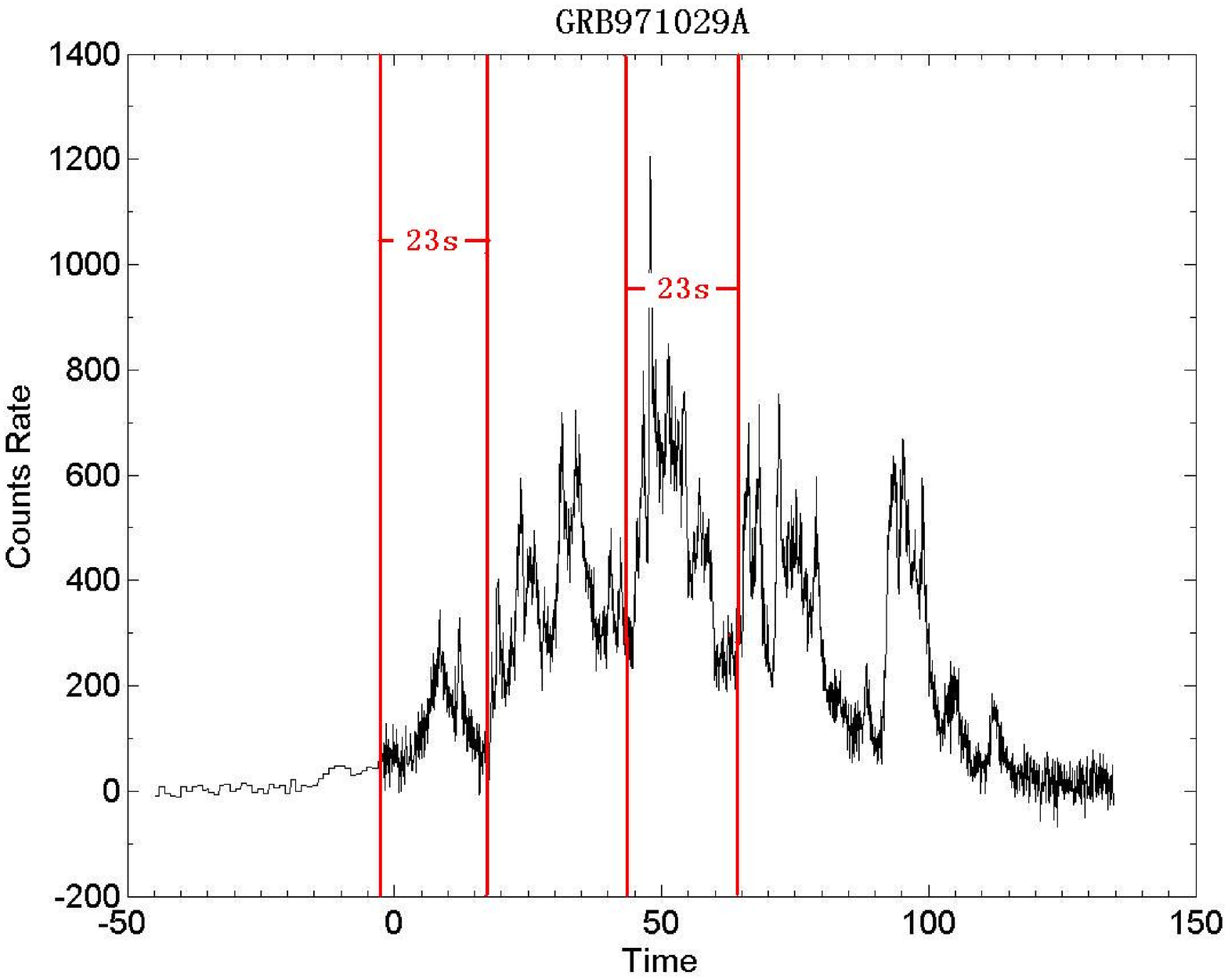}
\end{minipage}%
\begin{minipage}[b]{0.5\textwidth}
\centering
\includegraphics[width=2.5in]{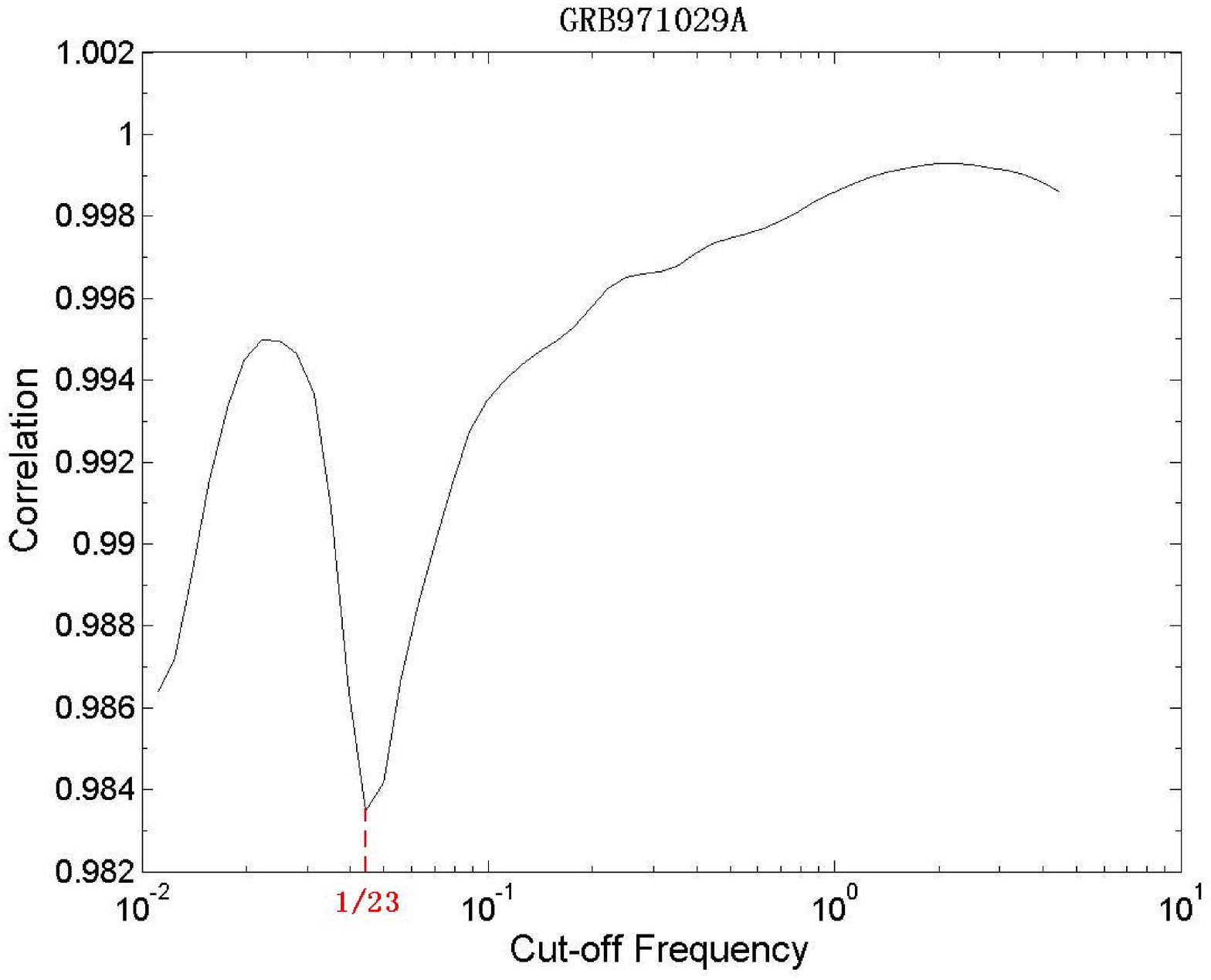}
\end{minipage}\\
       \caption{Examples for the one-dip only bursts. The left panel are the lightcurves, and the right panel are the correlation curves. The pulses that correspond to the identified frequencies are marked in the lightcurves. The time scales are rounded to the nearest 0.5.}
%for 31 GRBs whose $Coef_{\rm i}$ versus $Cutf_{\rm i}$ curve presents one dip. The left panel is LCs, the right panel
%        is $Coef_{\rm i}$ versus $Cutf_{\rm i}$ figures. We denote the dip corresponding slow component in the LCs with red lines and
%         the time resolution for the dip is 0.5s.}
           \label{Fig2}
            \end{figure}

\clearpage
\begin{figure}[h!!!]
\begin{minipage}[b]{0.5\textwidth}
\centering
\includegraphics[width=2.5in]{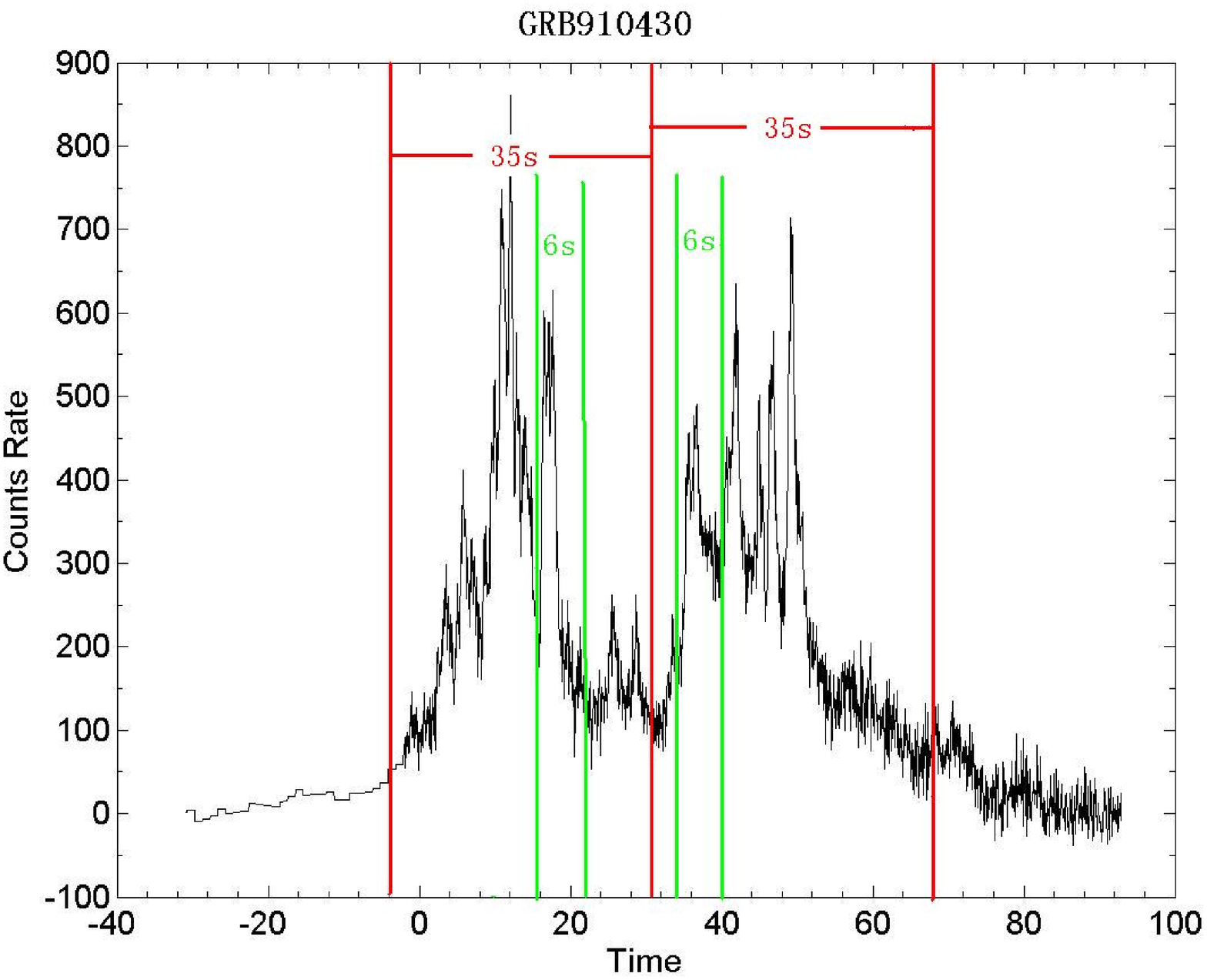}
\end{minipage}%
\begin{minipage}[b]{0.5\textwidth}
\centering
\includegraphics[width=2.5in]{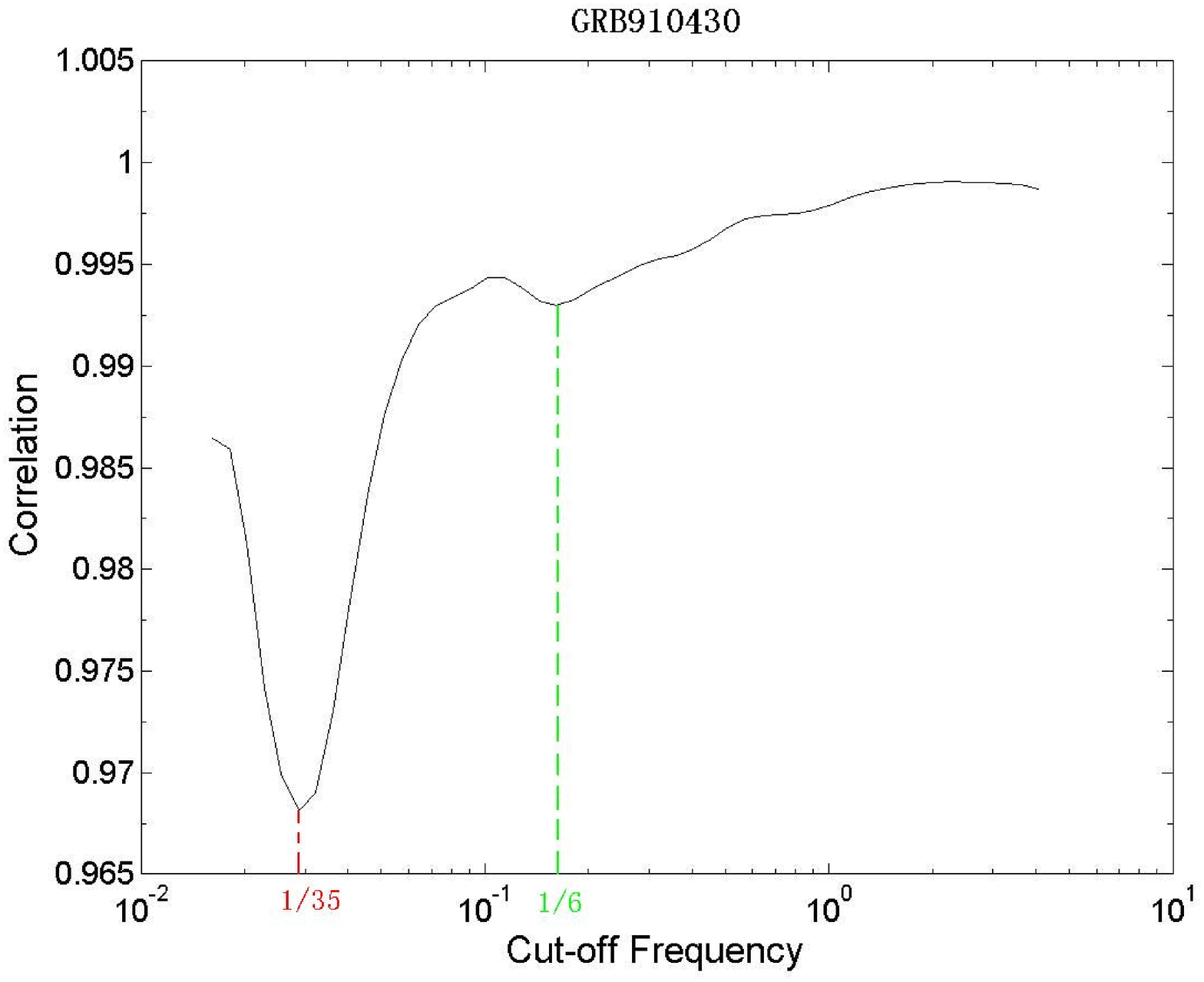}
\end{minipage}\\
\begin{minipage}[b]{0.5\textwidth}
\centering
\includegraphics[width=2.5in]{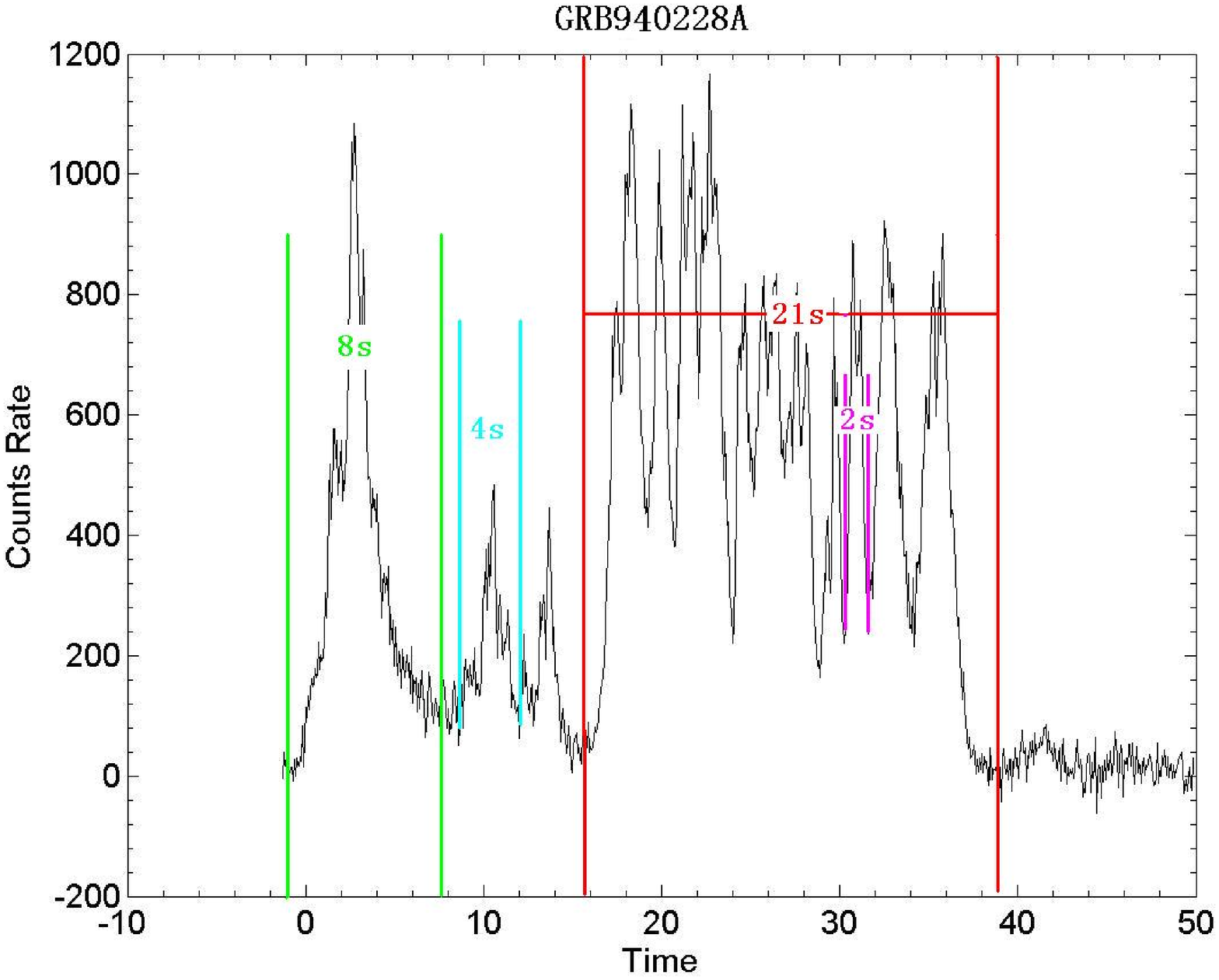}
\end{minipage}%
\begin{minipage}[b]{0.5\textwidth}
\centering
\includegraphics[width=2.5in]{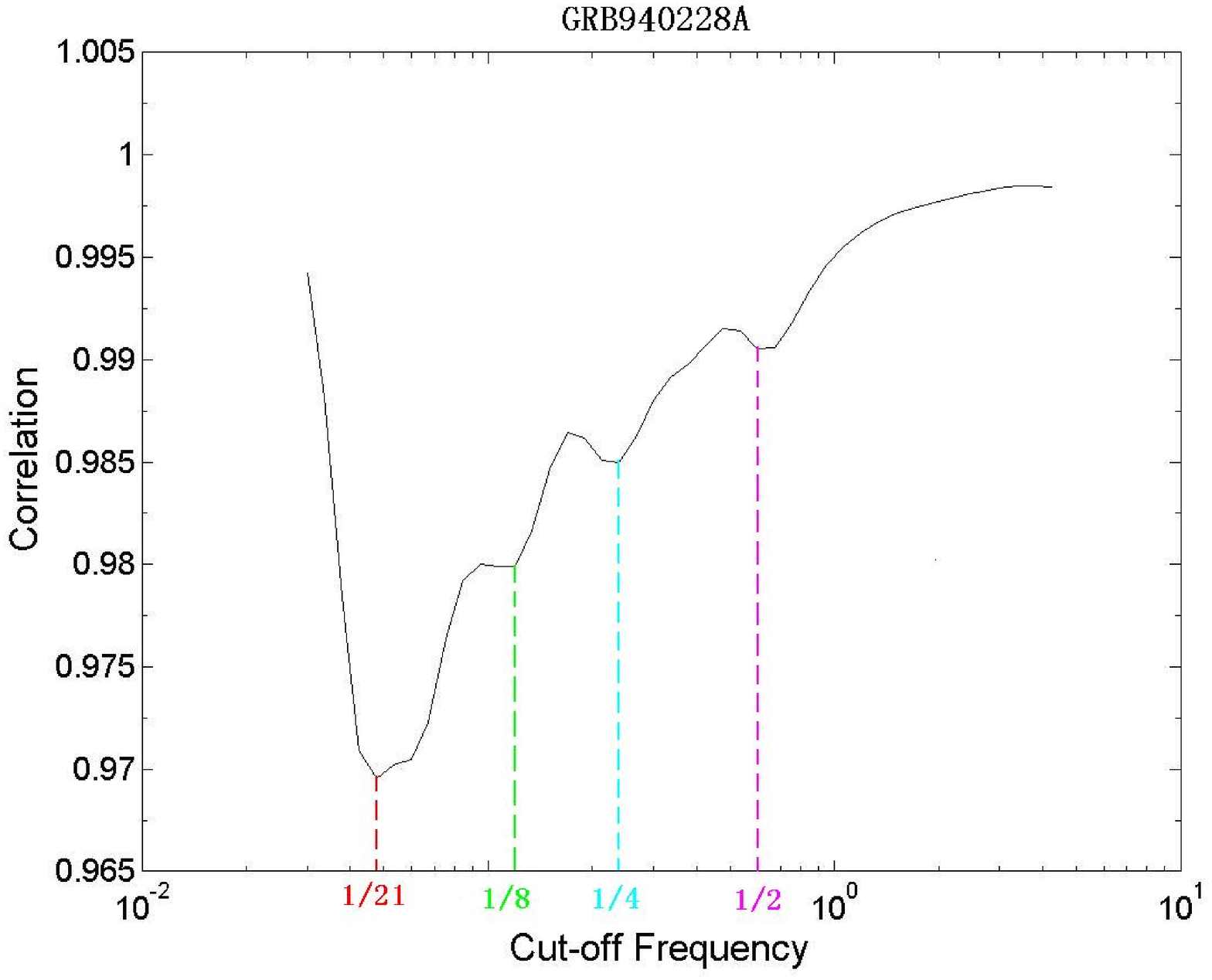}
\end{minipage}\\
\begin{minipage}[b]{0.5\textwidth}
\centering
\includegraphics[width=2.5in]{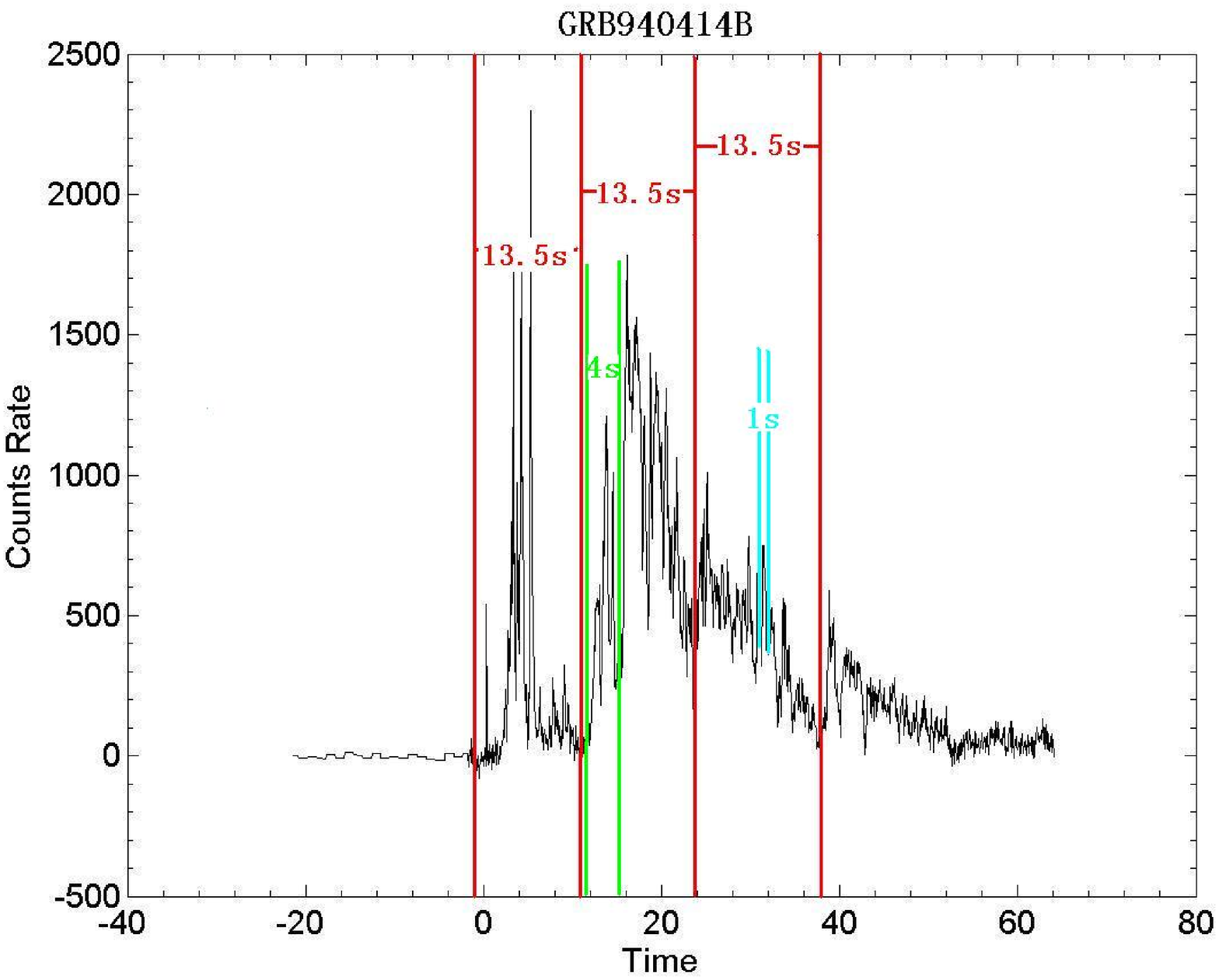}
\end{minipage}%
\begin{minipage}[b]{0.5\textwidth}
\centering
\includegraphics[width=2.5in]{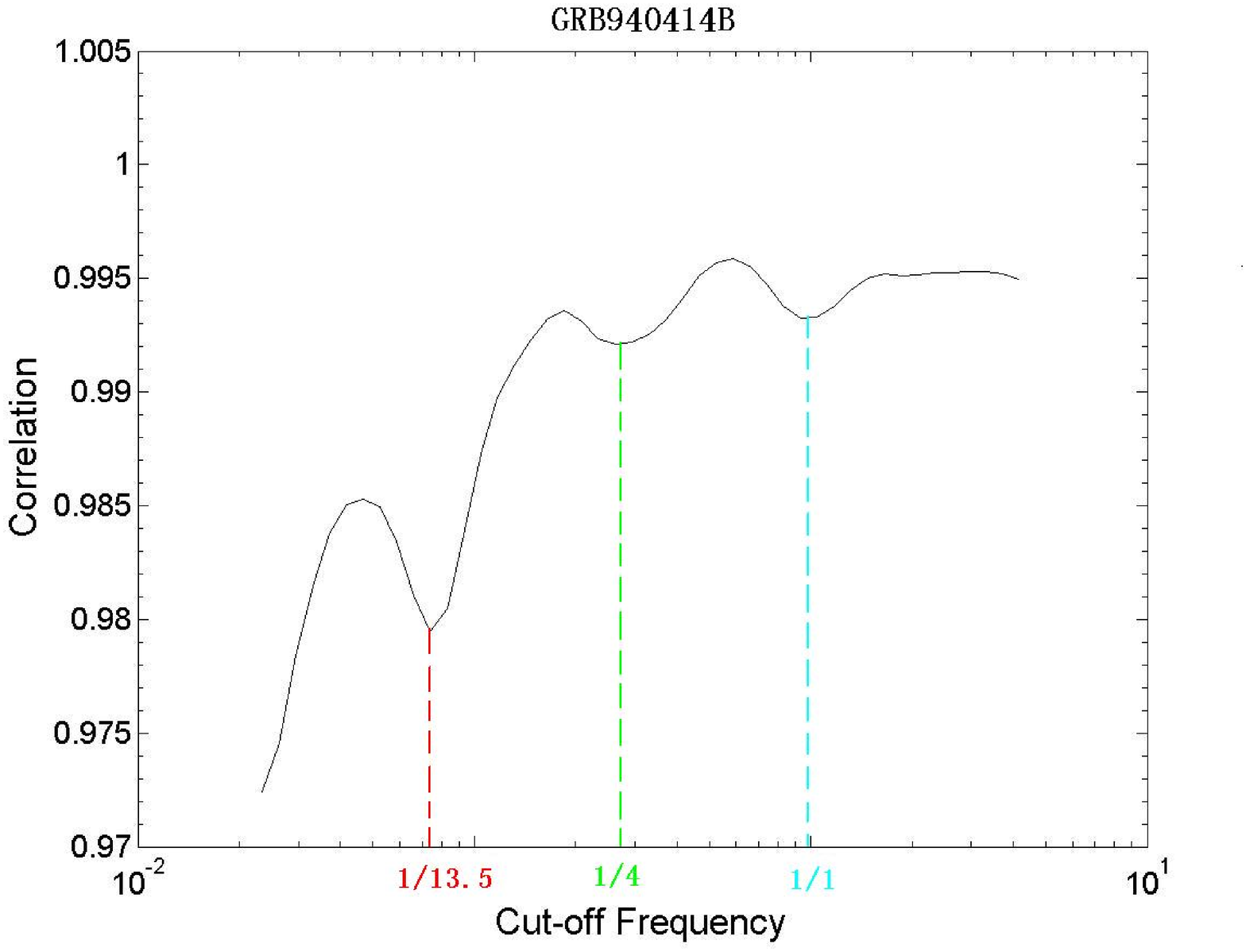}
\end{minipage}%
      \caption{Examples for the multi-dip bursts. The left panel are the lightcurves, and the right panel are the correlation curves. The pulses that correspond to the identified frequencies are marked in different colors in the lightcurves. The time scales are rounded to the nearest 0.5.}
           \label{Fig3}
            \end{figure}

\clearpage
\begin{figure}[h!!!]
\begin{minipage}[b]{1.0\textwidth}
\centering
\includegraphics[width=3in]{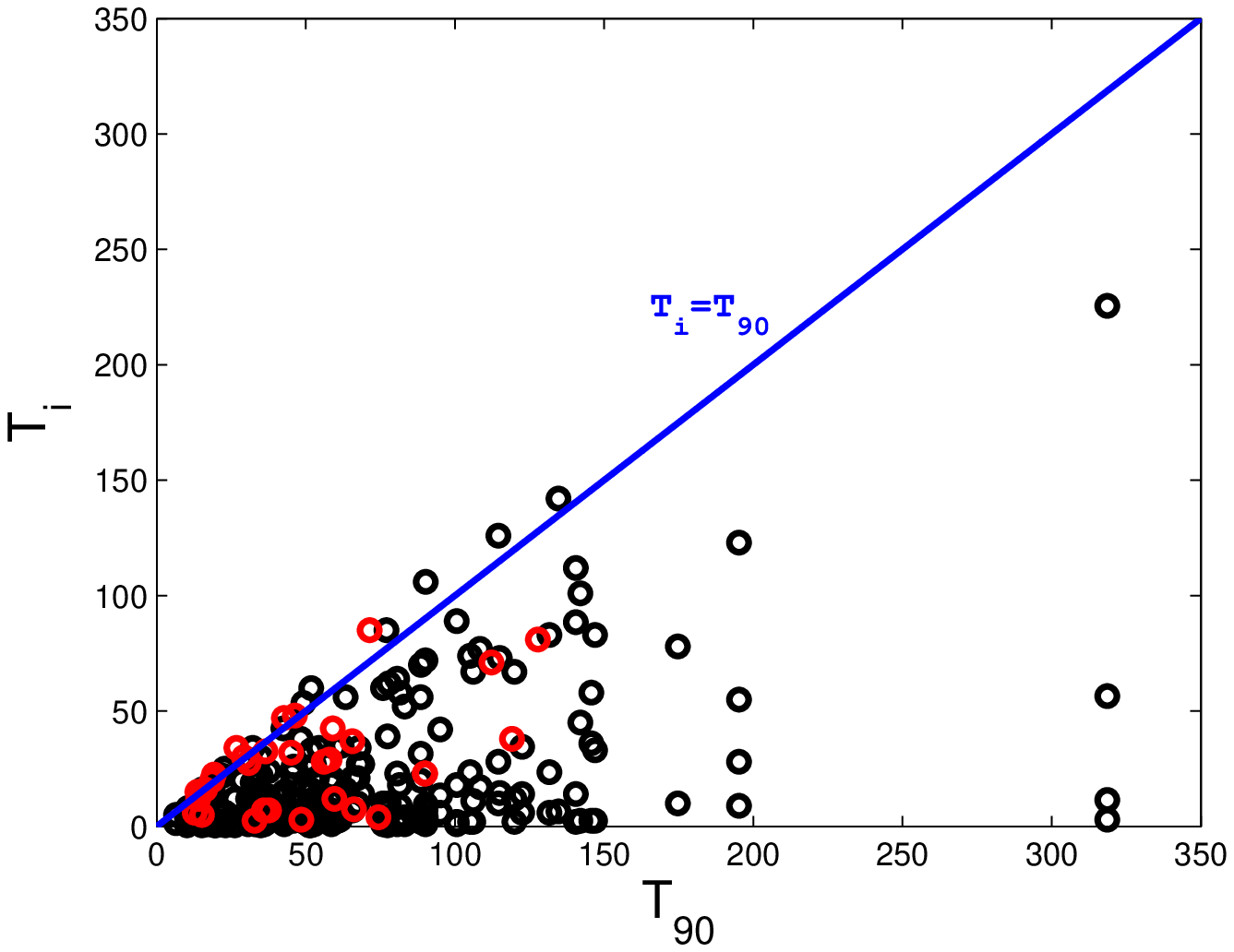}
\end{minipage} \\%
\begin{minipage}[b]{1.0\textwidth}
\centering
\includegraphics[width=3in]{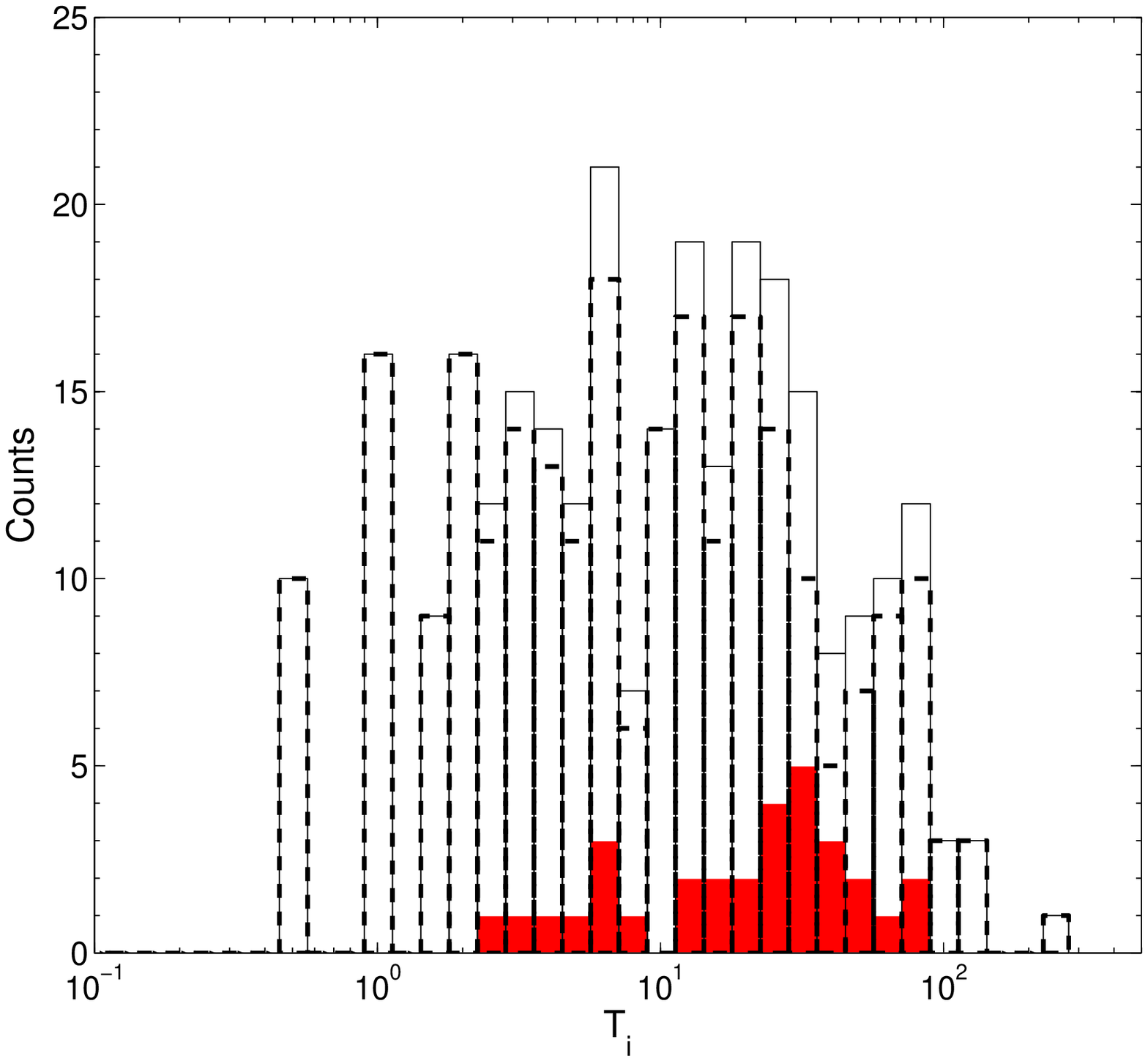}
\end{minipage} \\%
\begin{minipage}[b]{1.0\textwidth}
\centering
\includegraphics[width=3in]{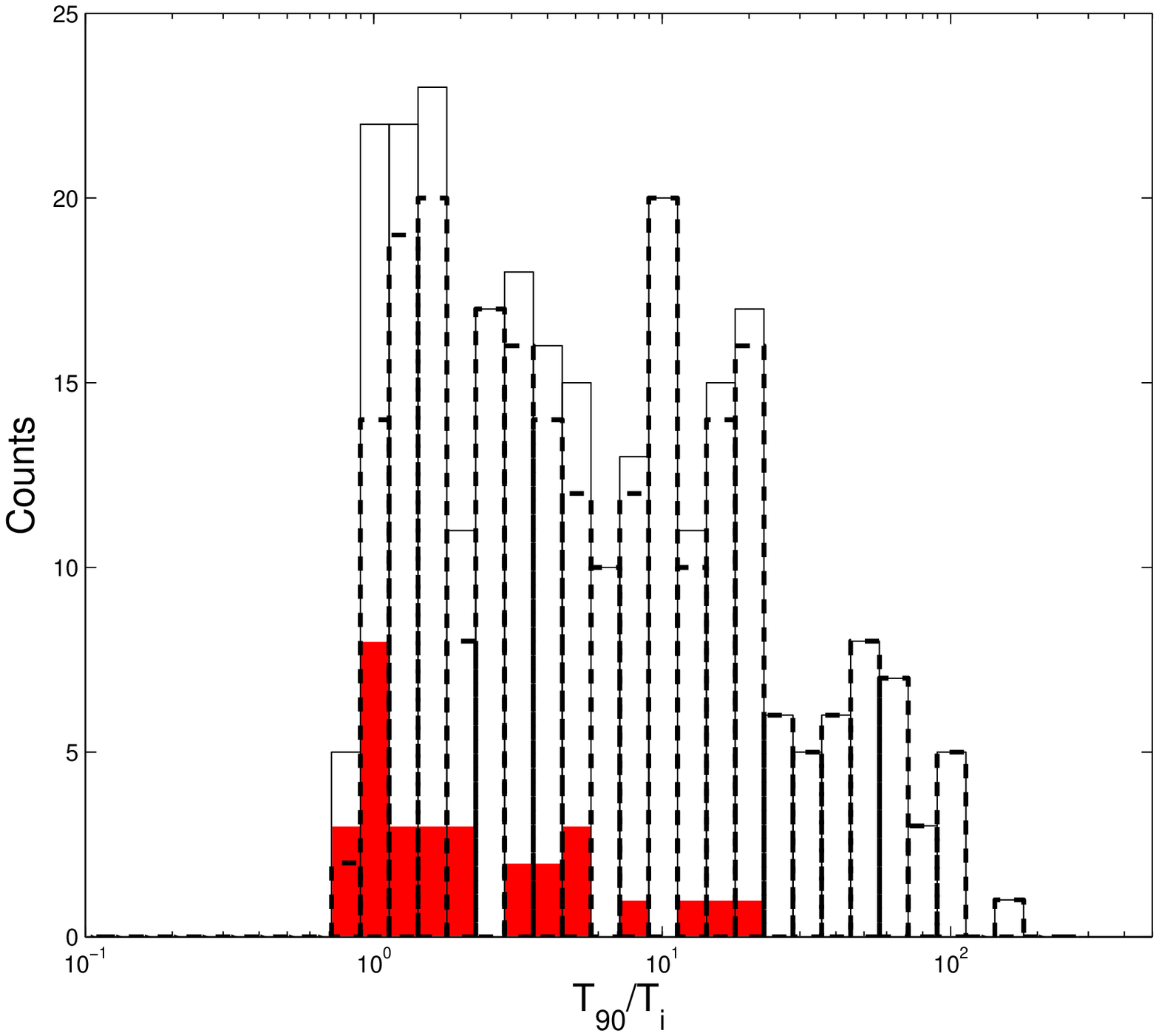}
\end{minipage}%
       \caption{Statistical results of the identified characteristic frequencies. (a) The $T_i - T_{90}$ distribution; (b) histogram of $T_i$; and (c) histogram of $T_{90}/T_i$. The time scales identified in the one-dip sample are marked in red. In (a) the black circles denote the time scales identified in multi-dip GRBs. In (b) and (c), the dashed histograms are for the multi-dip sample, and the final solid histograms are for the entire sample.}
% between $T_{\rm 90}$ and the first slow component timescale $T_{\rm 1}$; the middle and right panel
%        are distributions of $T_{90}/T_{1}$ and all the identified timescales respectively.}
           \label{Fig4}
            \end{figure}

\clearpage
\begin{figure}[h!!!]
\begin{minipage}[b]{1.0\textwidth}
\centering
\includegraphics[width=3in]{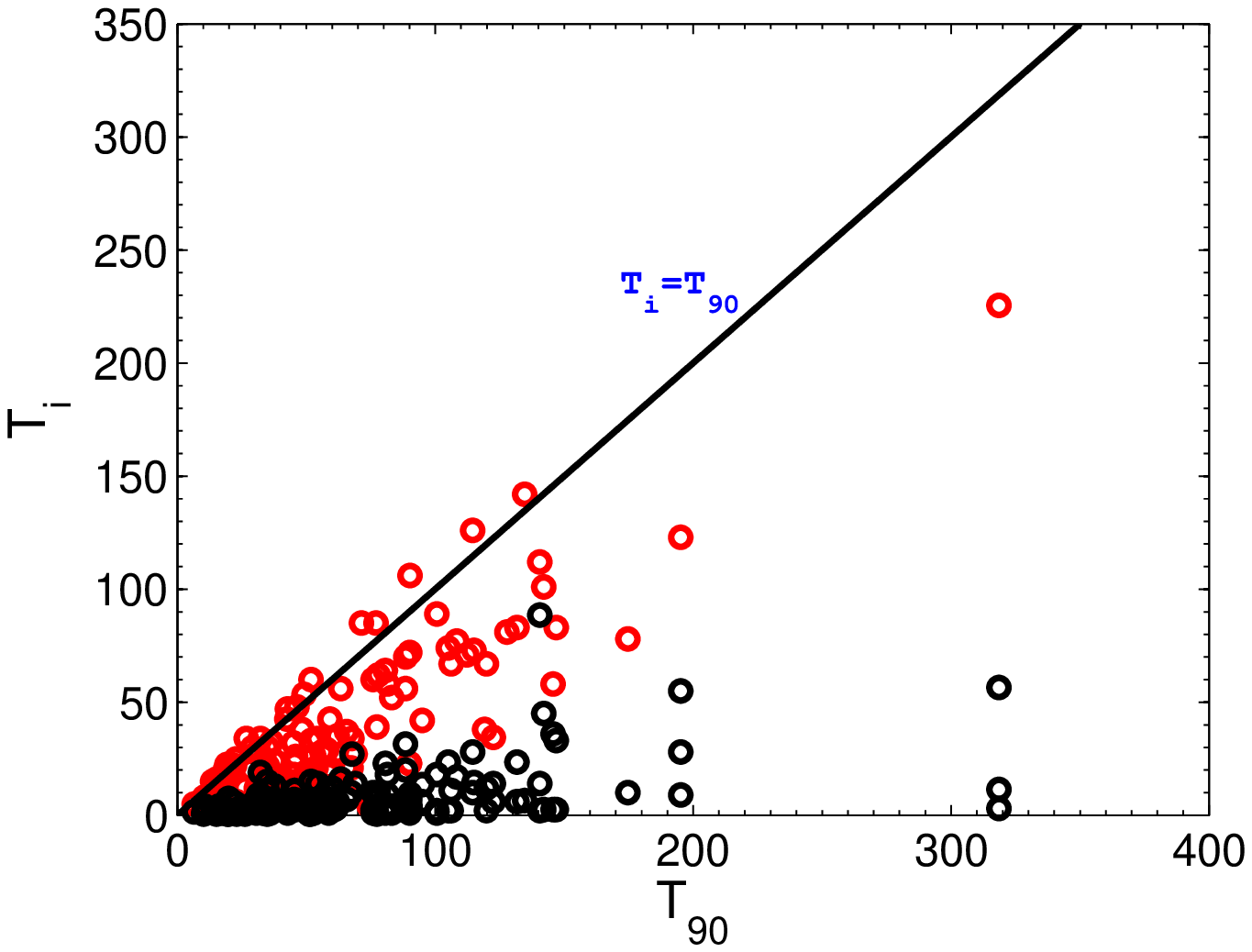}
\end{minipage} \\%
\begin{minipage}[b]{1.0\textwidth}
\centering
\includegraphics[width=3in]{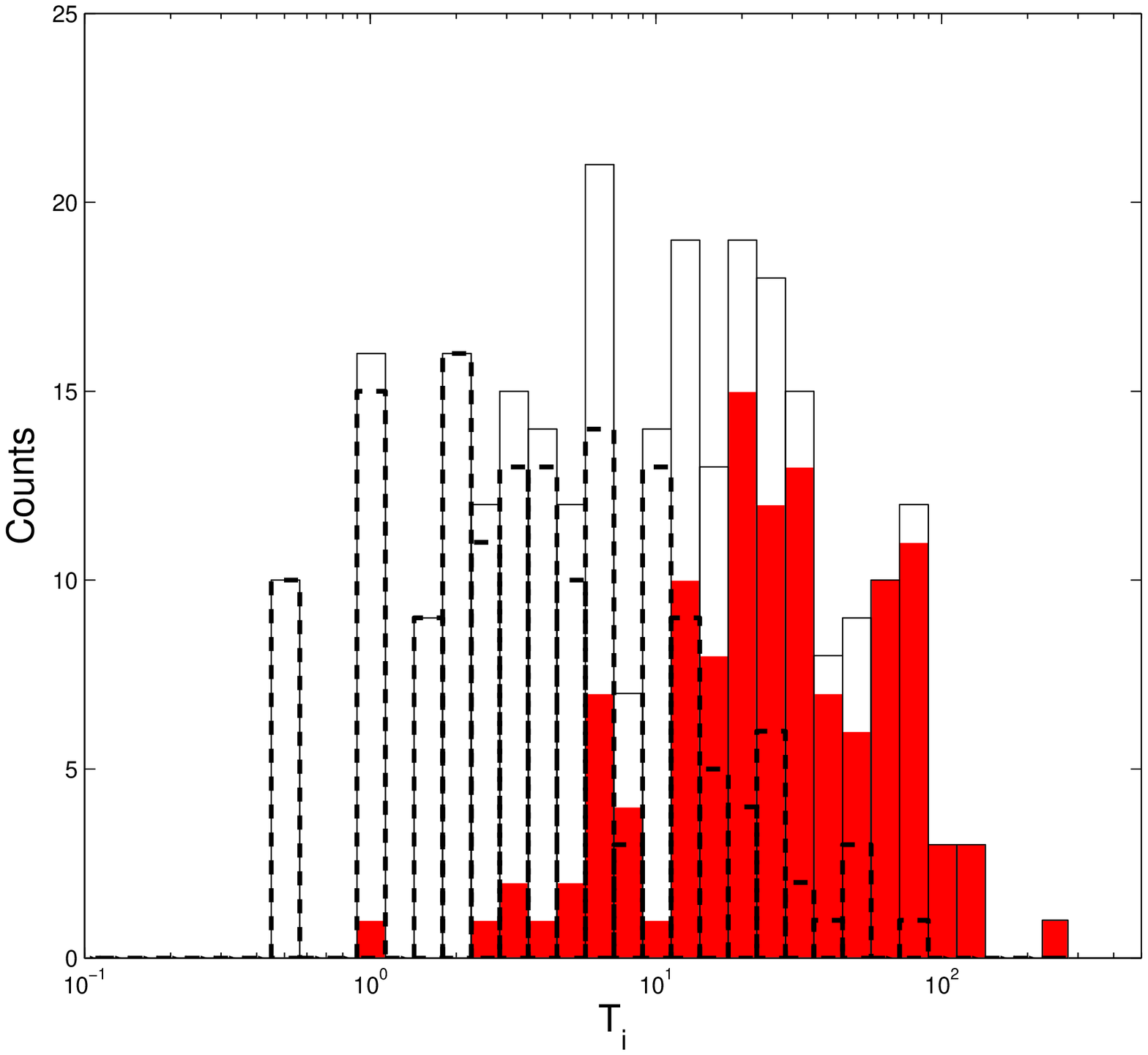}
\end{minipage} \\%
\begin{minipage}[b]{1.0\textwidth}
\centering
\includegraphics[width=3in]{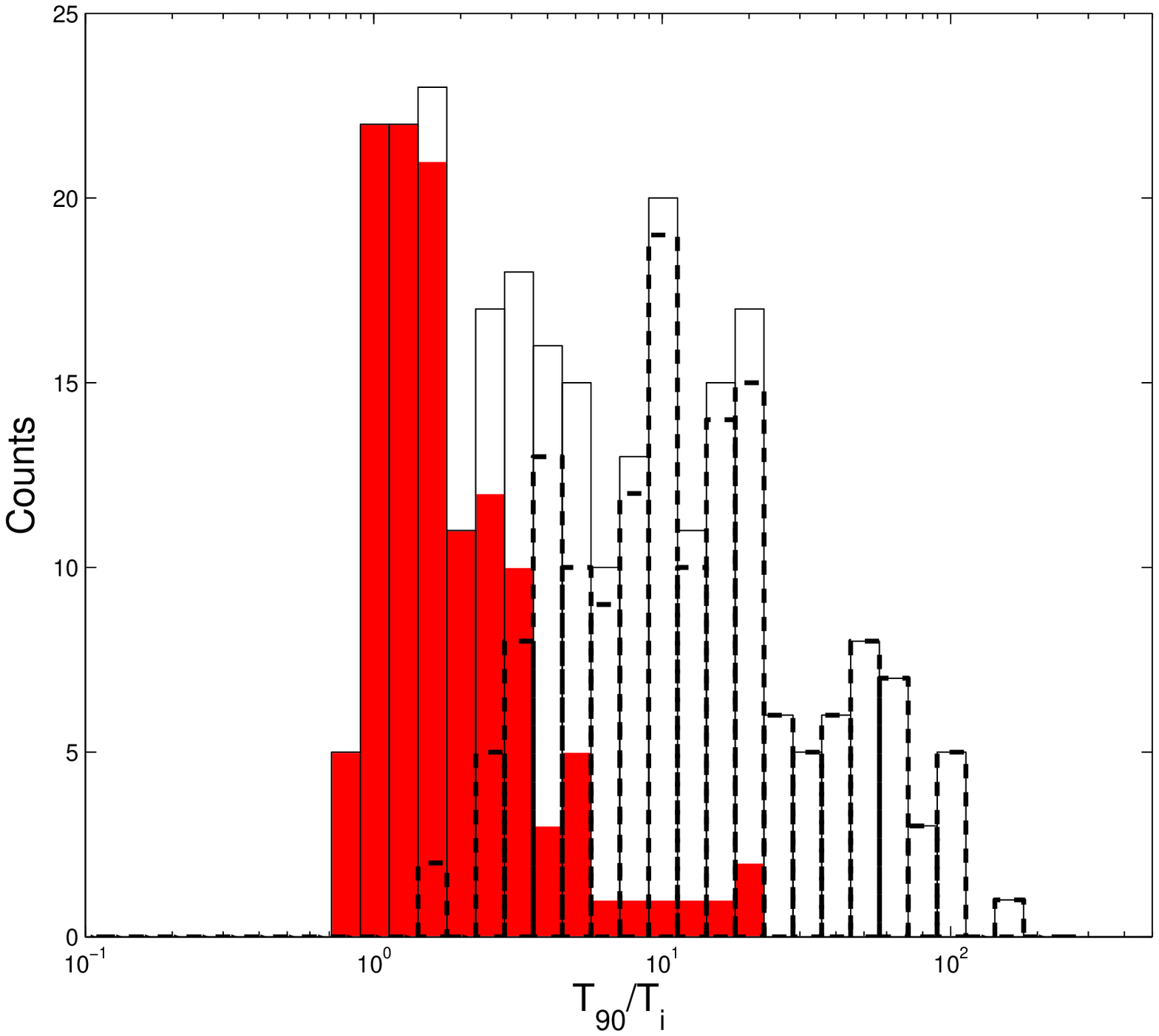}
\end{minipage}%
       \caption{The same as Fig.4, except the red color denotes the slowest time scale in all bursts, and black denotes the rest.}
           \label{Fig5}
            \end{figure}

\clearpage

\appendix
\section{1. Butterworth Low-pass Filter}

%\section{Butterworth Low-pass filter}

For a time series signal $S(t)$ passing an ideal low-pass filter
with cutoff angular frequency $\omega_c=2\pi f_c$, the residual
signal would read \citep{opp98}
\begin{eqnarray}
S(\tau,\omega_c)=\frac{1}{\pi}\int_{-\infty}^{\infty}{\frac{S(t)\sin[\omega_c(\tau-t)]}{\tau-t}}dt.
\end{eqnarray}

For example, if $S(t)=\sin(At)$, then one has
\begin{eqnarray}
   S(\tau,\omega_c)=\frac{\sin(A\tau)}{2}\times[
   {\rm Sign}(1-\frac{A}{\omega_c})+{\rm Sign}(1+\frac{A}{\omega_c})]~,
\end{eqnarray}
where ``Sign'' is the sign symbol of the expression. This formula
can be translated to
\begin{eqnarray}
S(\tau,\omega_c)=\left\{
                        \begin{array}{ll}
                                0 ,& \mbox{$\omega_c<A$}\\
                                \frac{\sin(A\tau)}{2} ,& \mbox{$\omega_c=A$}\\
                                \sin(A\tau) ,& \mbox{$\omega_c>A$}
                                \end{array}
                                \right. \
                                \end{eqnarray}
which shows that the high-frequency signal is attenuated.

For a signal as the sum of two periodic components, e.g.,
$S(t)=\sin(At)+\sin(Bt)$ with $A<B$, one can derive
\begin{eqnarray}
   S(\tau,\omega_c)=\left\{
                        \begin{array}{ll}
                                0 ,& \mbox{$\omega_c<A$}\\
                                \frac{\sin(A\tau)}{2} ,& \mbox{$\omega_c=A$}\\
                                \sin(A\tau) ,&
                                \mbox{$A<\omega_c<B$}\\
                                \sin(A\tau)+\frac{\sin(B\tau)}{2} ,& \mbox{$\omega_c=B$}\\
                                \sin(A\tau)+\sin{B\tau} ,& \mbox{$\omega_c>B$}
                                \end{array}
                                \right. \
                                \end{eqnarray}

It is obvious to see how the two signals are screened when a
progressively lower angular cutoff frequency is applied. If one
chooses two angular cutoff frequencies
%If we stepwise set the cutoff frequency, and the step length satisfy
that satisfy $\omega_{\rm c,i}-\omega_{\rm c,i-1}<B-A$, one would
get a correlation coefficient between two RLCs to be $R_{\rm i}=1$
if
%$S(\tau,\omega_{c,i})$ and
%$S(\tau,\omega_{c,i-1})$ should be equal to 1 when
$A \& B\nsubseteq(\omega_{c,i}\sim\omega_{c,i-1})$, or $R_{\rm i}
\ll 1$
%otherwise should be lower than 1 when
if $A | B\subseteqq(\omega_{c,i}\sim\omega_{c,i-1})$.

Similar results can be obtained if one sets $S(t)=\cos(At)$ or
$S(t)=\cos(At)+\cos(Bt)$ with $A<B$.

For a more complicated time series, one can always decompose it into
the summation of many $Sine$ or $Cosine$ functions through Fourier
transforms. For any angular cutoff frequency $\omega_{\rm c,i}$ (and
the corresponding cutoff frequency $f_{\rm c,i}=\omega_{\rm
c,i}/2\pi$), the low-pass filter then attenuates the signal above
this frequency.

%As we know, with Fourier transformation, every signal could be
%transformed into the summation of signals with $Sine$ function and
%signals with $Cosine$ function. For real signals there should be
%several frequencies for the $Sine$ or $Cosine$ function, and the
%value of the correlation coefficient between $S(\tau,\omega_{c,i})$
%and $S(\tau,\omega_{c,i-1})$ will depend on whether these
%frequencies fall into the region of $\omega_{c,i}\sim\omega_{c,i-1}$
%and how many of them do, that is the mathematic proof of the
%motivation of this stepwise filter method.

%\end{appendix}

%\appendix
\section{2. Additional simulation tests of the SFC algorithm}

In order to better understand the SFC algorithm, we perform a set of
additional simulations.

%Besides proving the validity of SFC method, we also need to figure
%out this new algorithm's characteristic feature before we apply it
%to real GRB data. A Mock Catalog of light curves with different
%pulse profiles, amplitude of pulses, pulse duration distributions,
%pulse duration separations, lags duration distributions, and total
%number of pulses has been generated, shown in Fig.6, and we find out
%following scenarios based on it.

\begin{figure}
  \centering
  \subfigure[]{
    \label{fig:subfig:a} %% label for first subfigure
    \includegraphics[width=1.7in]{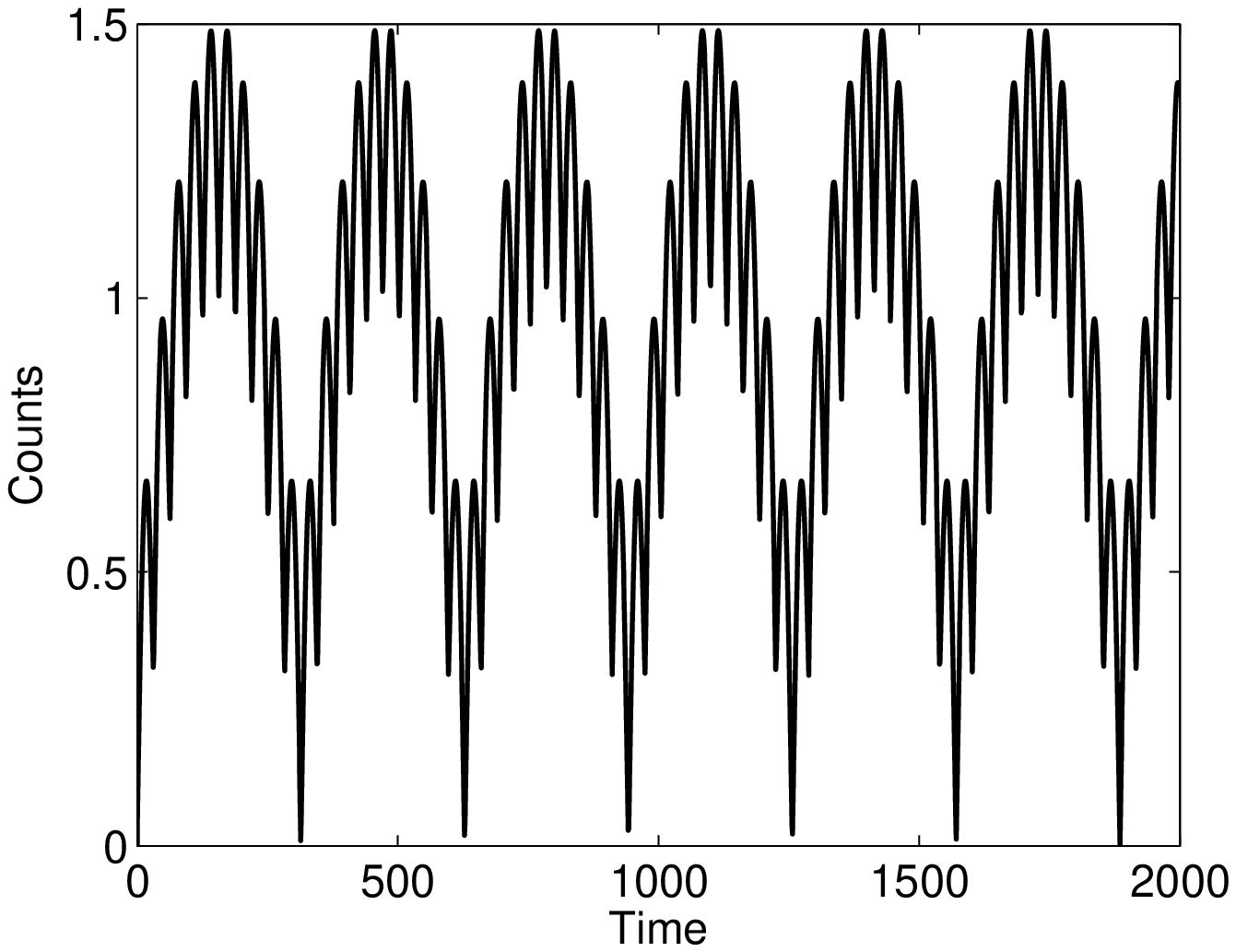}}
  %\hspace{1in}
  \subfigure[]{
    \label{fig:subfig:b} %% label for second subfigure
    \includegraphics[width=1.7in]{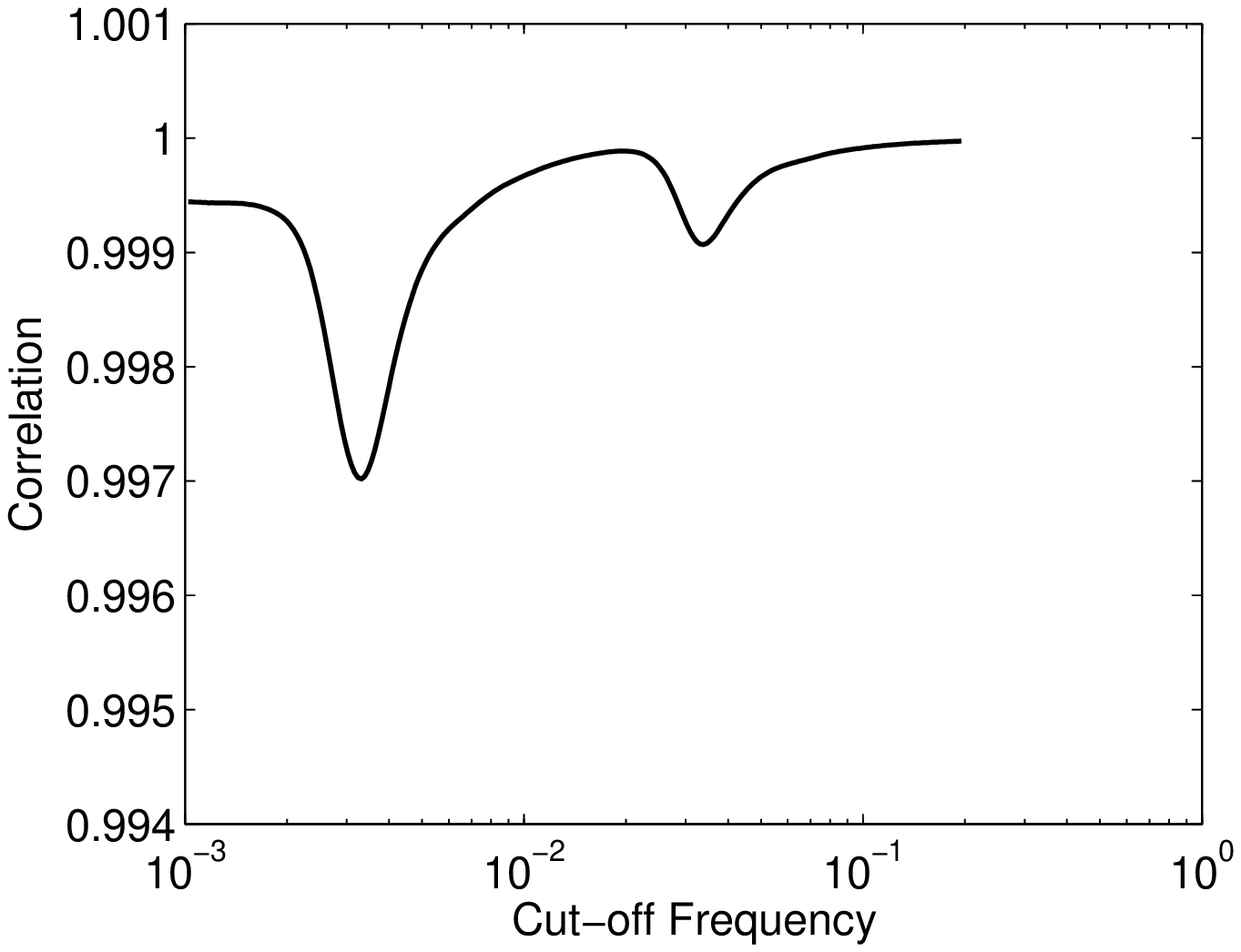}}
\centering
  \subfigure[]{
    \label{fig:subfig:c} %% label for first subfigure
    \includegraphics[width=1.7in]{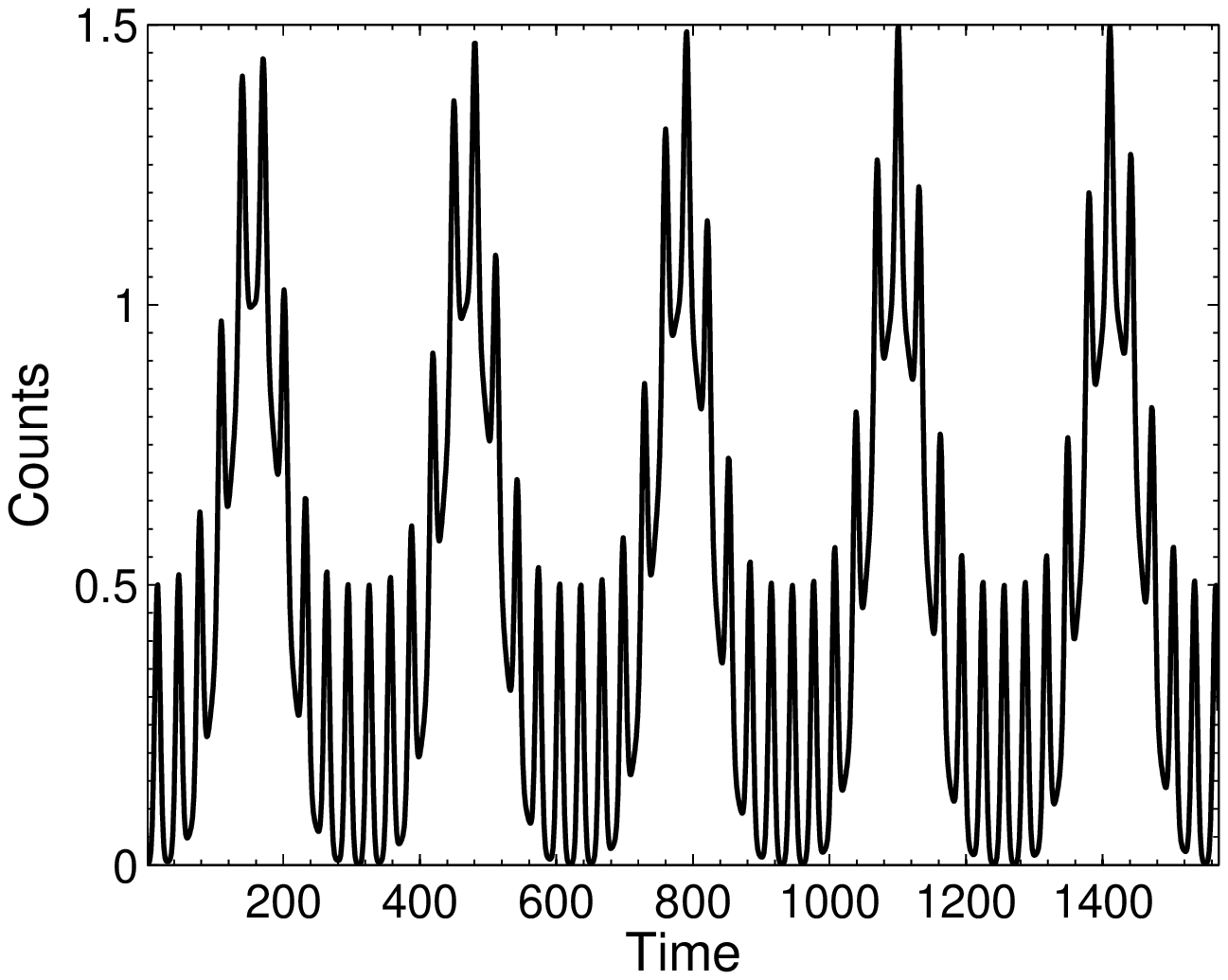}}
  %\hspace{1in}
  \subfigure[]{
    \label{fig:subfig:d} %% label for second subfigure
    \includegraphics[width=1.7in]{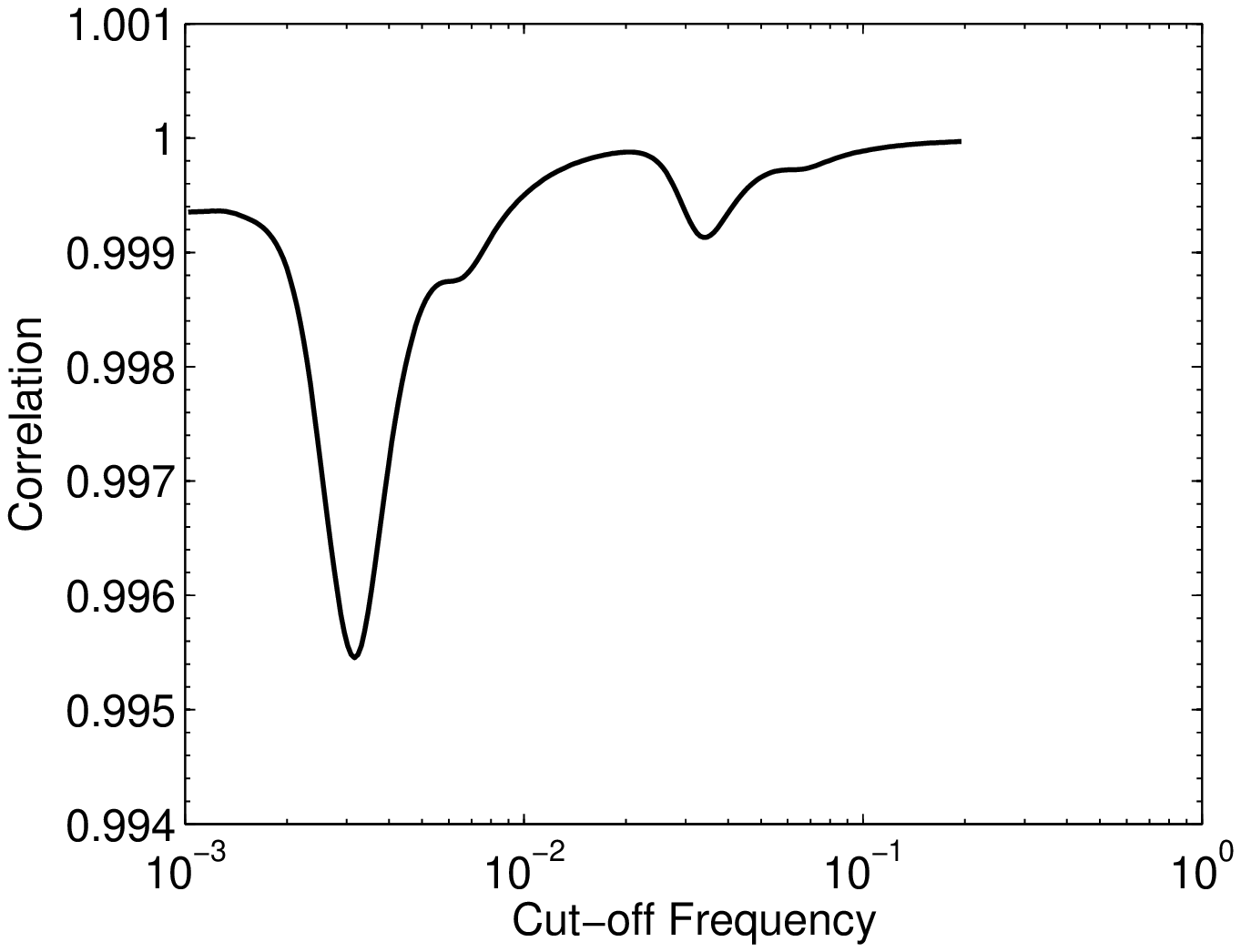}}
\centering
  \subfigure[]{
    \label{fig:subfig:e} %% label for first subfigure
    \includegraphics[width=1.7in]{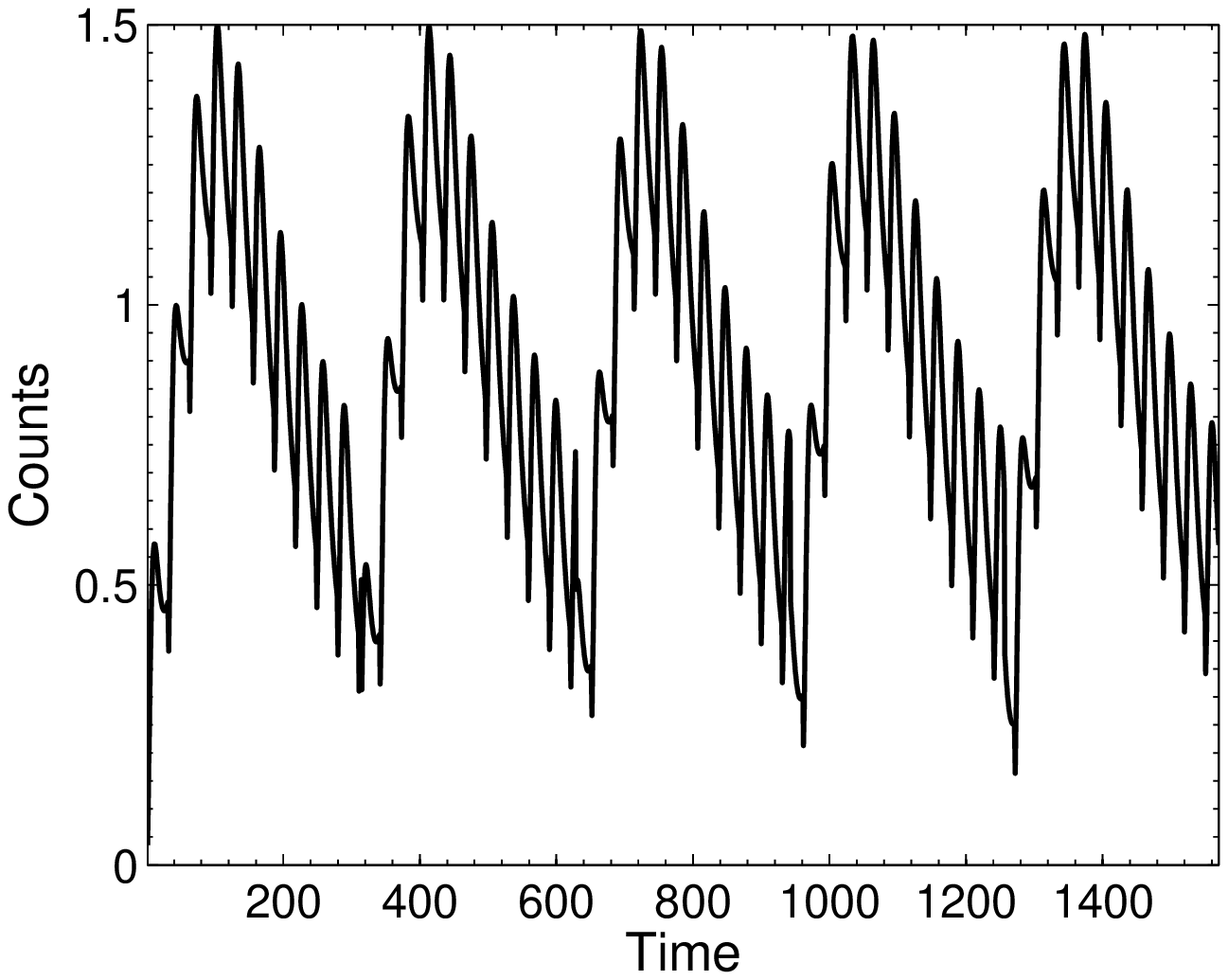}}
  %\hspace{1in}
  \subfigure[]{
    \label{fig:subfig:f} %% label for second subfigure
    \includegraphics[width=1.7in]{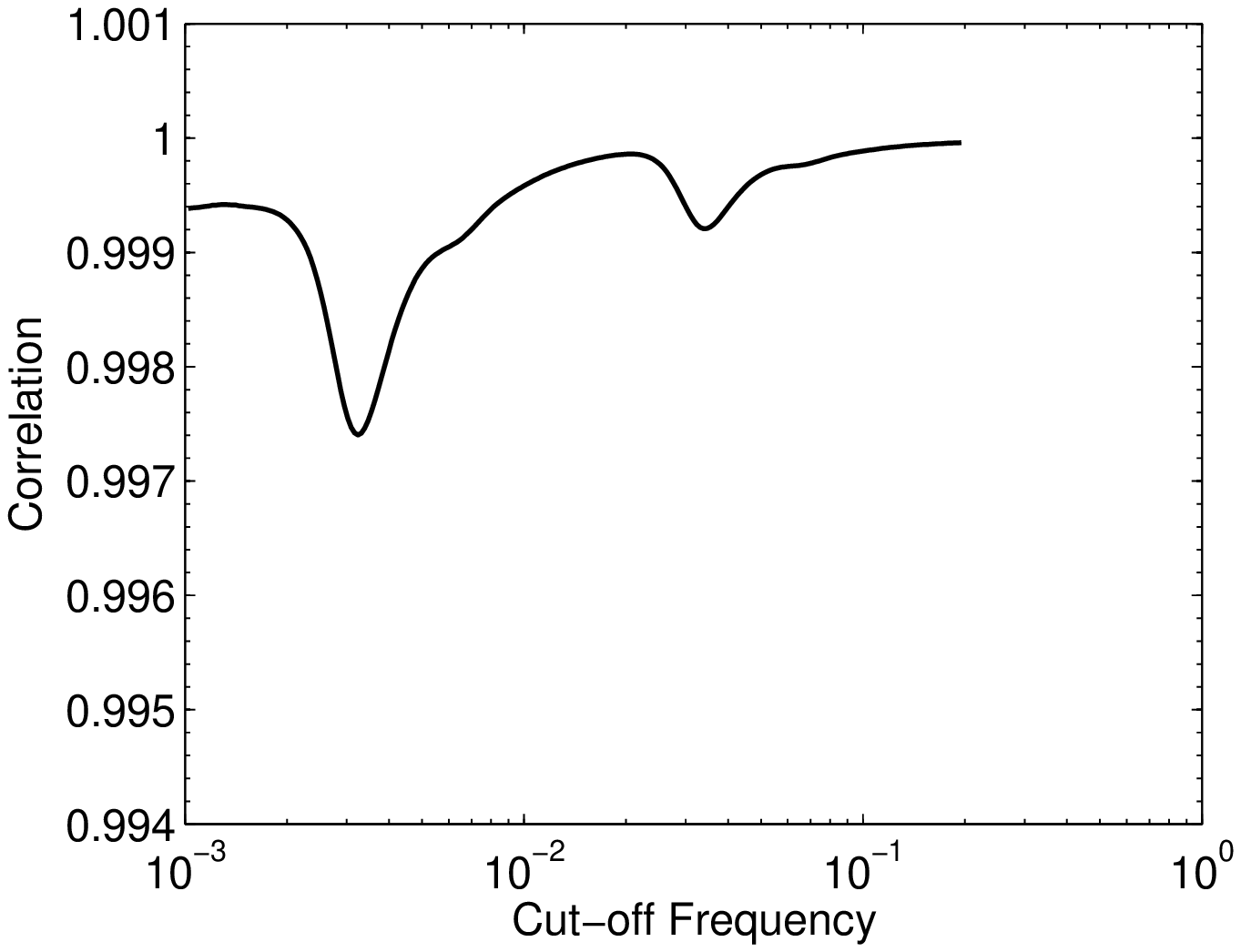}}
\centering
  \subfigure[]{
    \label{fig:subfig:g} %% label for first subfigure
    \includegraphics[width=1.7in]{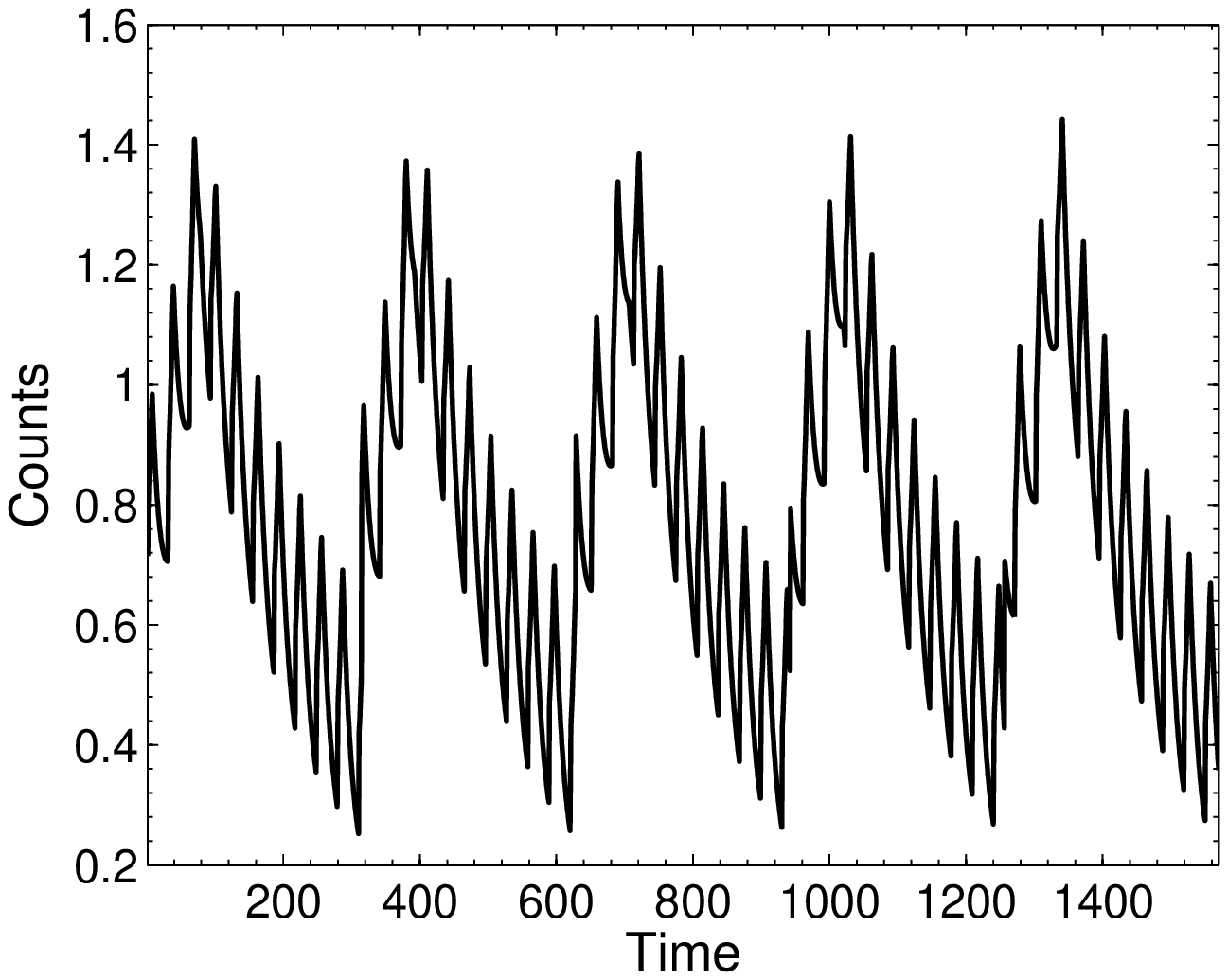}}
  %\hspace{1in}
  \subfigure[]{
    \label{fig:subfig:h} %% label for second subfigure
    \includegraphics[width=1.7in]{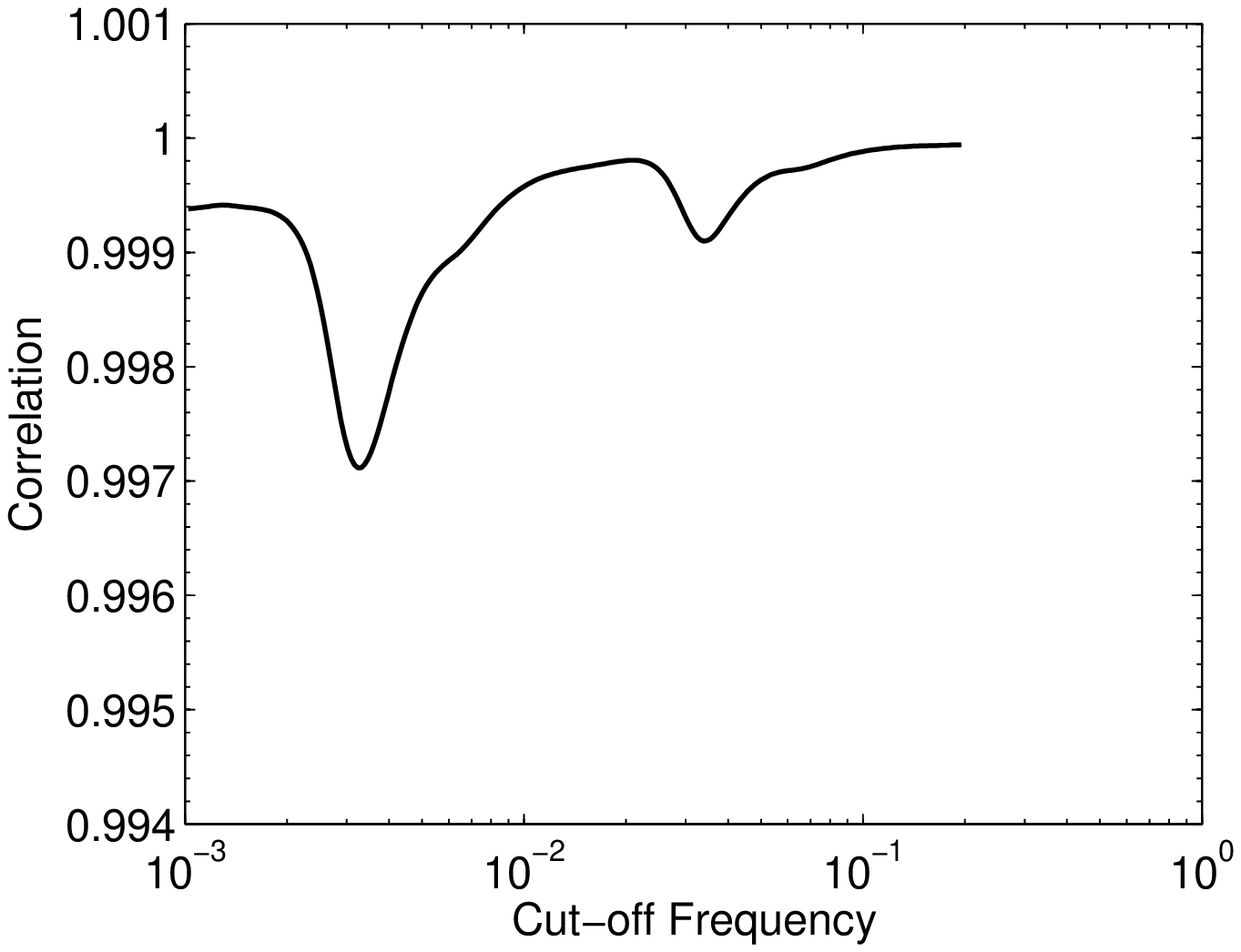}}
        \centering
  \subfigure[]{
    \label{fig:subfig:i} %% label for first subfigure
    \includegraphics[width=1.7in]{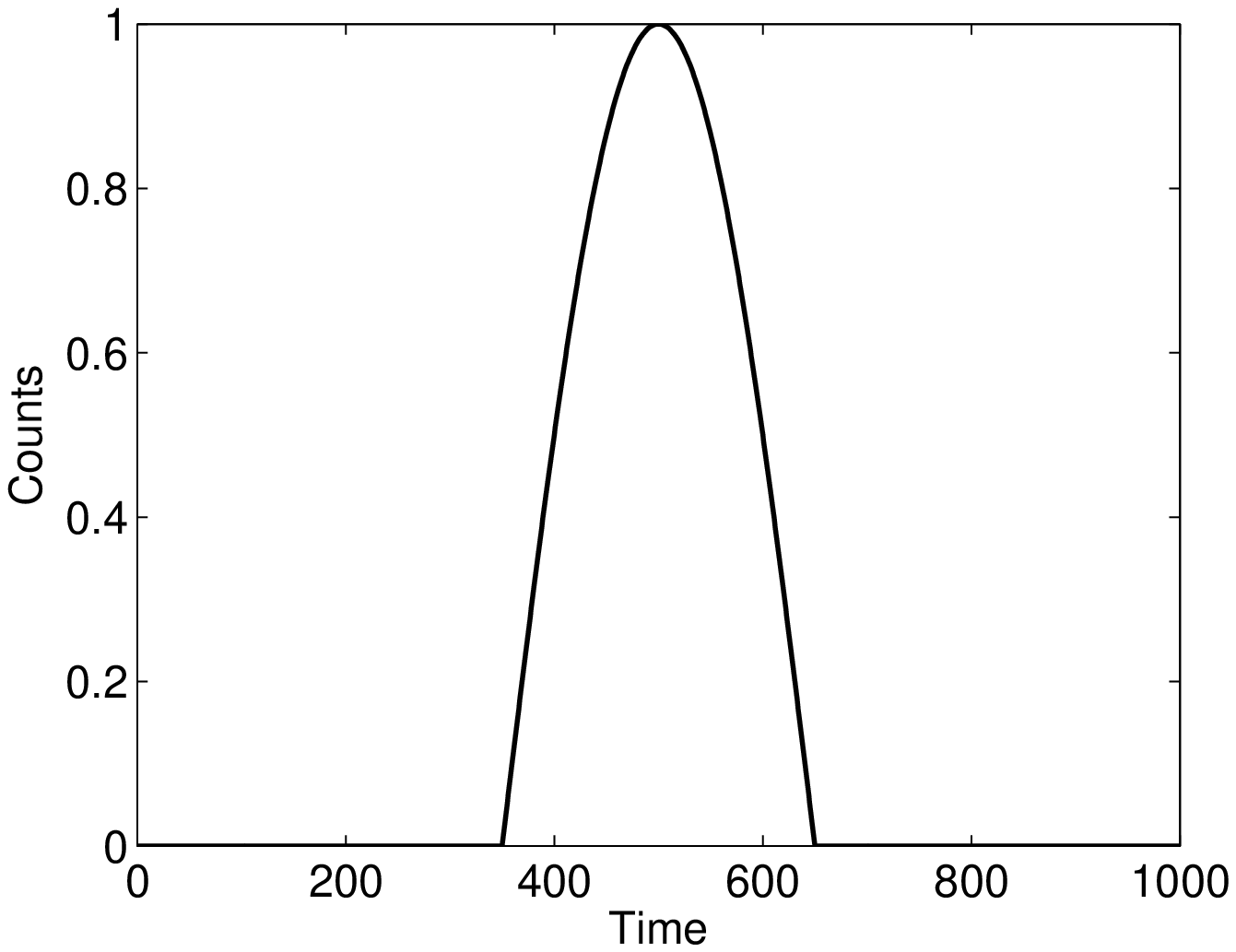}}
  %\hspace{1in}
  \subfigure[]{
    \label{fig:subfig:j} %% label for second subfigure
    \includegraphics[width=1.7in,height=1.25in]{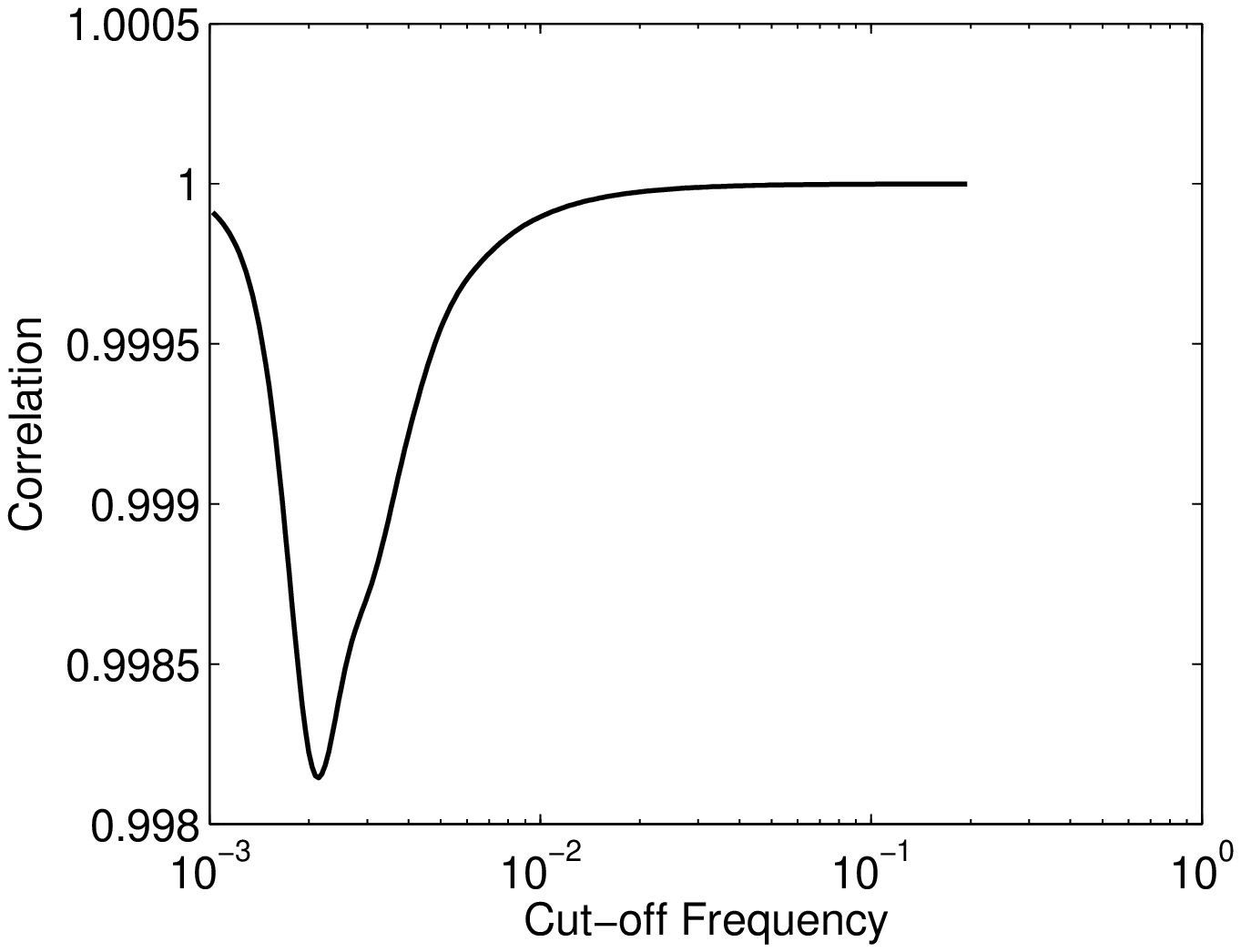}}
\centering
  \subfigure[]{
    \label{fig:subfig:k} %% label for first subfigure
    \includegraphics[width=1.7in]{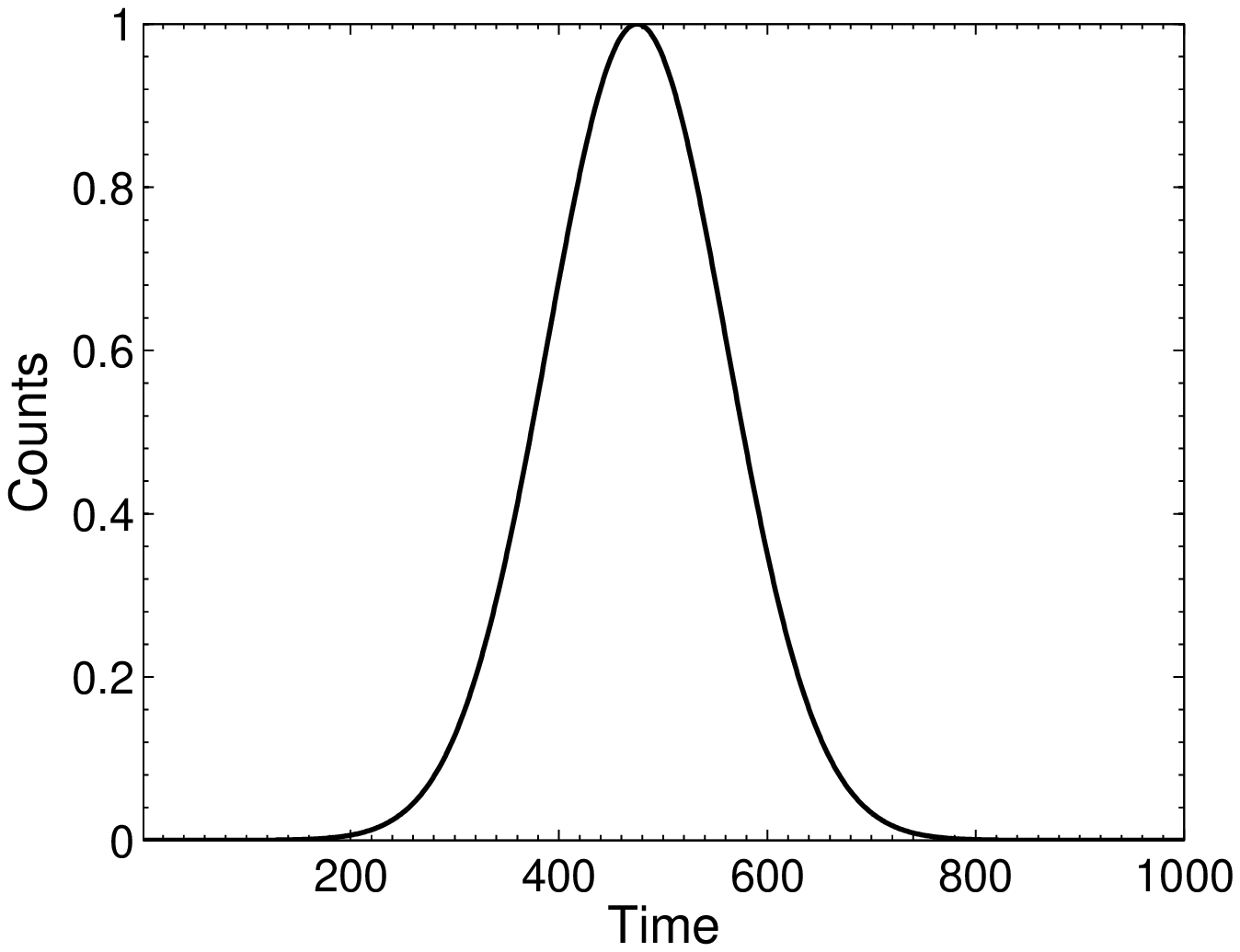}}
  %\hspace{1in}
  \subfigure[]{
    \label{fig:subfig:l} %% label for second subfigure
    \includegraphics[width=1.7in,height=1.25in]{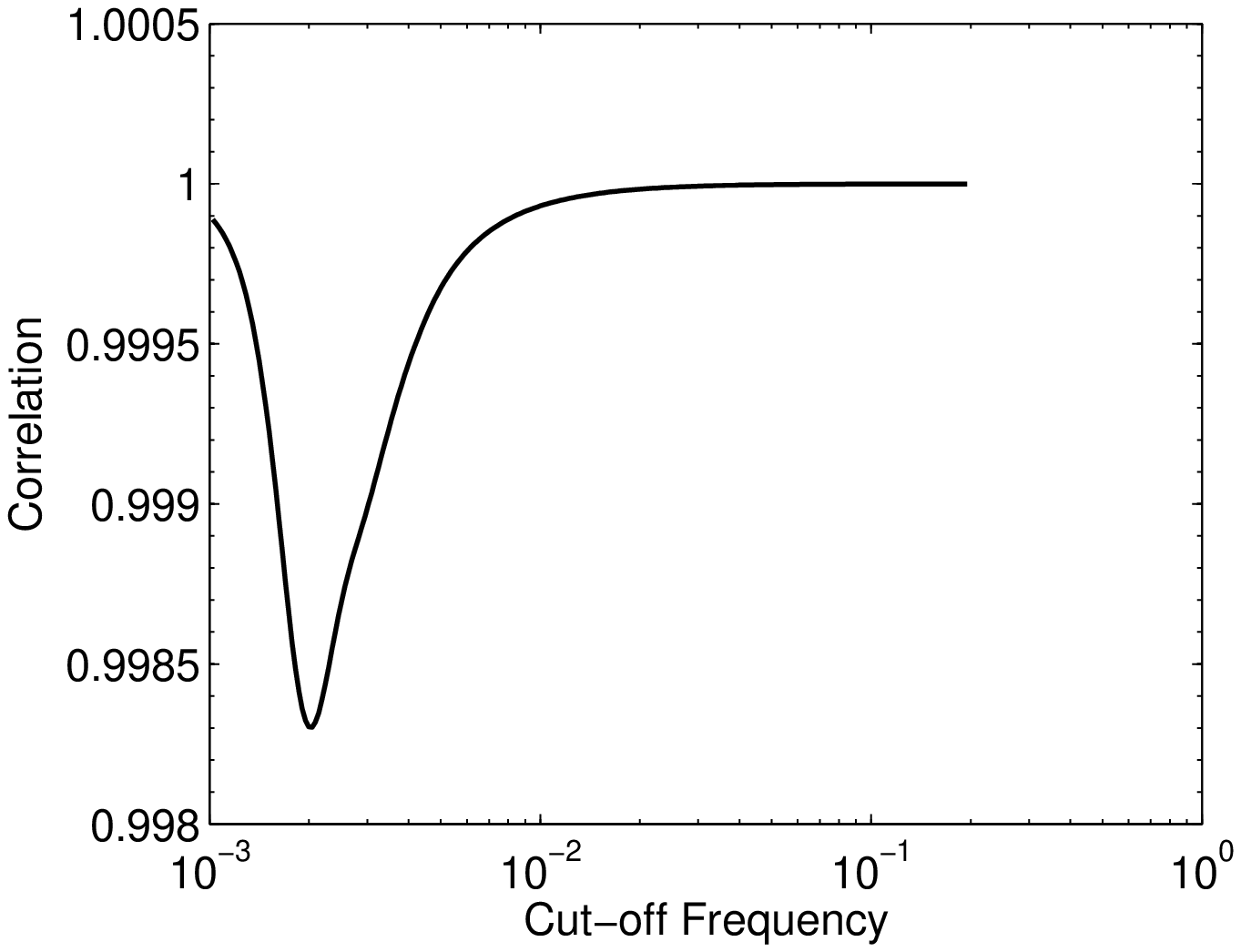}}
\centering
  \subfigure[]{
    \label{fig:subfig:m} %% label for first subfigure
    \includegraphics[width=1.7in]{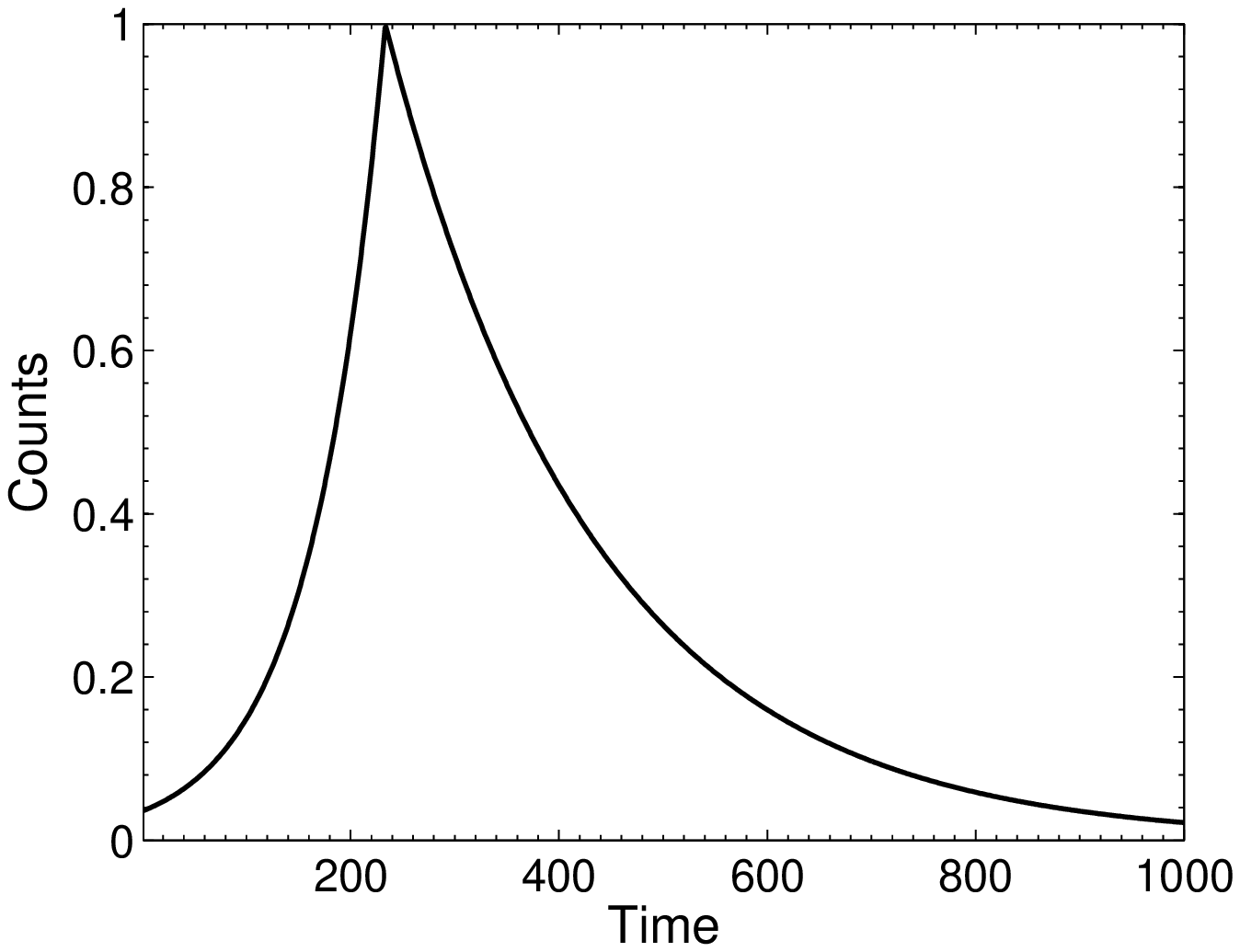}}
  %\hspace{1in}
  \subfigure[]{
    \label{fig:subfig:n} %% label for second subfigure
    \includegraphics[width=1.7in,height=1.25in]{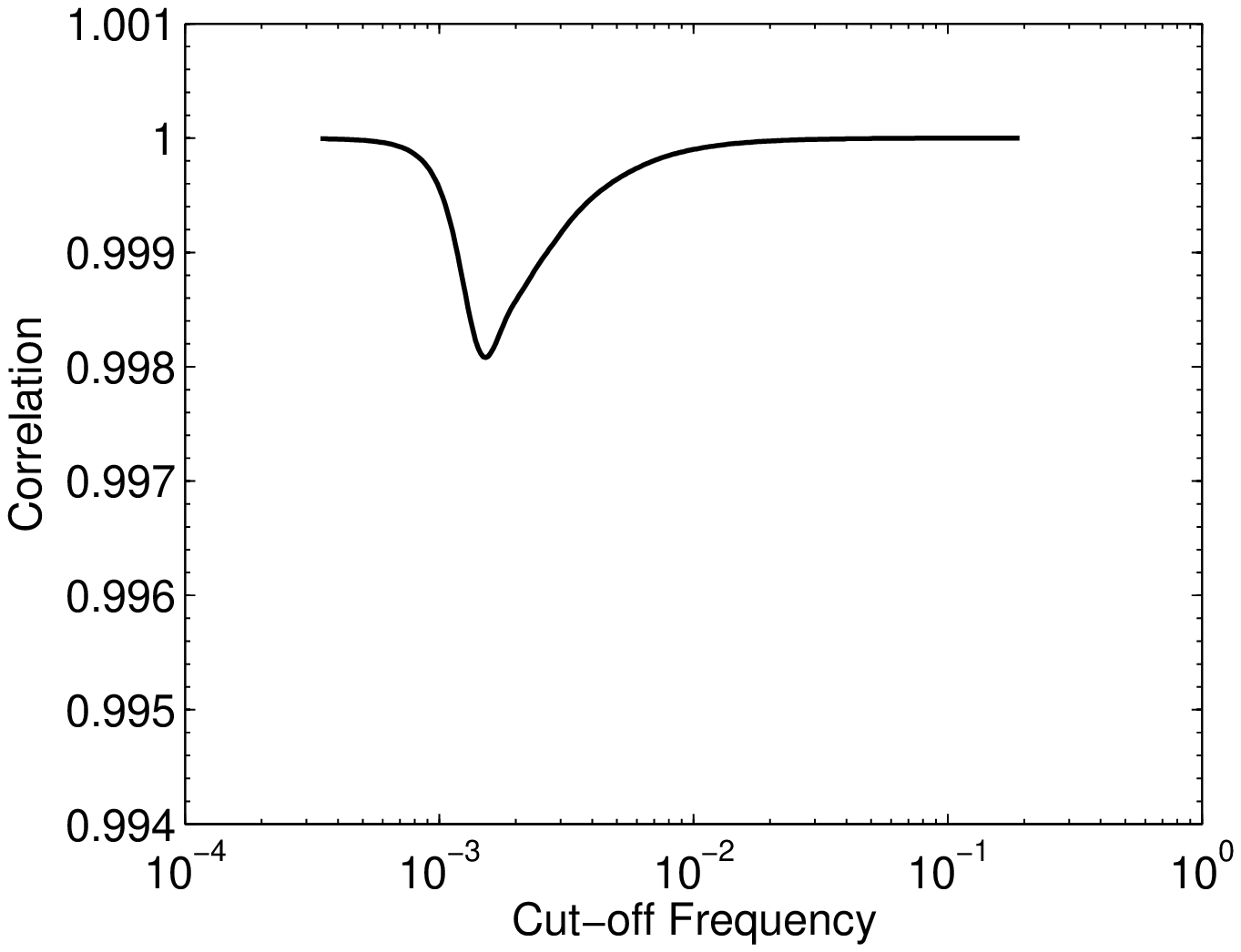}}
\centering
  \subfigure[]{
    \label{fig:subfig:o} %% label for first subfigure
    \includegraphics[width=1.7in]{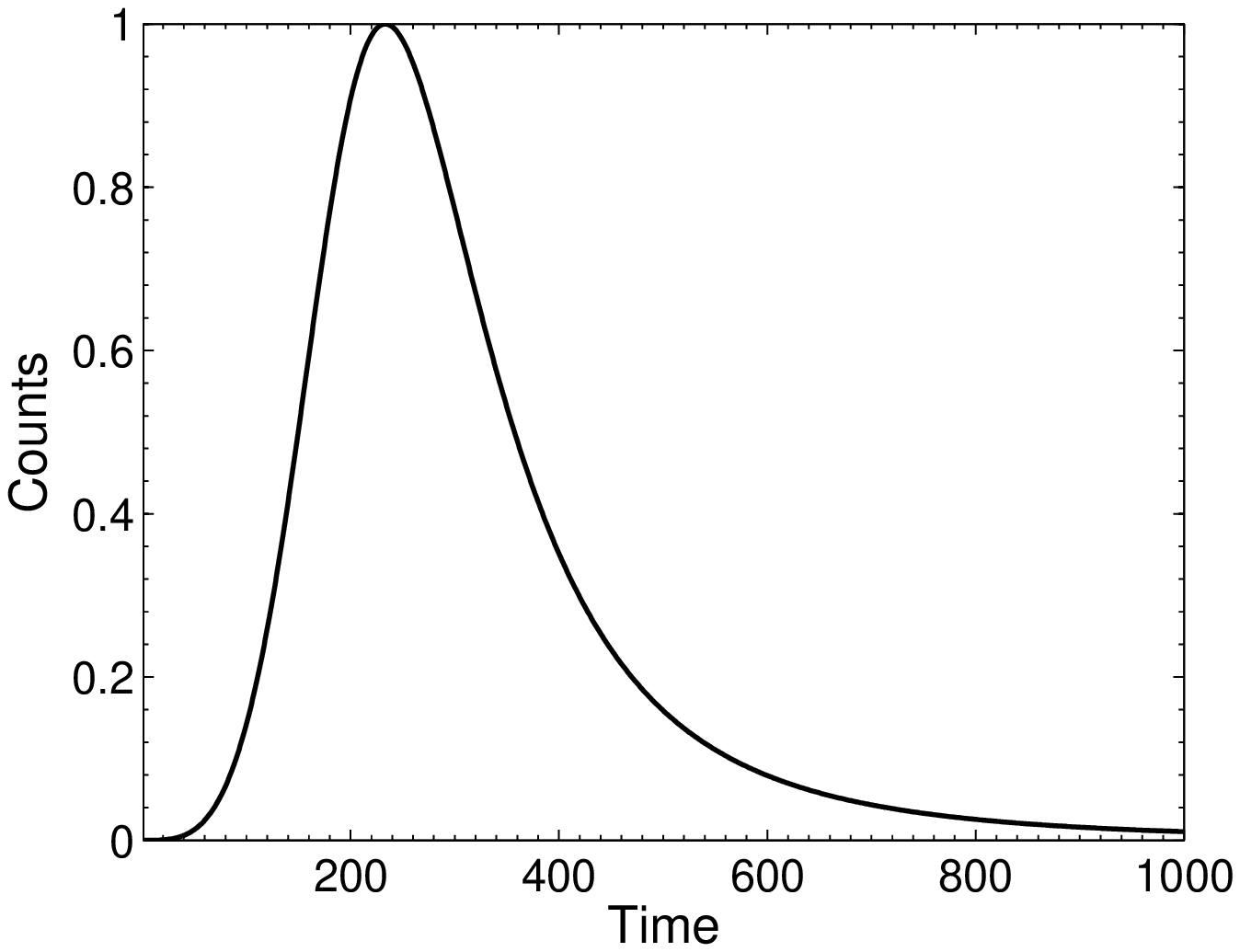}}
  %\hspace{1in}
  \subfigure[]{
    \label{fig:subfig:p} %% label for second subfigure
    \includegraphics[width=1.7in,height=1.25in]{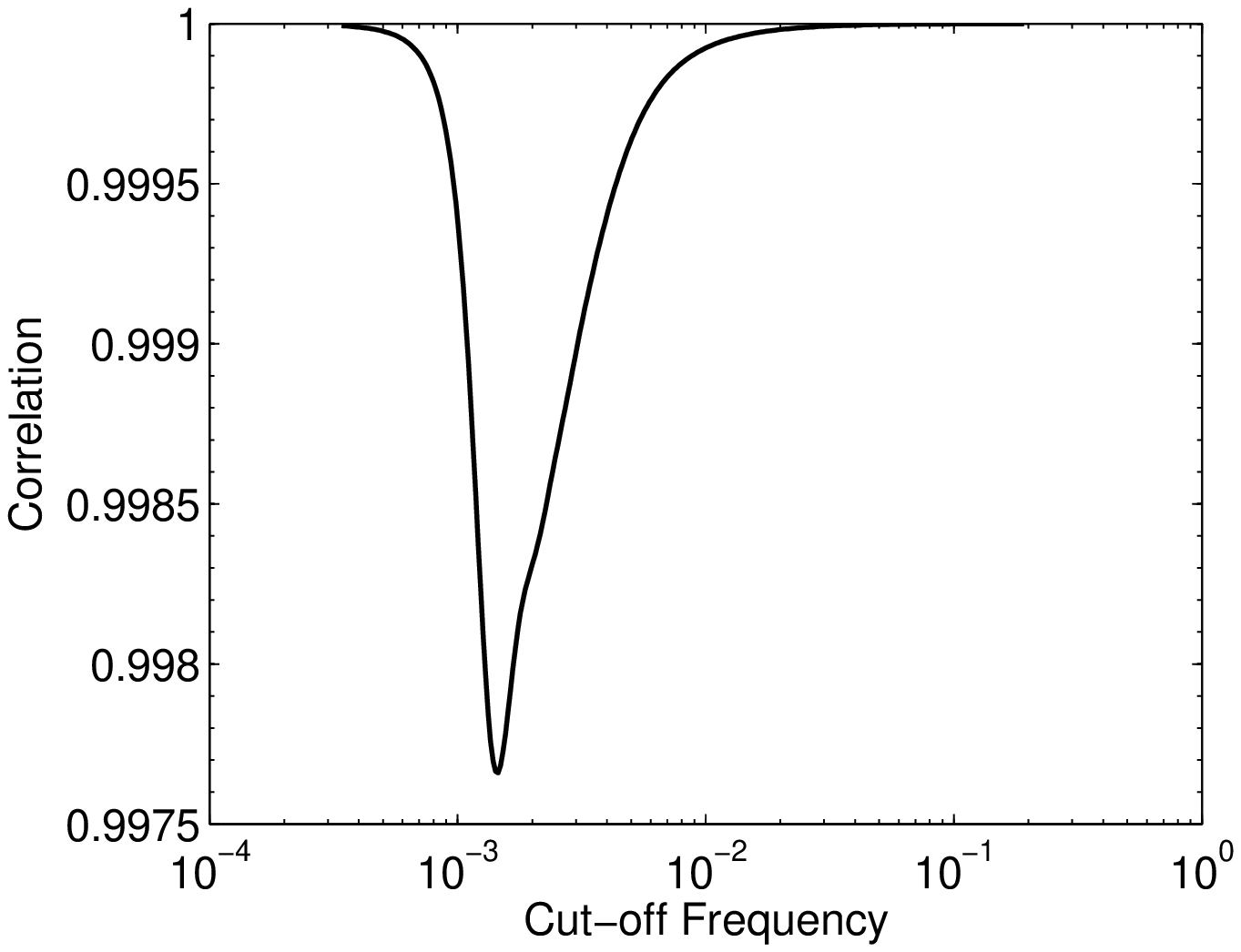}}
\caption{Mock catalog of lightcurves with different pulse profile
and their relevant correlation curves.}
  \label{fig:subfig} %% label for entire figure
\label{profile}
\end{figure}

1) Pulse profiles: We test four different pulse profile functions,
$A| \sin (\pi t/T) |$ (sine) function, Gaussian function, and two
``FRED" profiles proposed by \cite{Kocevski03} and \cite{Norris96}.
First, we generate multiple pulses lying side-by-side with a fast
component superposed on the slow component. For all four different
pulse functions, the pulse durations of the slow and fast components
are fixed to $100\pi$ and $10\pi$, respectively, with the amplitude
ratio between the two components fixed as $A_{\rm s}:A_{\rm f}=2:1$.
As shown in Fig.\ref{profile}(a-h), we can see that SFC is not
sensitive to the pulse profile function in the multi-pulse case.
In the rest of the simulation tests invoking multi-pulse lightcurves,
we adopt the sine function as examples, and use Fig.\ref{profile}(a,b)
as our nominal test to be compared with others (see tests 2-6 below).
Since the SFC method can catch a frequency component even if only
one pulse exists (test 3 below) and since some GRBs indeed only
have one broad pulse, next we test the four pulse profile functions
for one pulse only. We fix the full width at half maximum (FWHM)
to $\sim 200$ s, and vary the function shapes. To our surprise,
it is found that the identified typical durations from the SFC
curve dip frequencies are very different for the four functions
(Fig.\ref{profile}(i-p)): $\sim 470$ s for the sine shape, $\sim
500$ for the Gaussian shape, $\sim 676$ s for the Norris' shape,
and $\sim 708$ s for the Kocevski's shape. A closer investigation
suggests that the longer durations for the FRED shapes are mostly
due to the extended tails for these profile functions
\citep{Norris96,Kocevski03}. This explains the identified long
durations for some FRED-like lightcurves in Sample II (gaps and
long tails), which can be longer than $T_{90}$ in some cases.

\begin{figure}
\centering
  \subfigure[]{
    \label{fig:subfig:a} %% label for first subfigure
    \includegraphics[width=1.7in]{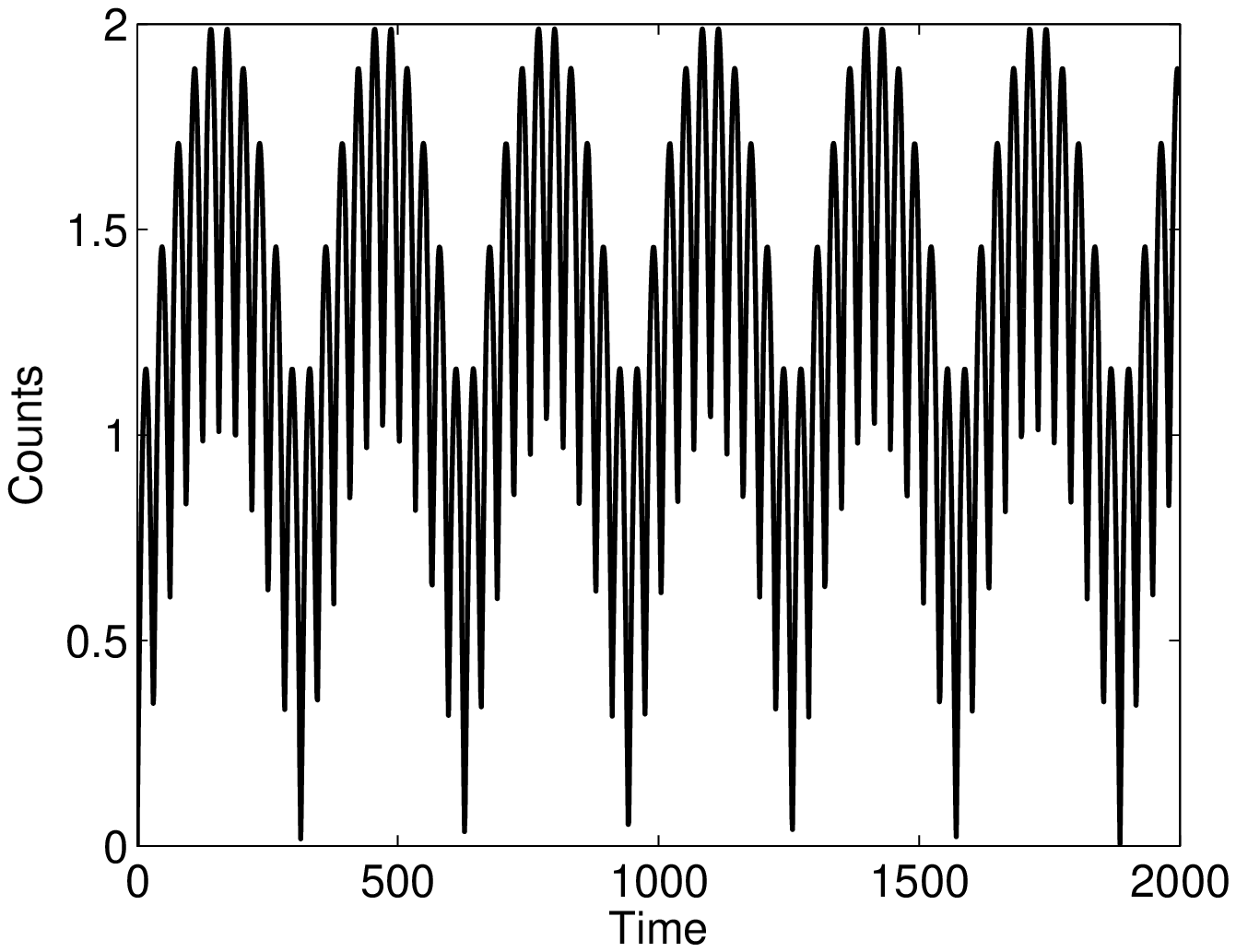}}
  %\hspace{1in}
  \subfigure[]{
    \label{fig:subfig:b} %% label for second subfigure
    \includegraphics[width=1.7in]{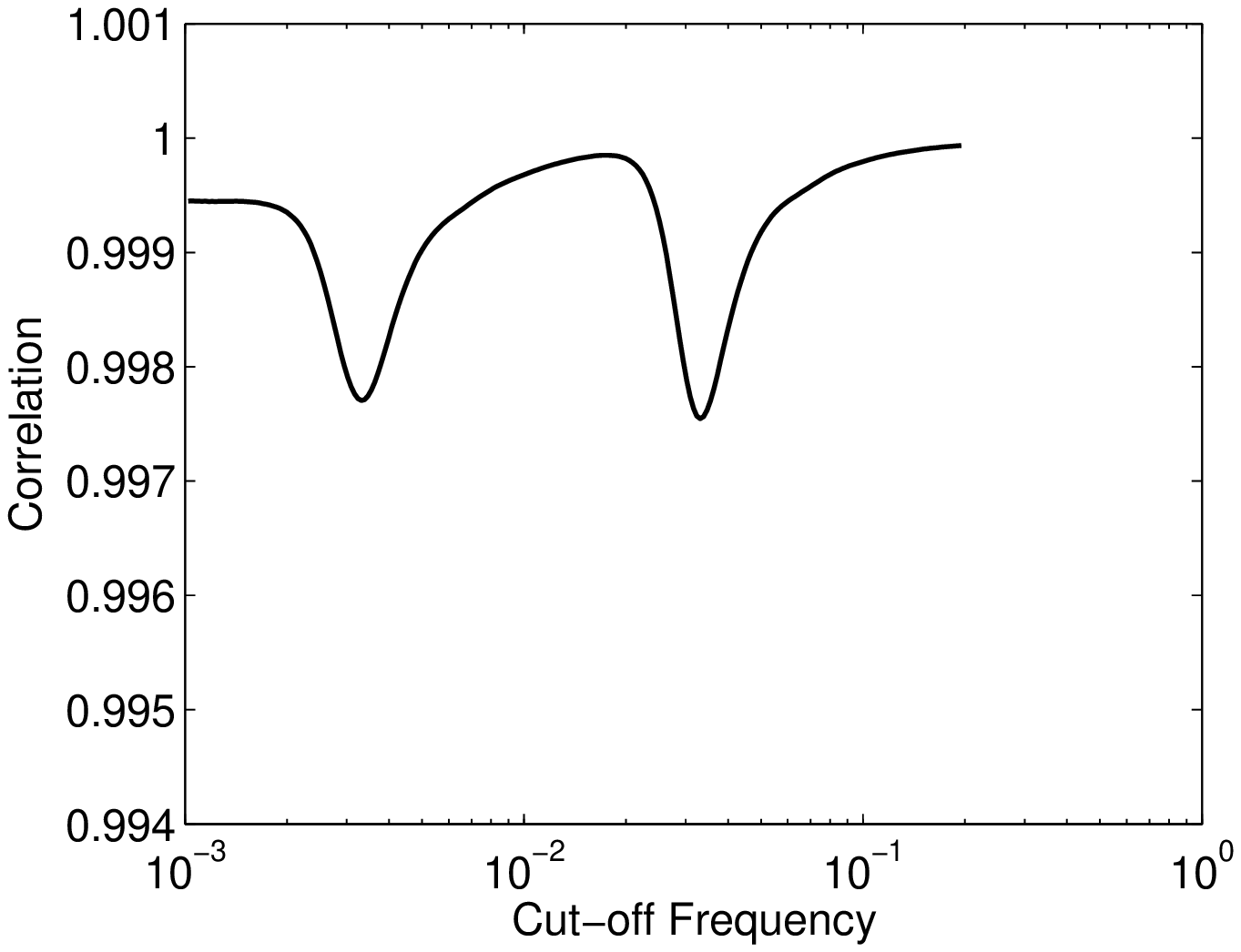}}
\centering
  \subfigure[]{
    \label{fig:subfig:c} %% label for first subfigure
    \includegraphics[width=1.7in]{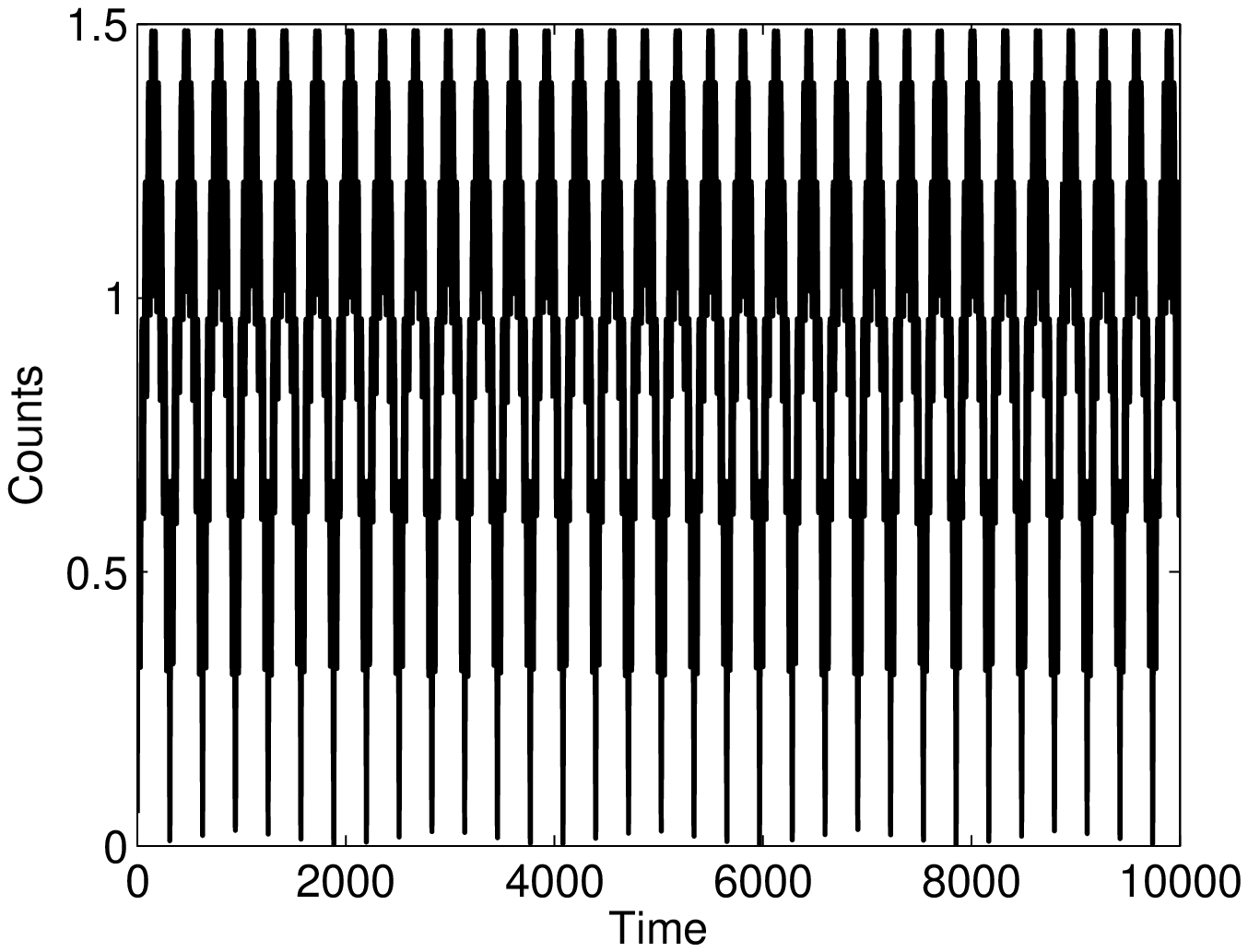}}
  %\hspace{1in}
  \subfigure[]{
    \label{fig:subfig:d} %% label for second subfigure
    \includegraphics[width=1.7in]{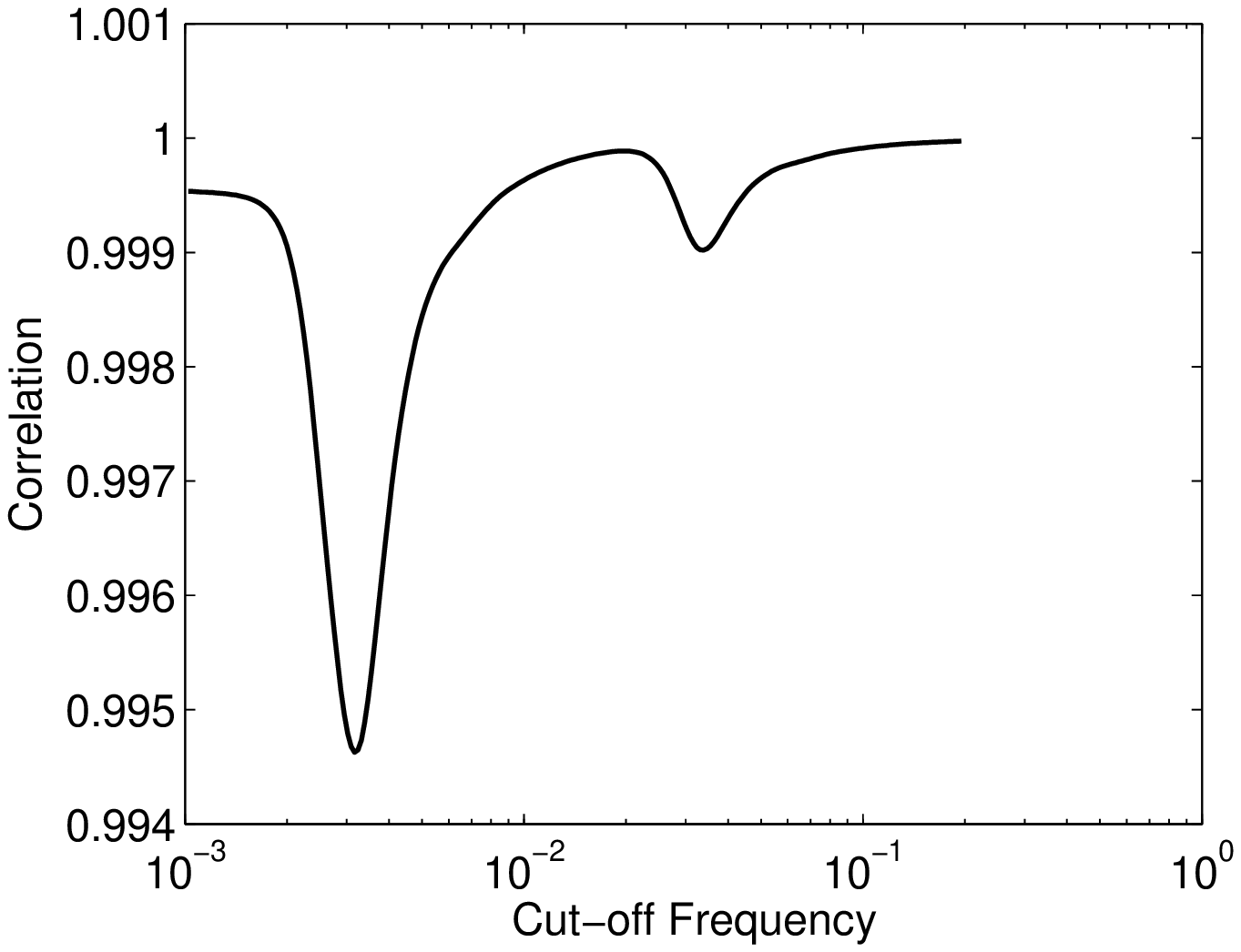}}
\centering
  \subfigure[]{
    \label{fig:subfig:e} %% label for first subfigure
    \includegraphics[width=1.7in]{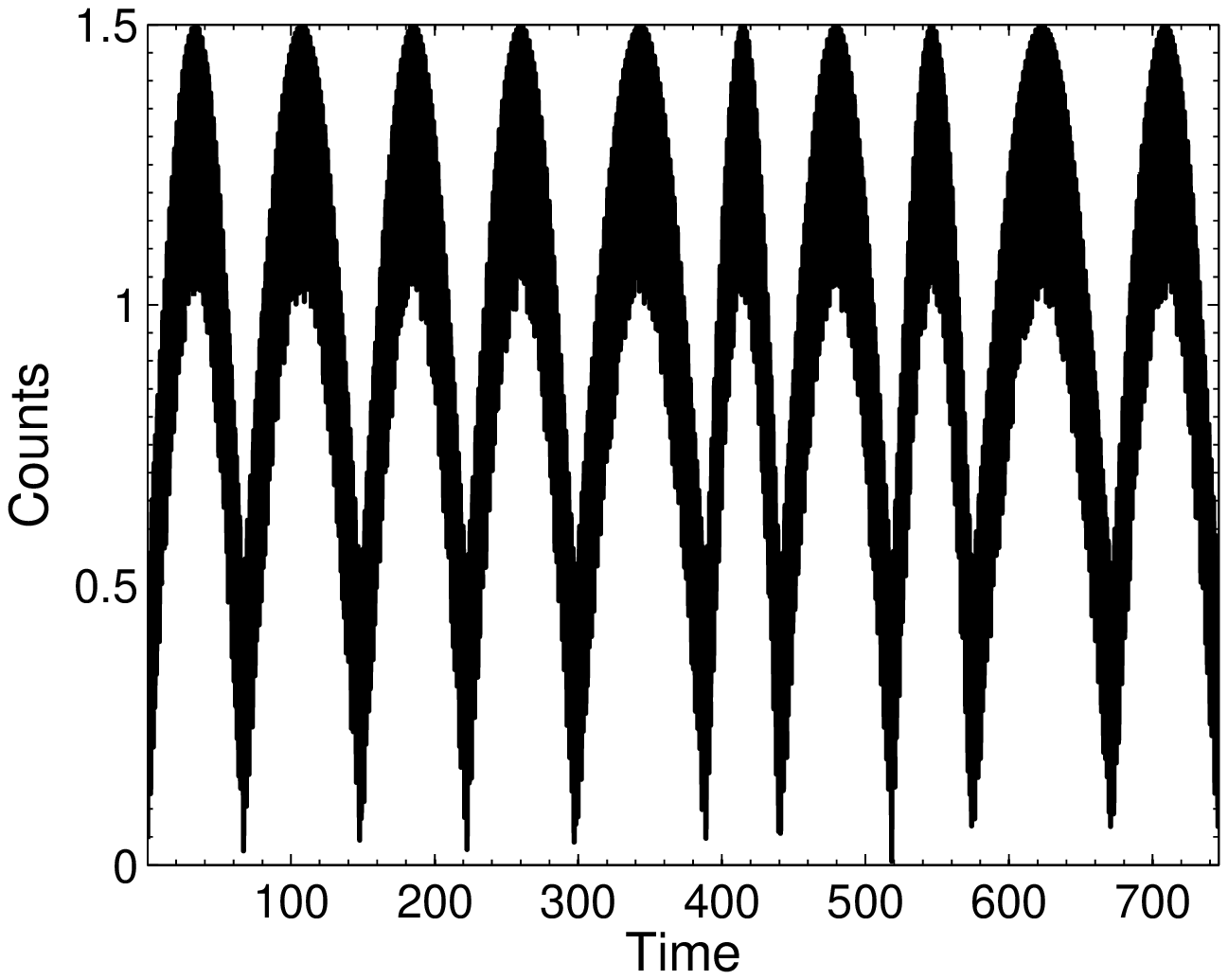}}
  %\hspace{1in}
  \subfigure[]{
    \label{fig:subfig:f} %% label for second subfigure
    \includegraphics[width=1.7in,height=1.25in]{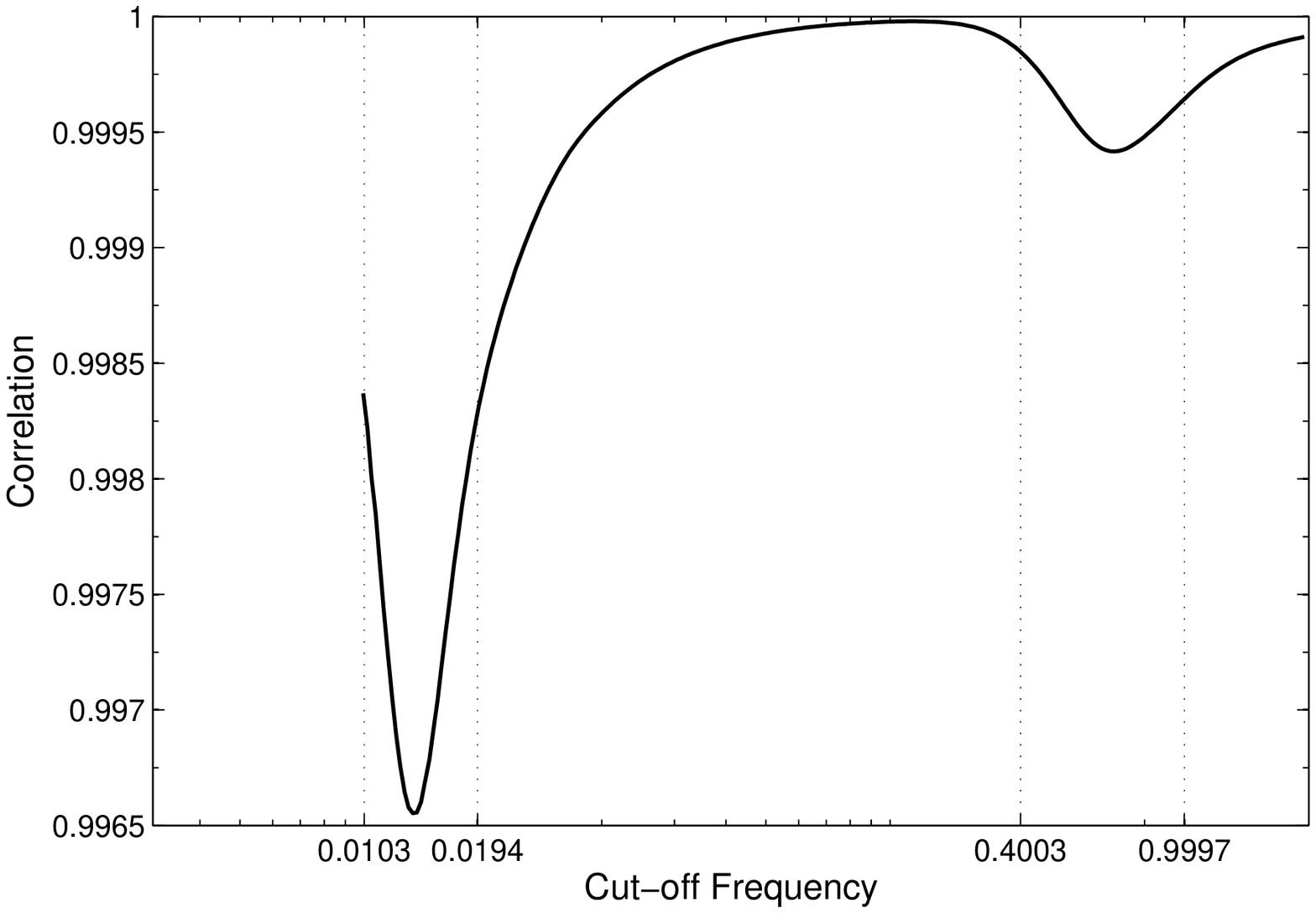}}
\centering
  \subfigure[]{
    \label{fig:subfig:g} %% label for first subfigure
    \includegraphics[width=1.7in]{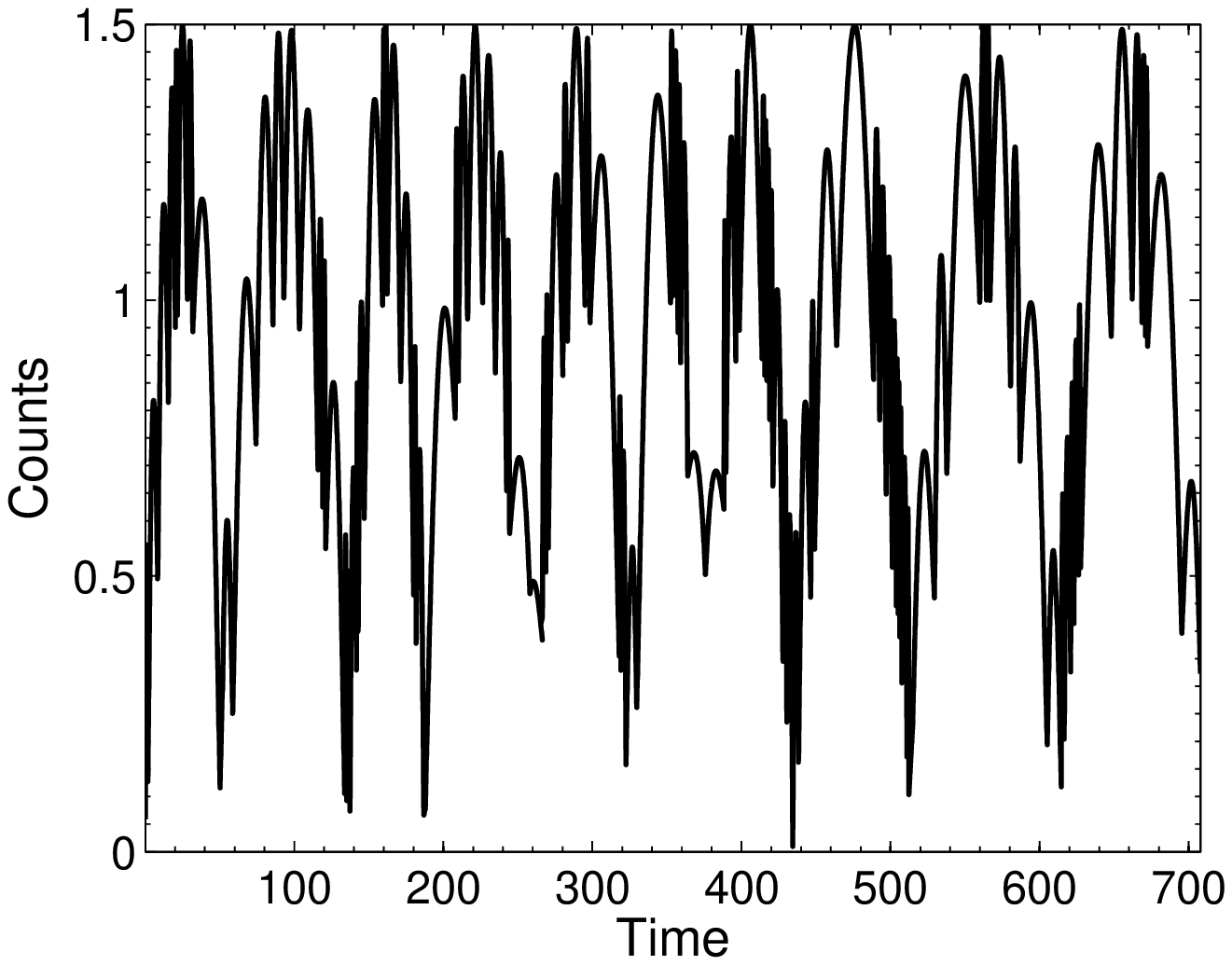}}
  %\hspace{1in}
  \subfigure[]{
    \label{fig:subfig:h} %% label for second subfigure
    \includegraphics[width=1.7in,height=1.25in]{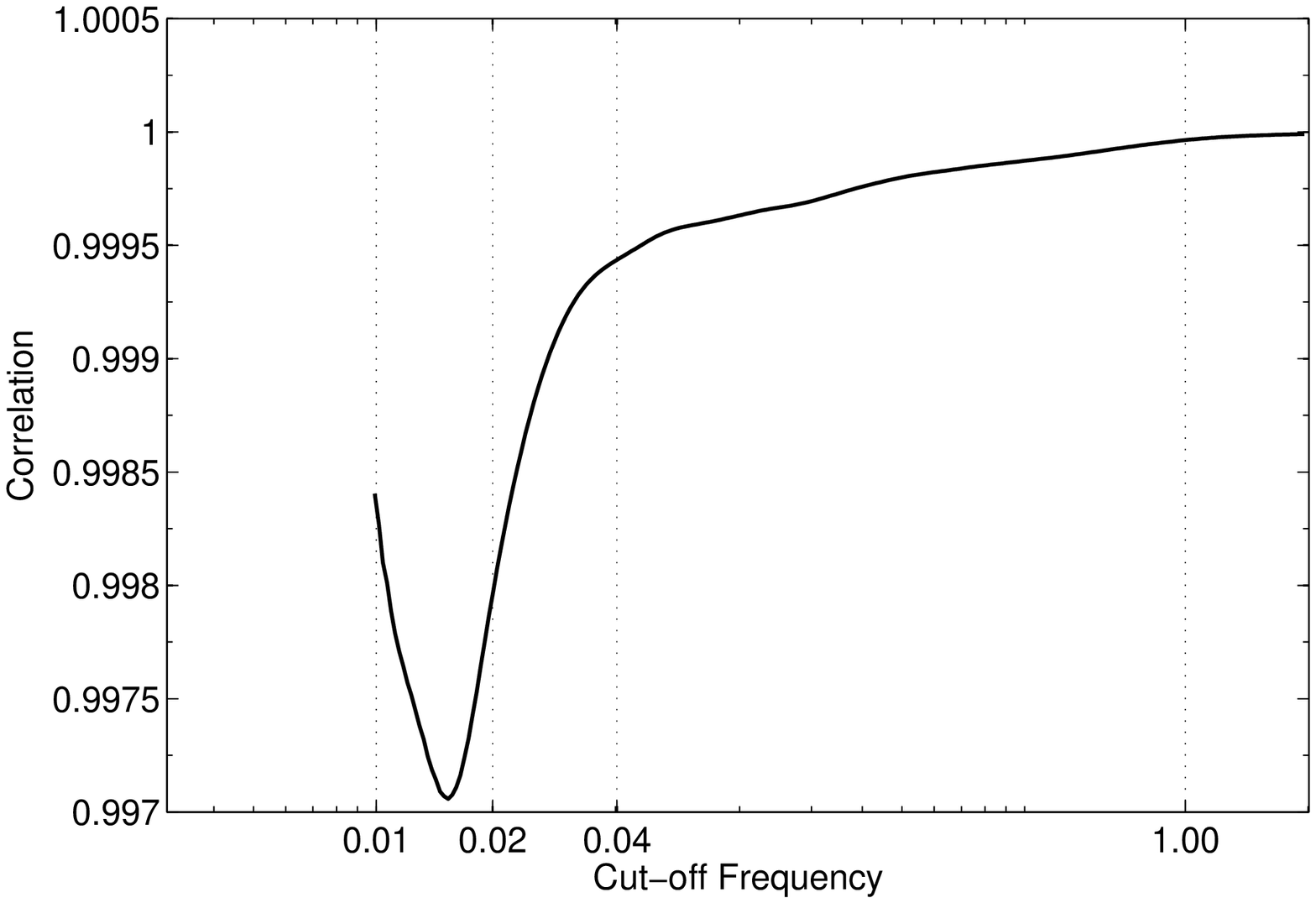}}
    \centering
  \subfigure[]{
    \label{fig:subfig:i} %% label for first subfigure
    \includegraphics[width=1.7in]{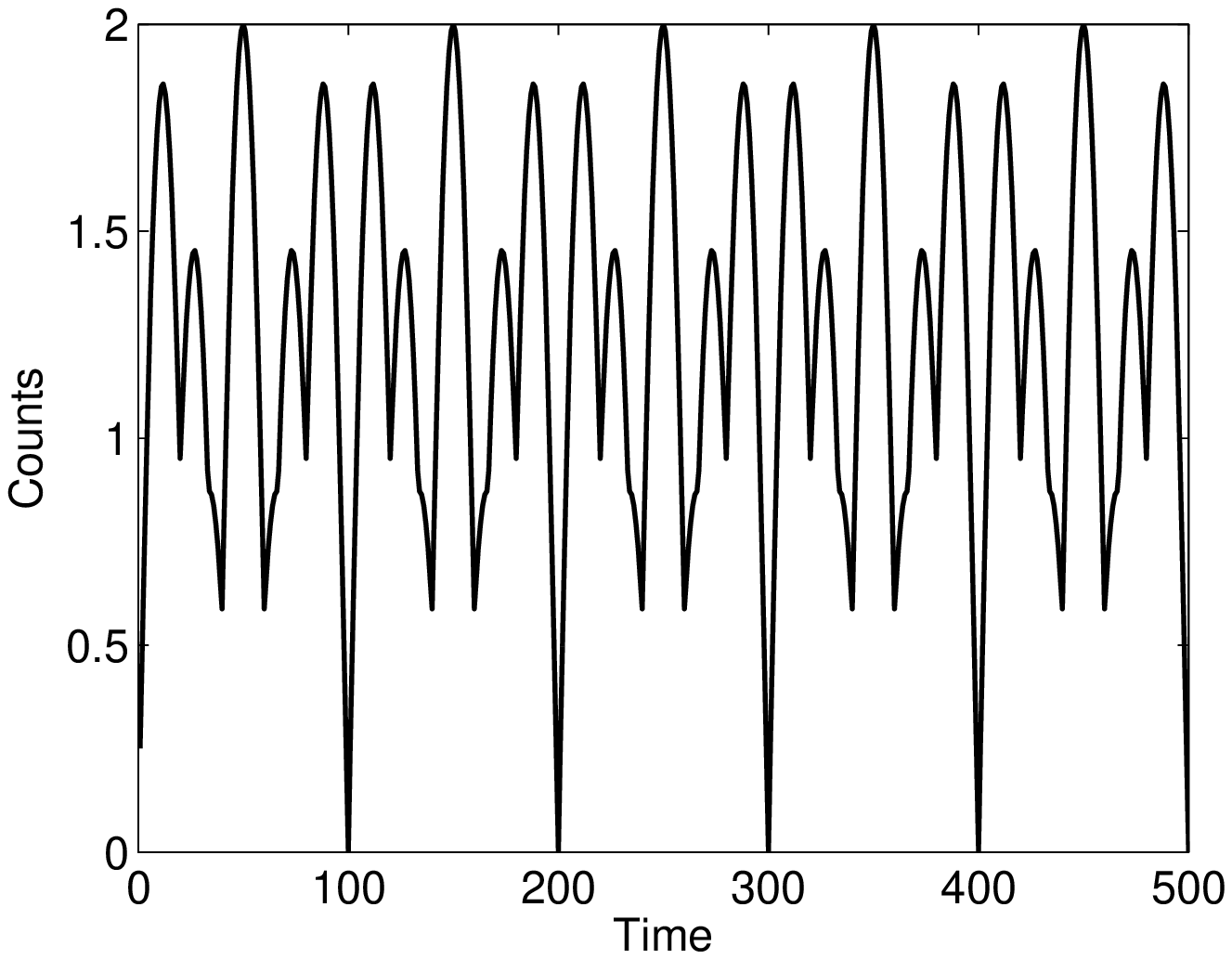}}
  %\hspace{1in}
  \subfigure[]{
    \label{fig:subfig:j} %% label for second subfigure
    \includegraphics[width=1.7in,height=1.25in]{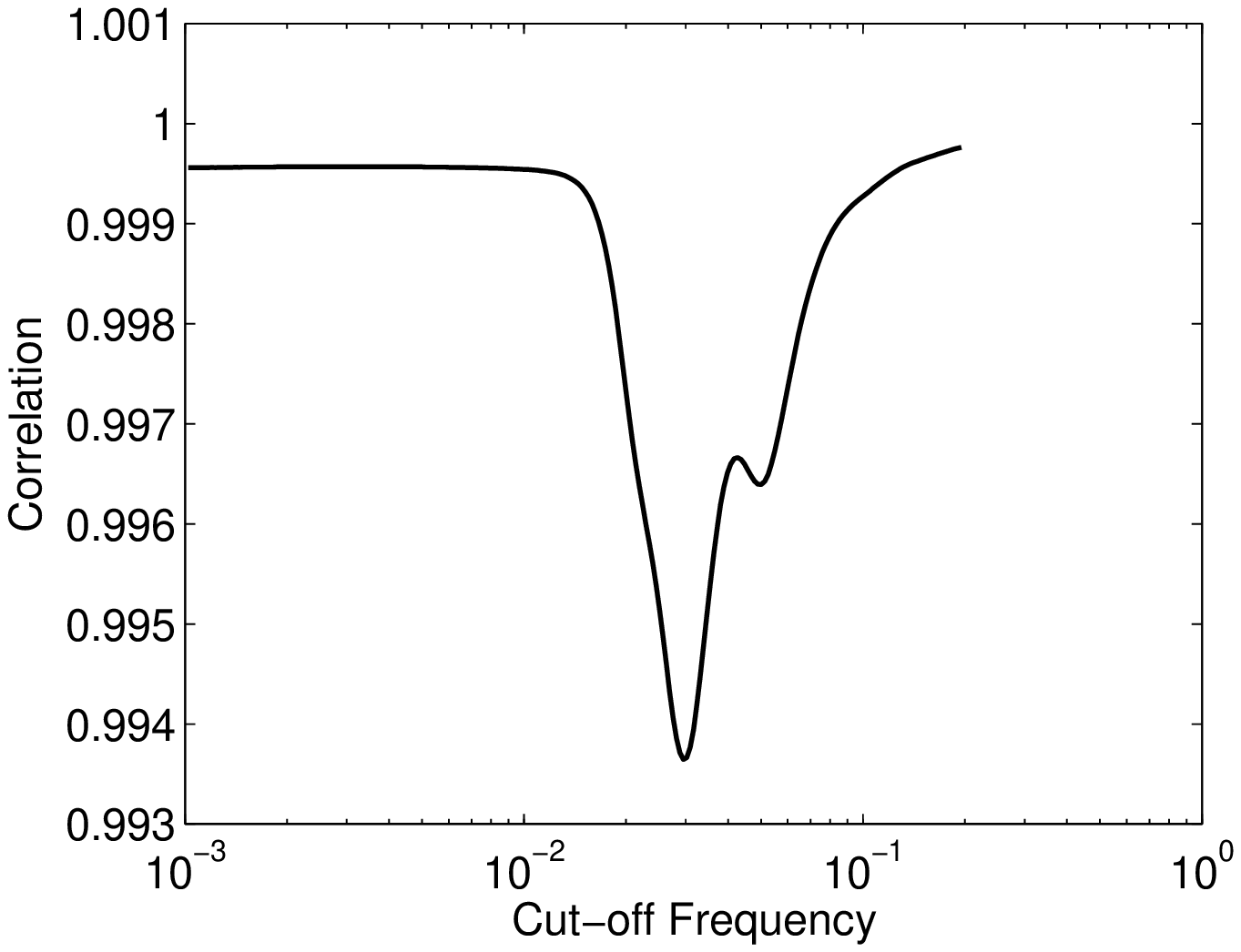}}
\centering
  \subfigure[]{
    \label{fig:subfig:k} %% label for first subfigure
    \includegraphics[width=1.7in]{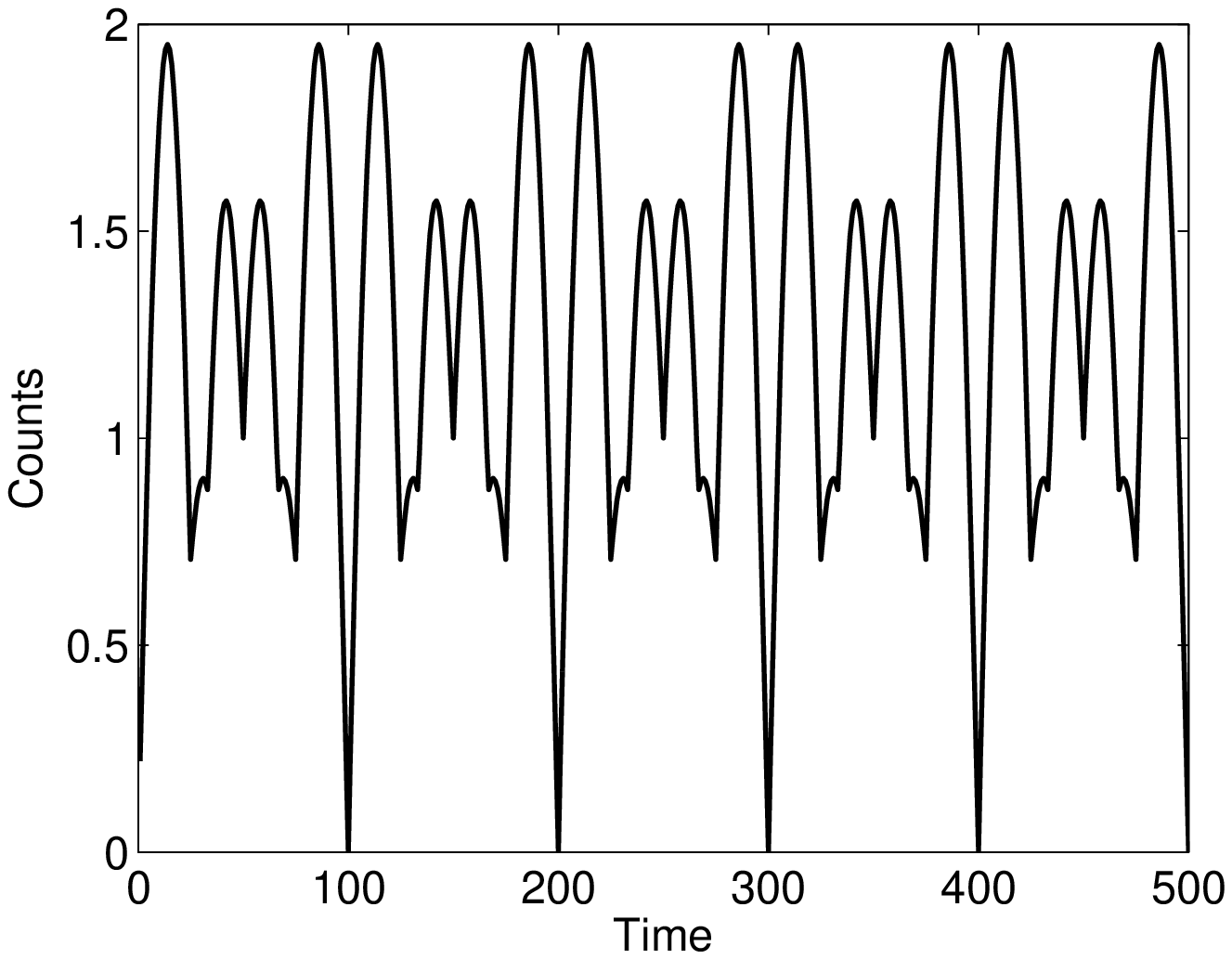}}
  %\hspace{1in}
  \subfigure[]{
    \label{fig:subfig:l} %% label for second subfigure
    \includegraphics[width=1.7in,height=1.25in]{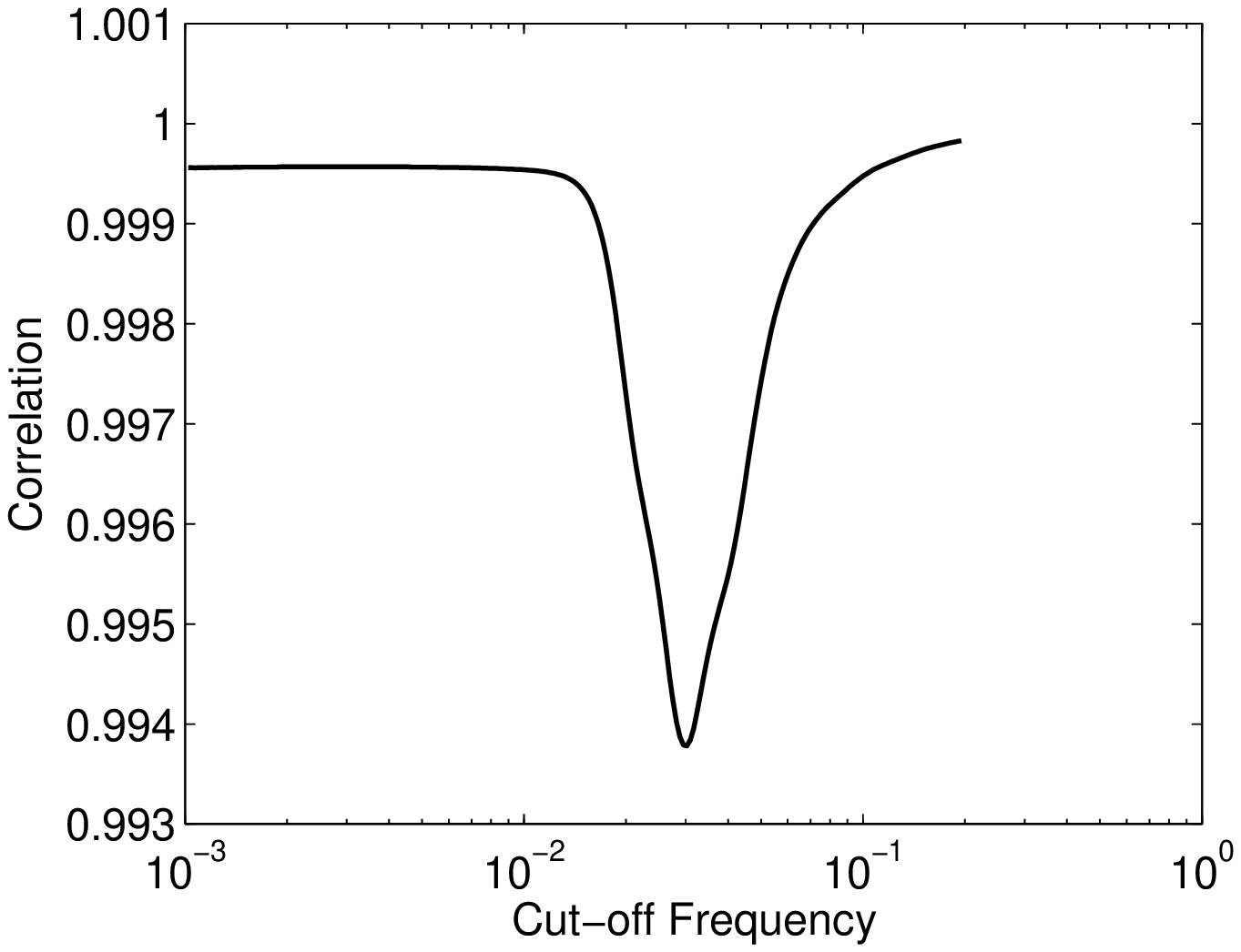}}
\centering
  \subfigure[]{
    \label{fig:subfig:m} %% label for first subfigure
    \includegraphics[width=1.7in]{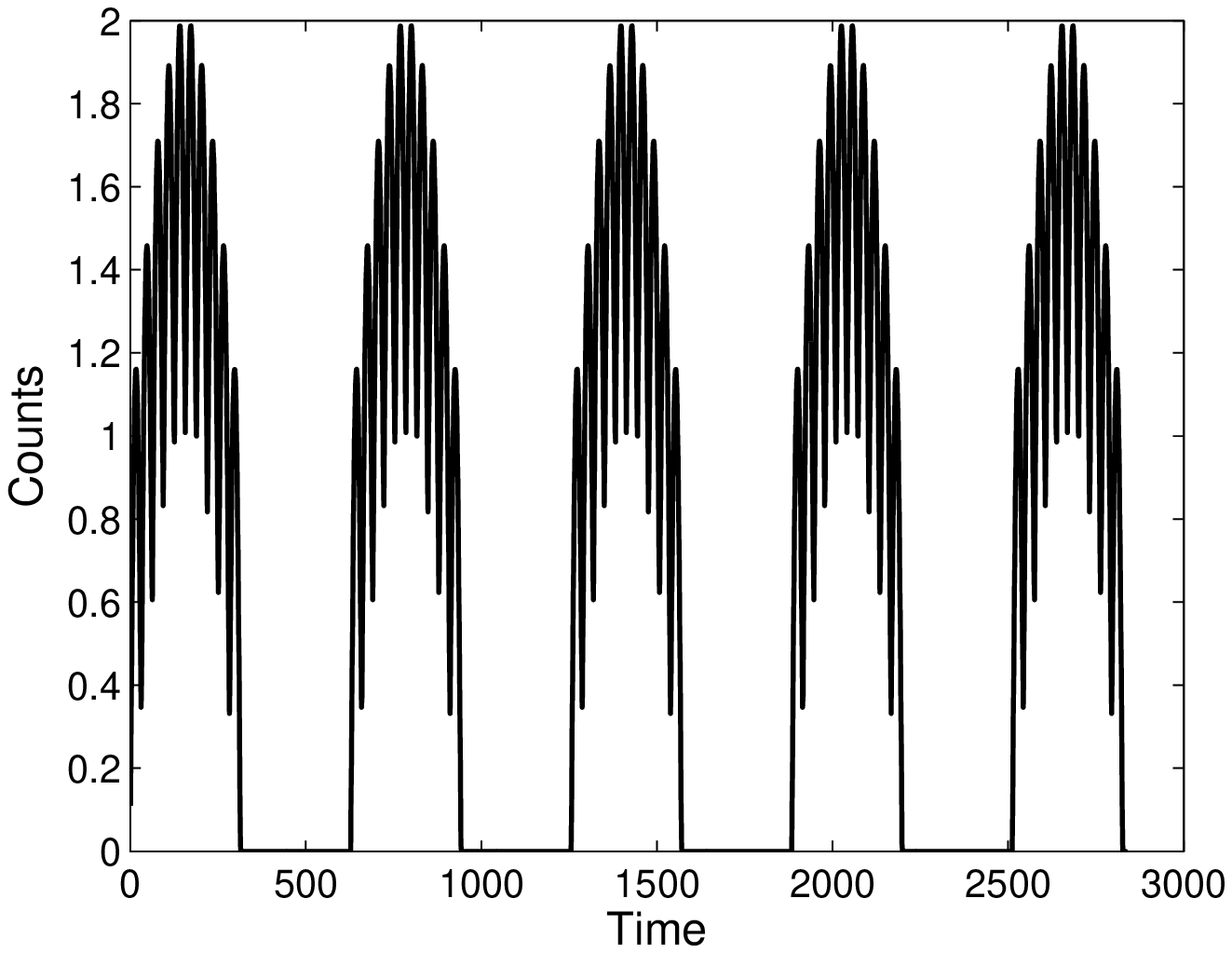}}
  %\hspace{1in}
  \subfigure[]{
    \label{fig:subfig:n} %% label for second subfigure
    \includegraphics[width=1.7in,height=1.25in]{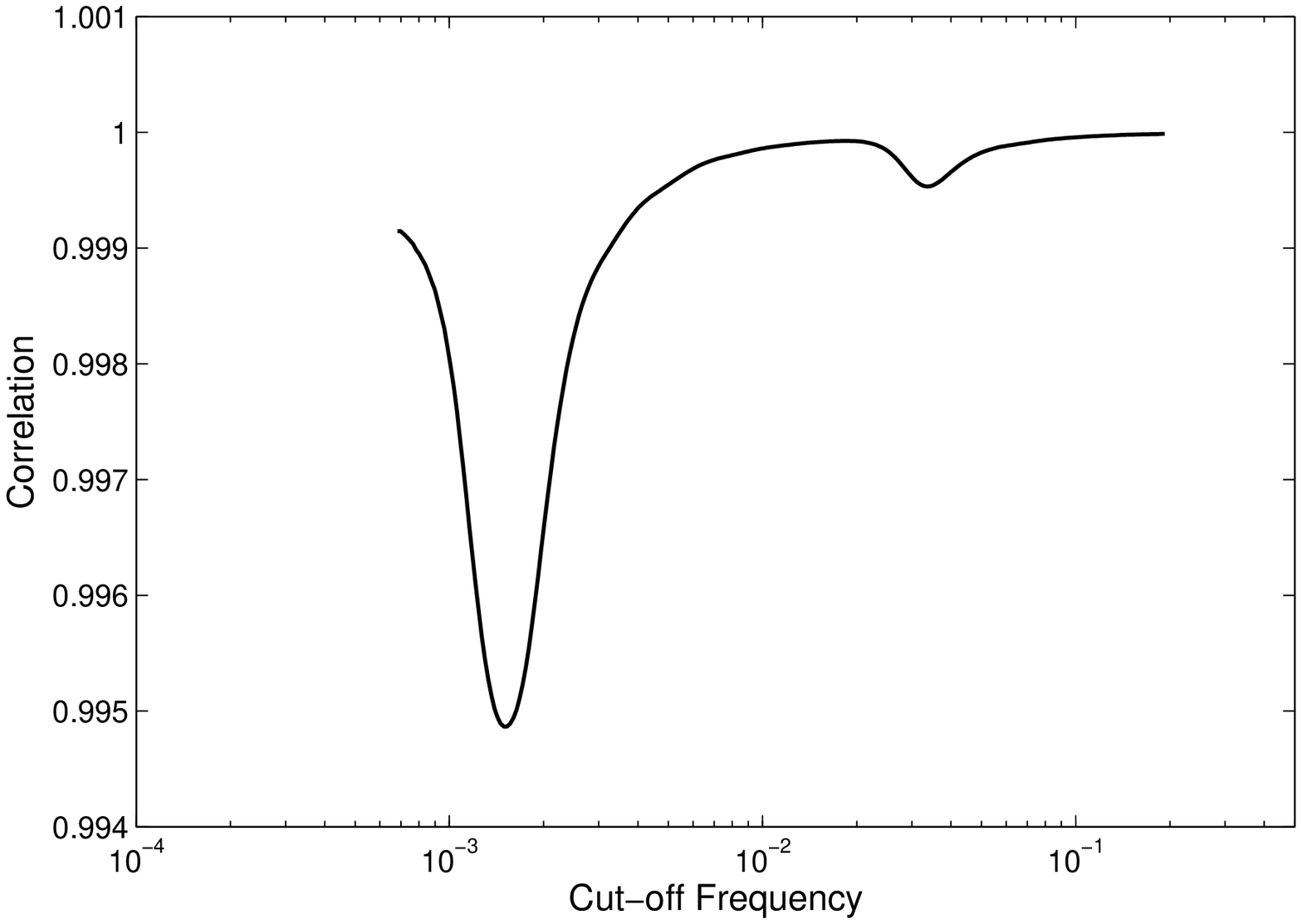}}
\centering
  \subfigure[]{
    \label{fig:subfig:o} %% label for first subfigure
    \includegraphics[width=1.7in]{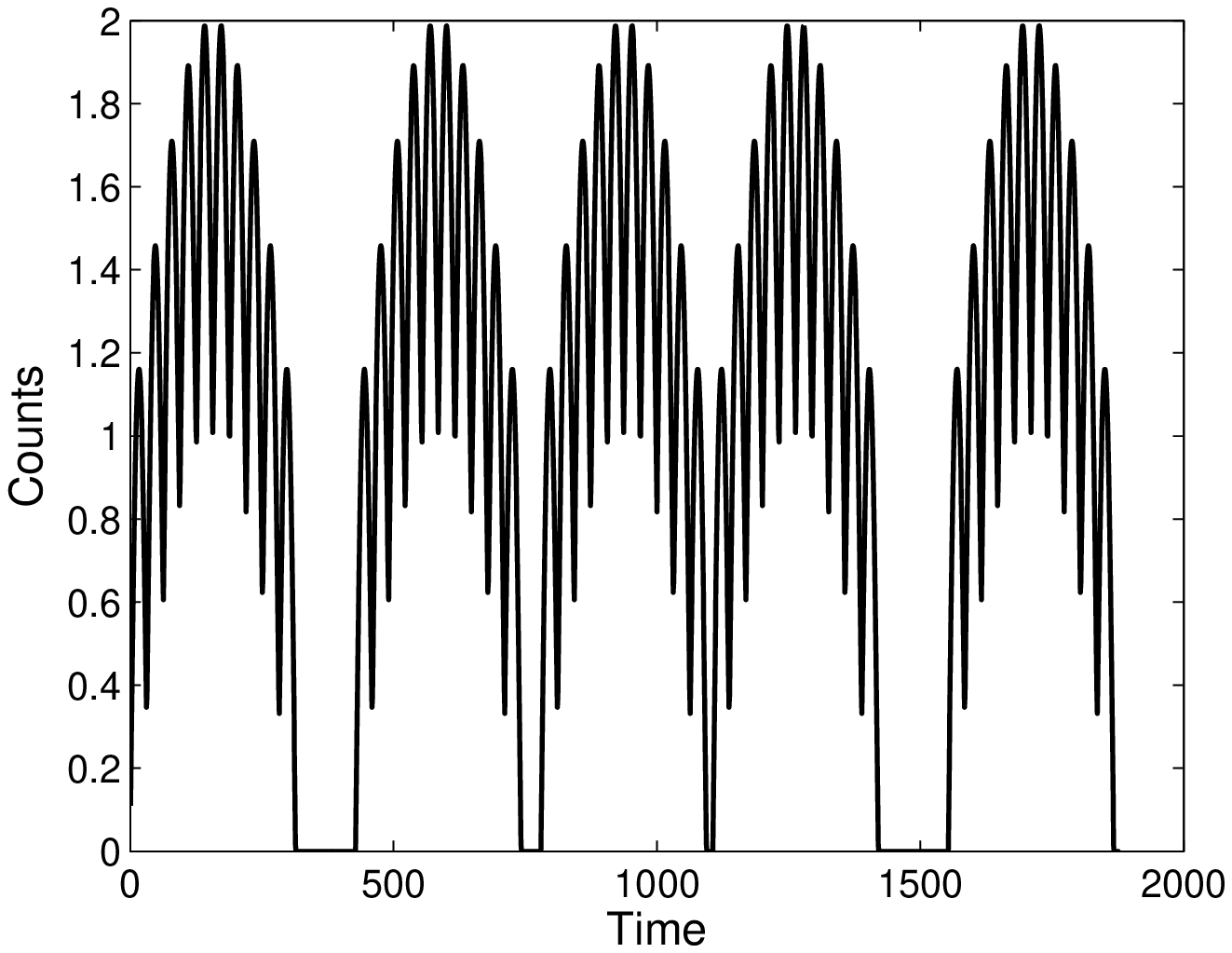}}
  %\hspace{1in}
  \subfigure[]{
    \label{fig:subfig:p} %% label for second subfigure
    \includegraphics[width=1.7in,height=1.25in]{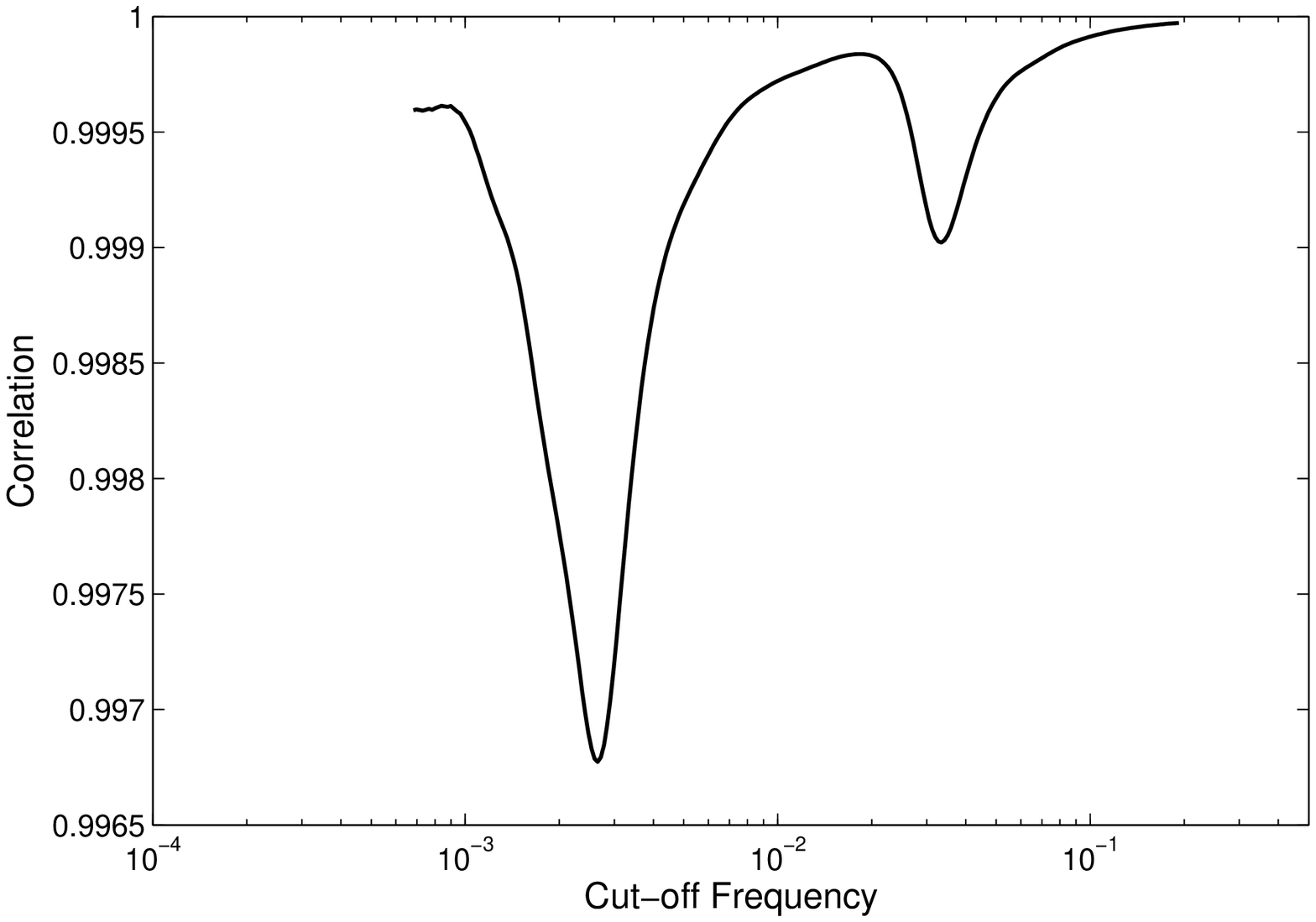}}
  \caption{Mock catalog of lightcurves with different pulse properties and their relevant correlation curves.}
  \label{fig:subfig} %% label for entire figure
\label{others}
\end{figure}

2) Amplitude of pulses: The amplitude of a frequency components is
an important factor. For our superposition tests, the relative
depths of the dips depend on the amplitude ratio of the slow
and fast components. This can be seen from the comparison
of Fig.\ref{profile}(a,b) for $A_{\rm s}:A_{\rm f}=2:1$
and Fig.\ref{others}(a,b) for $A_{\rm s}:A_{\rm f}=1:1$
(with the nominal parameters $T_s=100 \pi$ s, and $T_f = 10
\pi$ s). For this set of parameters, the
fast component dip disappears when  $A_{\rm s} : A_{\rm f}
> 15:1$, while the slow component dip disappears when
$A_{\rm s} : A_{\rm f} < 1:25$. The asymmetry is understandable
since a low pass filter favors the slow component.

%For the mock lightcurve as shown in Fig.6(a),
%we change the amplitude ratio from $A_{\rm s}:A_{\rm f}=2:1$ to
%$A_{\rm s}:A_{\rm f}=1:1$, and fix other parameters. Compare
%subfigures of Fig.6(i)$\sim$ Fig.6(j) with Fig.6(a) $\sim$ Fig.6(b),
%we can see that amplifying amplitude of pulses will greatly enhance
%the significance of the corresponding dip in the SFC $R_{\rm
%i}-f_{\rm c,i}$ curve.

3) Number of pulses:
%Similar with 2), we change the number of pulses
%rather than the amplitude ratio, and also fix other parameters.
Fixing the nominal parameters but increasing the number of pulses,
we find that the corresponding dips in the SFC $R_{\rm i}-f_{\rm c,i}$
curve become deeper. See Fig.\ref{others}(c,d) as compared with
Fig.\ref{profile}(a,b).
%Compare subfigures of Fig.6(k)$\sim$ Fig.6(l) with Fig.6(a)$\sim$
%Fig.6(b), we can see that changing the number of pulses will also
%influence the significance of the corresponding dip in the SFC
%$R_{\rm i}-f_{\rm c,i}$ curve. However, we note that the relevant
%dip of slow component is still apparent even if the number of slow
%component pulse is only 6 (see Fig.6(a)$\sim$ Fig.6(b)).
On the other hand, the slow component can be detected even with one
single broad pulse, as long as its amplitude is large enough.
See Fig.\ref{profile}(i-p).

4) Pulse duration spread: Similar to the mock light curve shown in
the middle panel of Fig.1, we generate a set of light curves whose
slow component duration range is fixed in $T_{\rm s}= (50 - 100)$ s
and the amplitude ratio is fixed to $A_{\rm s}:A_{\rm f}=2:1$. We
gradually spread the fast component duration range. We find that
the significance of the relevant dip of the fast component in the SFC
$R_{\rm i}-f_{\rm c,i}$ curve diminishes and eventually disappears
as the frequency spread is wide enough (Fig.\ref{others}(e-h)).
% and then disappear
%(see Fig.6(m)$\sim$ Fig.6(p)). We propose that the pulse duration
%spread is another important properties the SFC sensitive to.

5) Separation between two components: Back to the two-frequency case,
we fix the amplitude ratio as $A_{\rm s}:A_{\rm f}=1:1$ and the slow
component duration as $T_{\rm s}= 100$, and then gradually brings
the fast component duration closer and closer to the slow one. We
find that the significance of the relevant dip of fast component
in the SFC $R_{\rm i}-f_{\rm c,i}$ curve diminishes, and merges
with the slow frequency component when $\frac{f_{\rm f}-f_{\rm s}}
{f_{\rm s}} \leq 0.5$ is satisfied (Fig.\ref{others}(i-n)).
%Fig.6(o)$\sim$ Fig.6(t)).

6) Gaps between pulses: Here we still test the simple two-frequency
case. We fix the pulse duration of slow component as $100\pi$, fast
component as $10\pi$, and the amplitude ratio between the two
components as $A_{\rm s}:A_{\rm f}=1:1$.  We then add gaps in the
lightcurve between the slow component pulses. The gap duration is
fixed to $100\pi$ (Fig.\ref{others}(o,p)),
%Fig.6(u)$\sim$ Fig.6(v))
or is randomly distributed in the range $0 - 100\pi$
(Fig.\ref{others}(q,r)).  We find that the SFC curve
still shows two components. However, the corresponding period for
slow component is larger than $100\pi$, indicating that part
of the gap duration is added to the pulse. We therefore draw
the conclusion that one should be careful to perform SFC analysis
when substantial gaps exist in a lightcurve. Indeed, only when
the gap is manually removed, can the original pulse width restored
(see Appendix 3 for a case study).

%\clearpage

%\appendix
\section{3. Case studies of GRBs not belonging to the good sample}

\begin{figure}
  \centering
  \subfigure[]{
    \label{fig:subfig:a} %% label for first subfigure
    \includegraphics[width=1.7in]{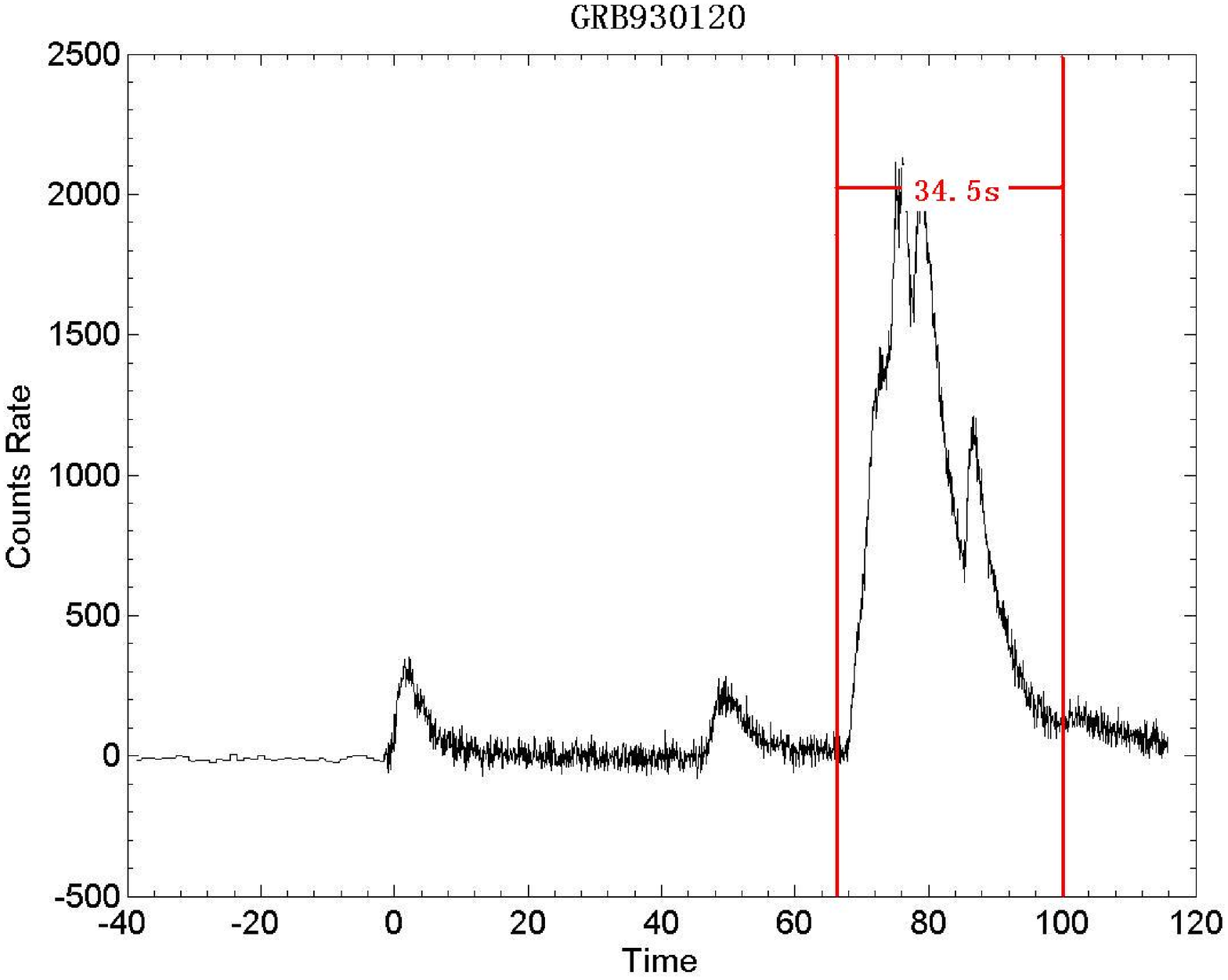}}
  %\hspace{1in}
  \subfigure[]{
    \label{fig:subfig:b} %% label for second subfigure
    \includegraphics[width=1.7in]{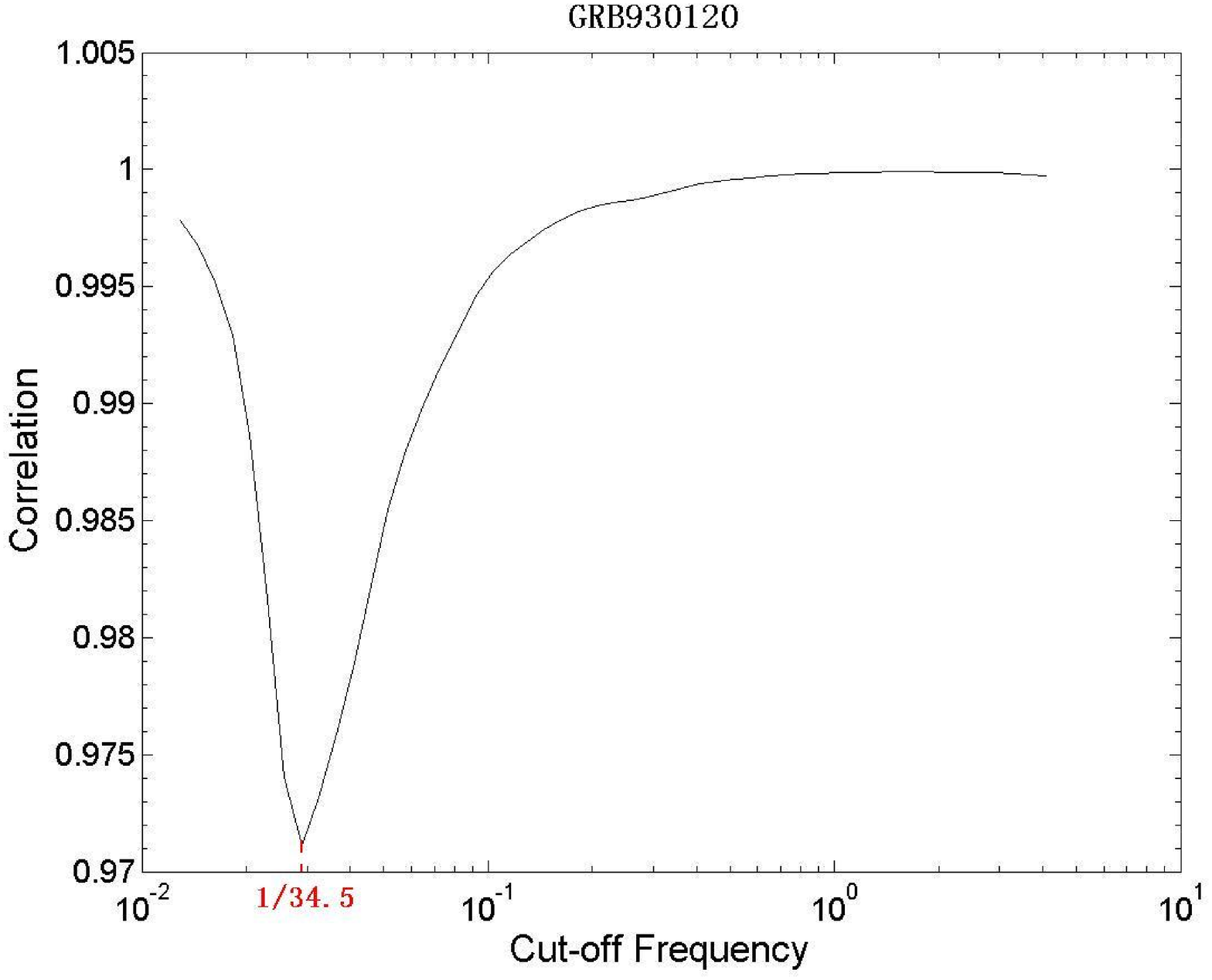}}
\centering
  \subfigure[]{
    \label{fig:subfig:a} %% label for first subfigure
    \includegraphics[width=1.7in]{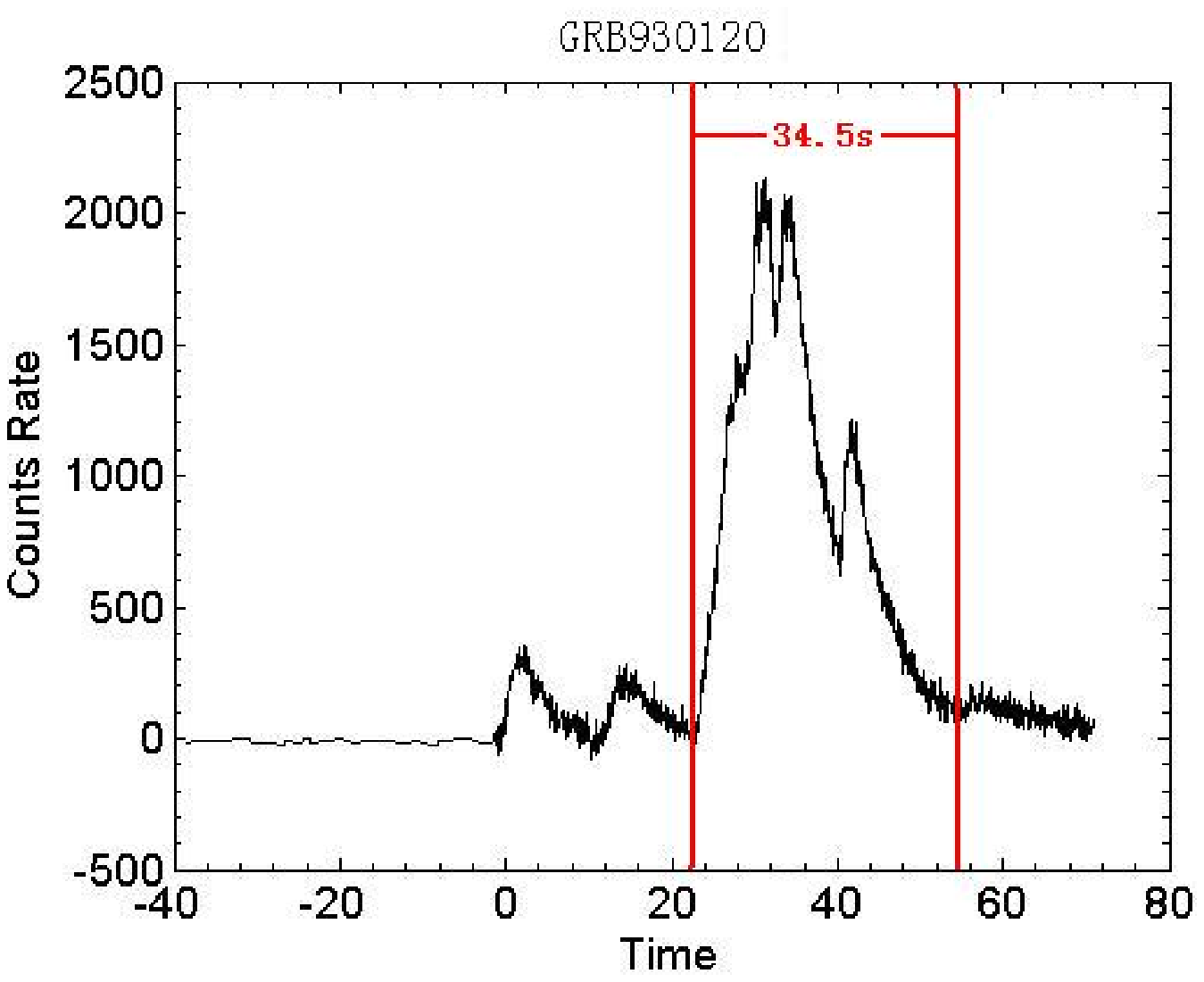}}
  %\hspace{1in}
  \subfigure[]{
    \label{fig:subfig:b} %% label for second subfigure
    \includegraphics[width=1.7in]{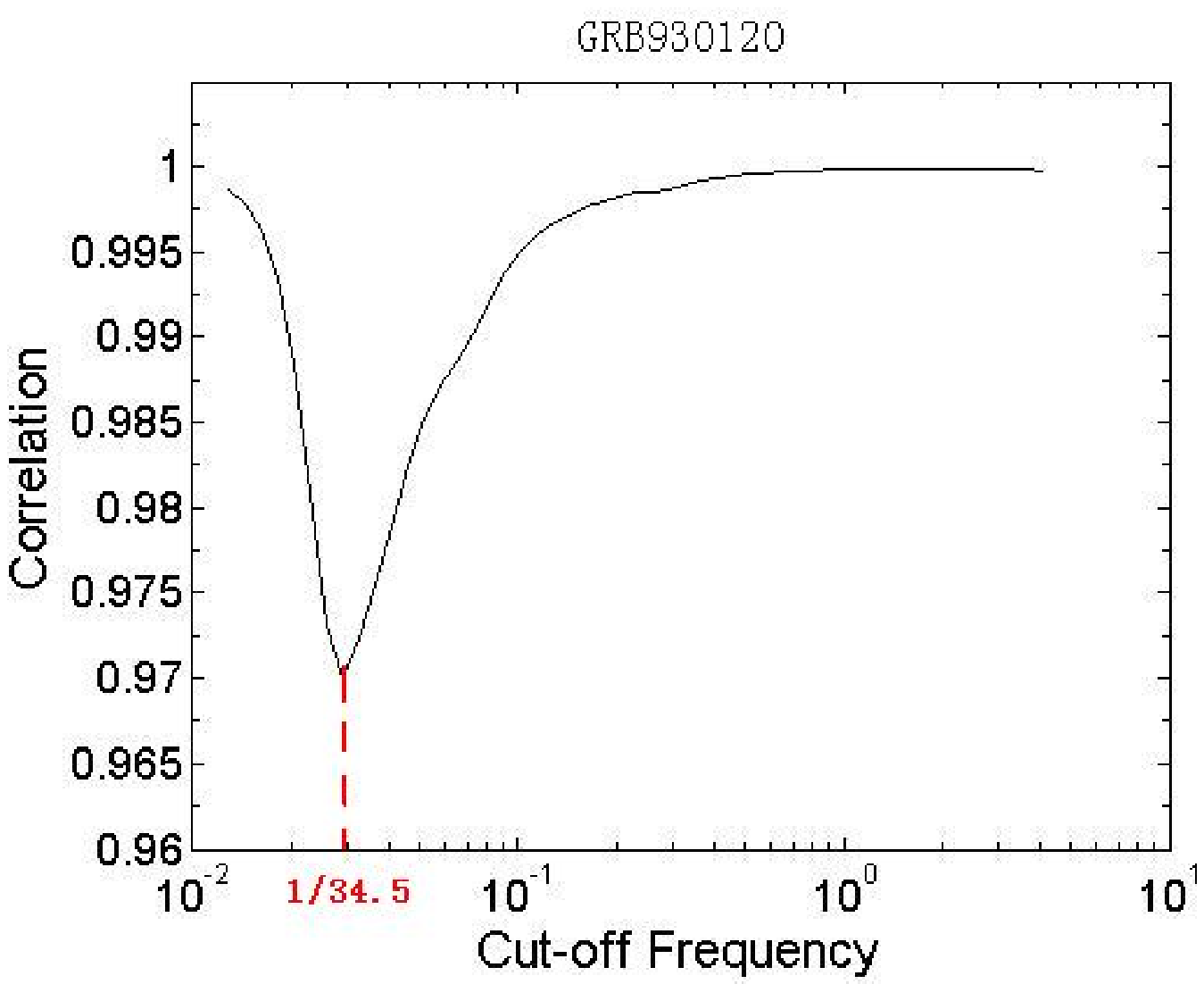}}
\centering
  \subfigure[]{
    \label{fig:subfig:a} %% label for first subfigure
    \includegraphics[width=1.7in]{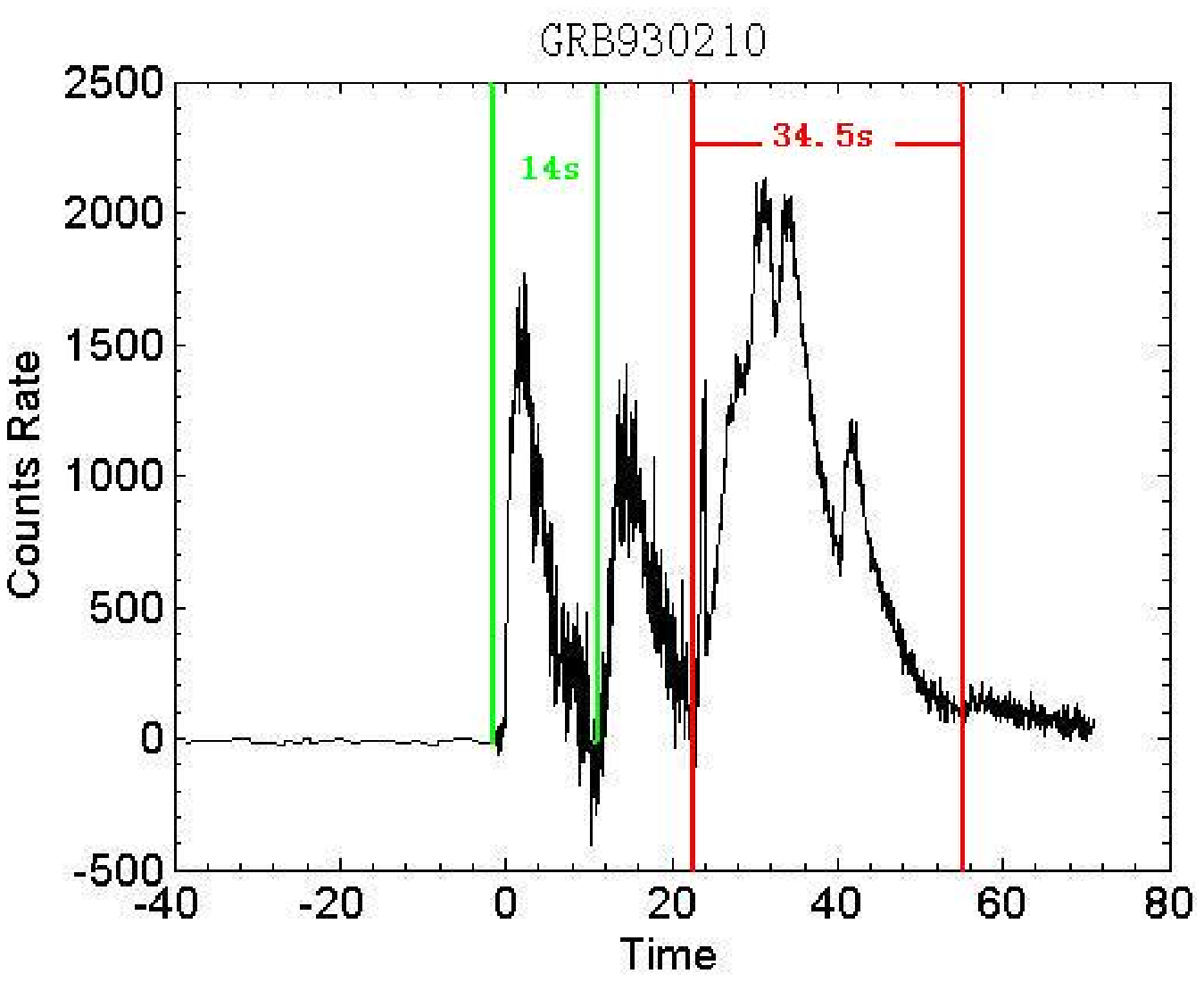}}
  %\hspace{1in}
  \subfigure[]{
    \label{fig:subfig:b} %% label for second subfigure
    \includegraphics[width=1.7in]{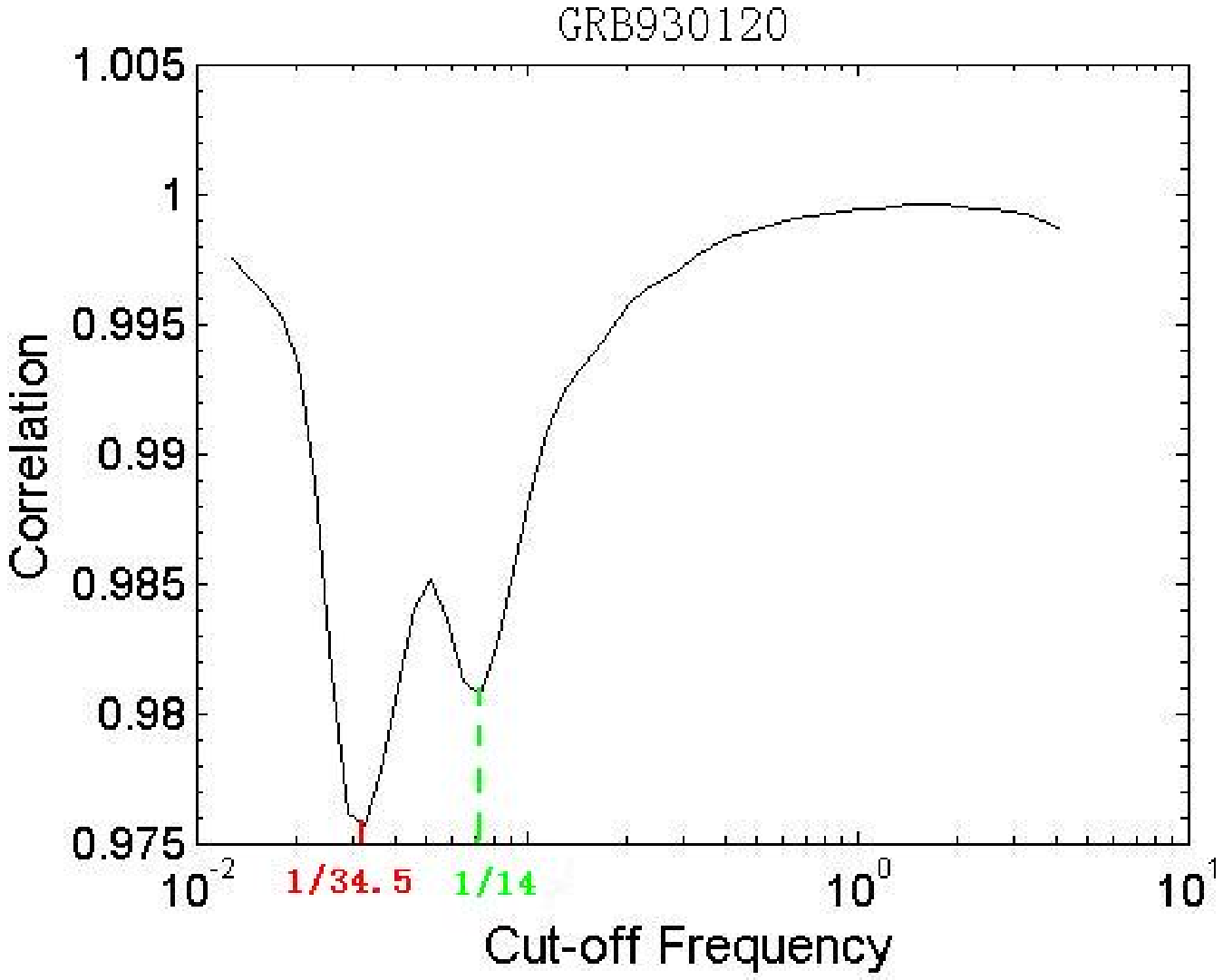}}
\centering
  \subfigure[]{
    \label{fig:subfig:a} %% label for first subfigure
    \includegraphics[width=1.7in]{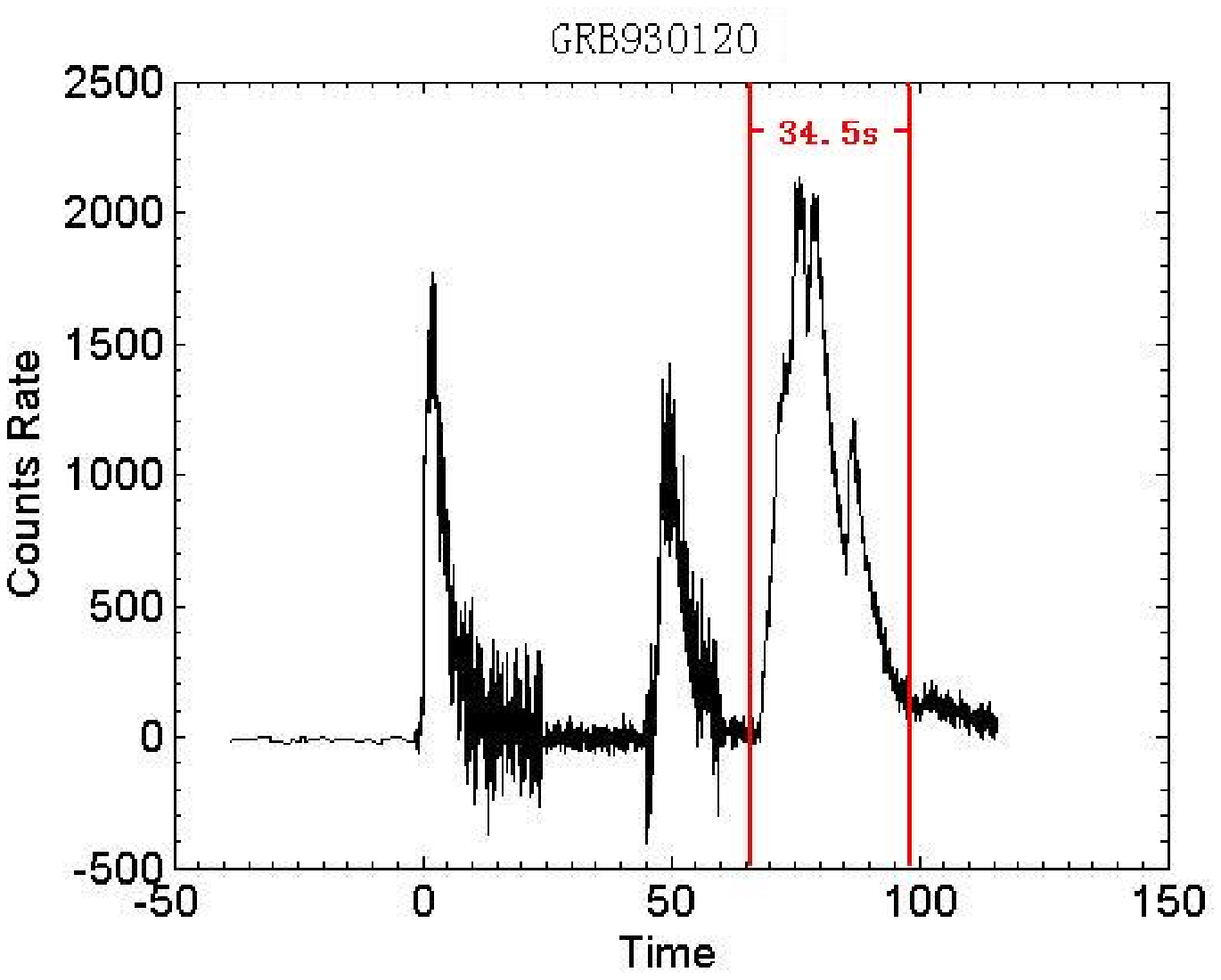}}
  %\hspace{1in}
  \subfigure[]{
    \label{fig:subfig:b} %% label for second subfigure
    \includegraphics[width=1.7in]{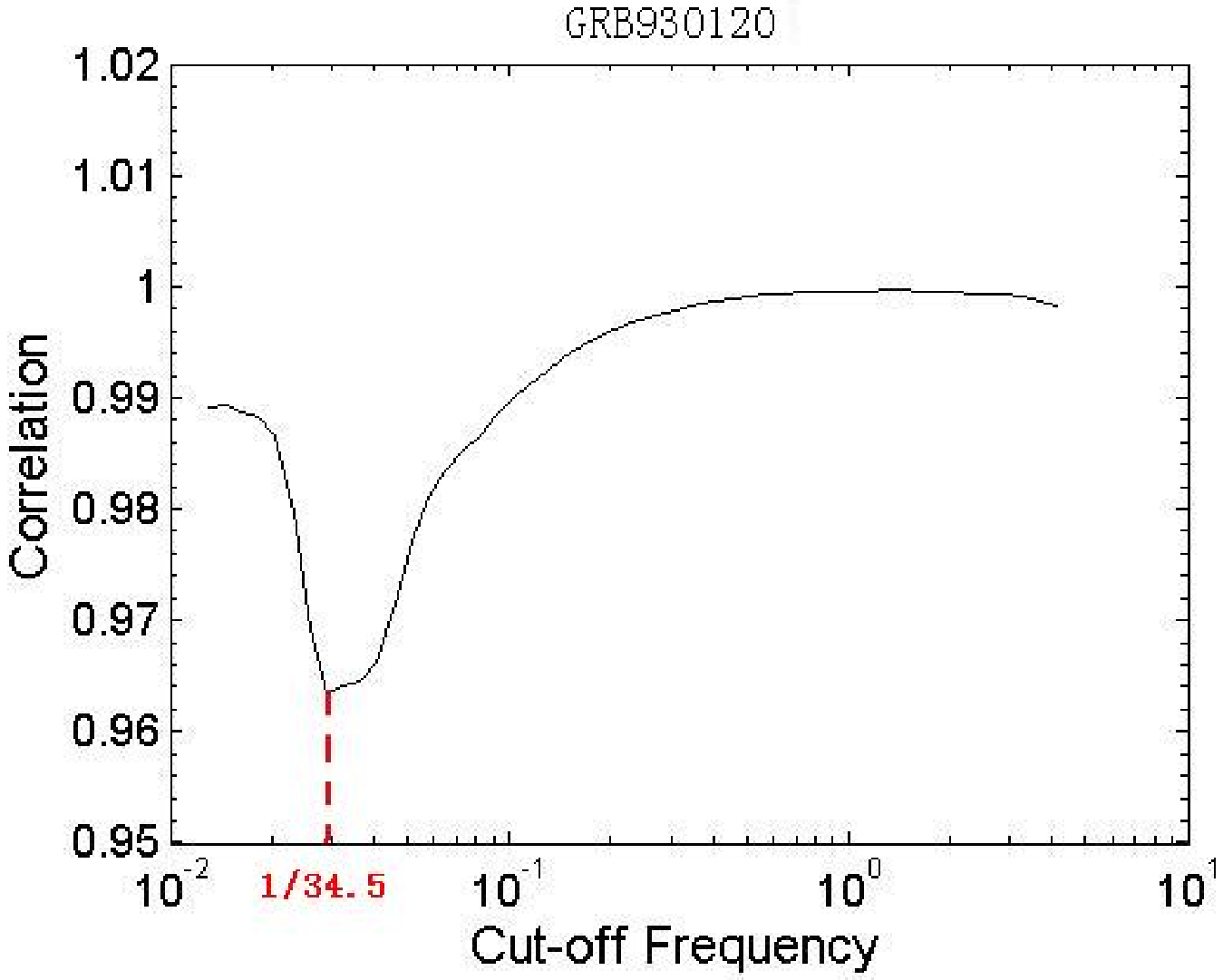}}
    \centering
  \subfigure[]{
    \label{fig:subfig:a} %% label for first subfigure
    \includegraphics[width=1.7in]{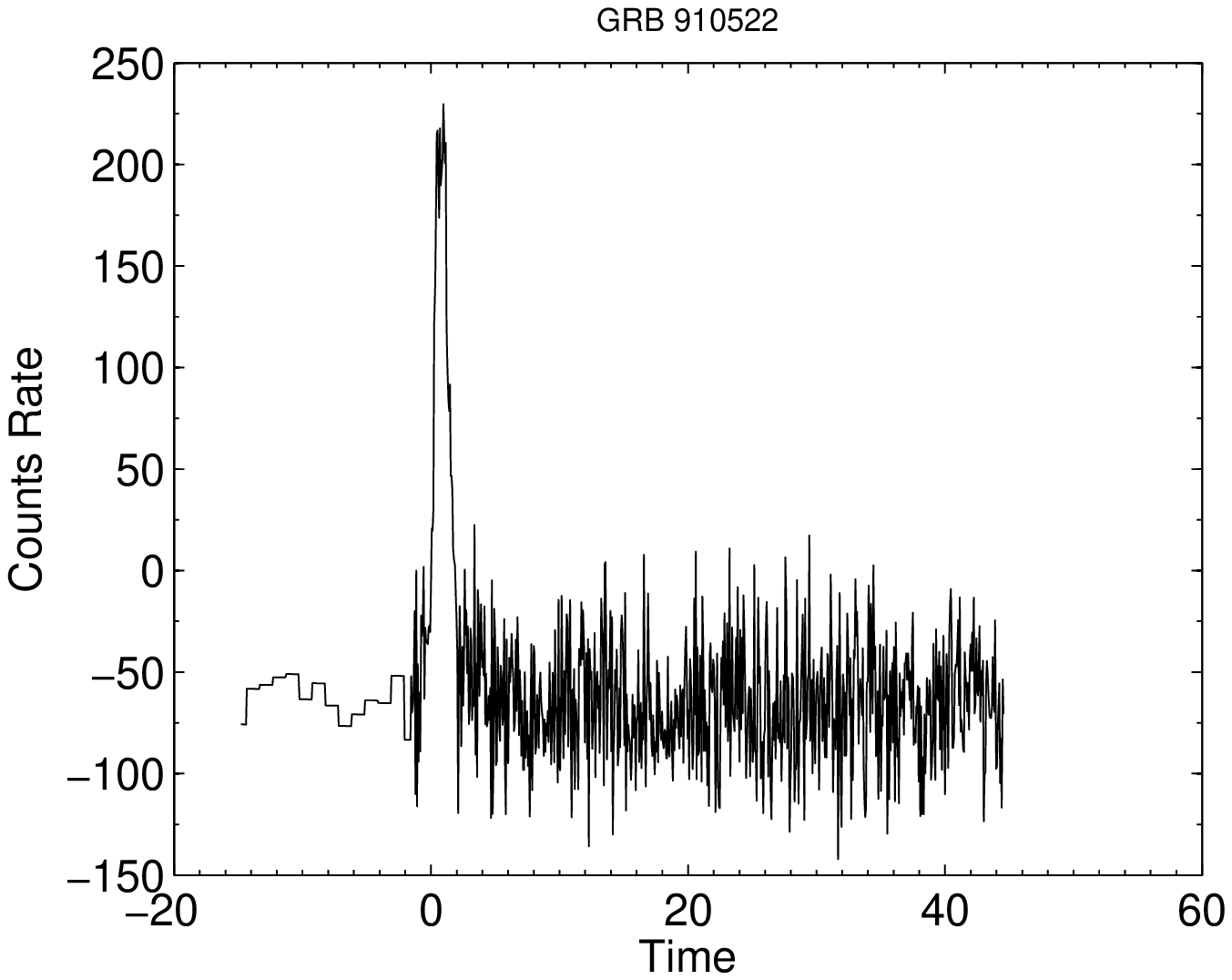}}
  %\hspace{1in}
  \subfigure[]{
    \label{fig:subfig:b} %% label for second subfigure
    \includegraphics[width=1.7in]{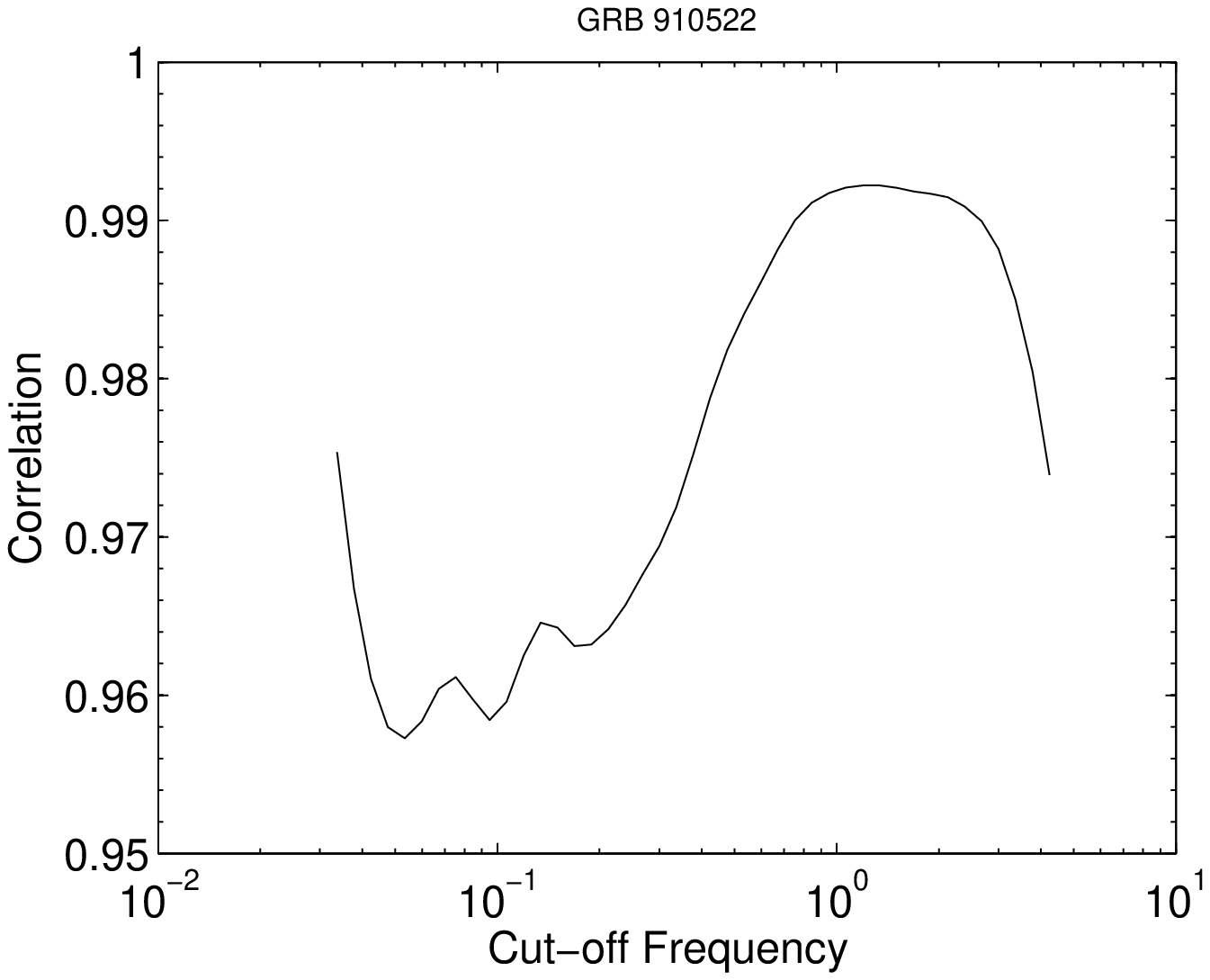}}
\centering
  \subfigure[]{
    \label{fig:subfig:a} %% label for first subfigure
    \includegraphics[width=1.7in]{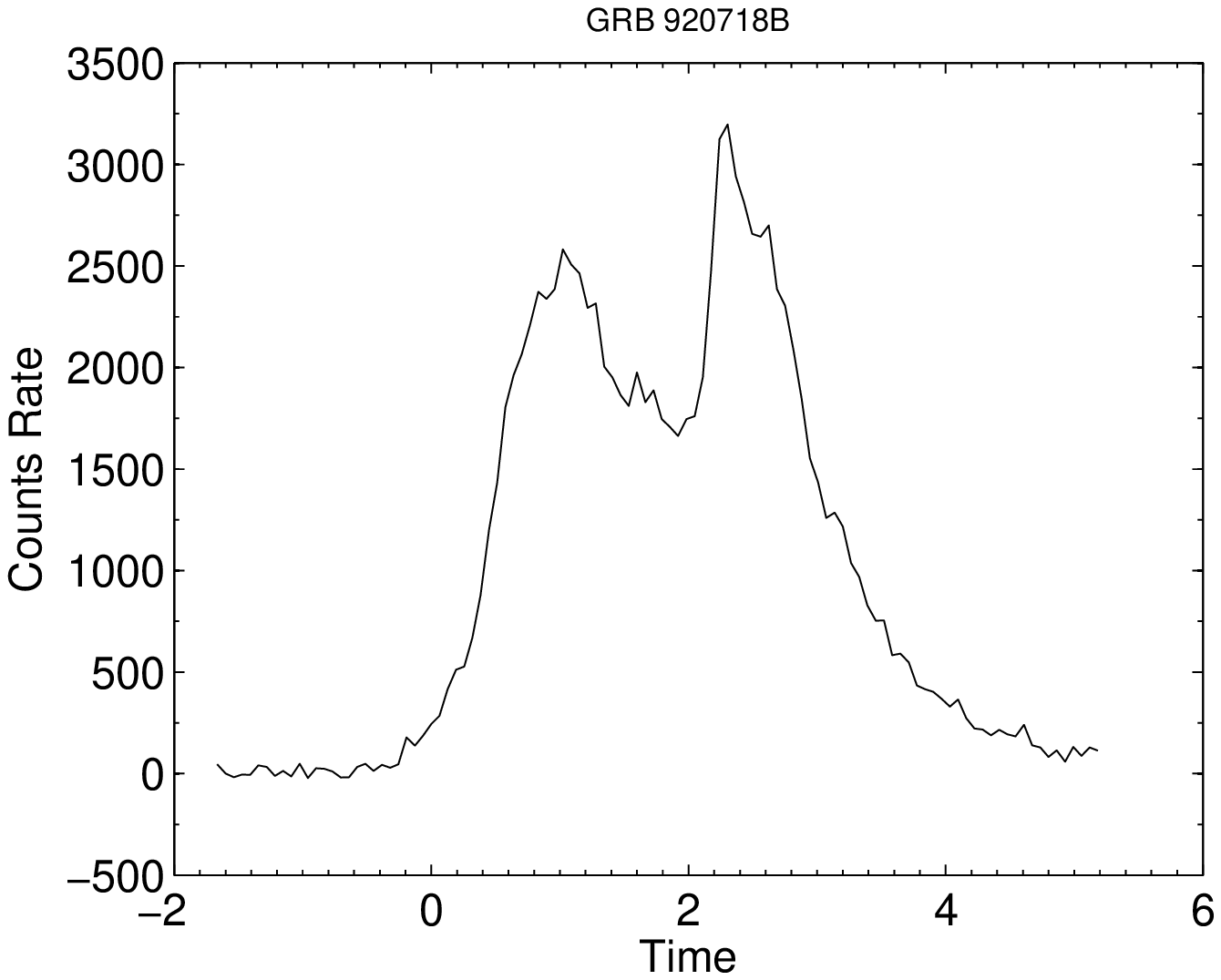}}
  %\hspace{1in}
  \subfigure[]{
    \label{fig:subfig:b} %% label for second subfigure
    \includegraphics[width=1.7in]{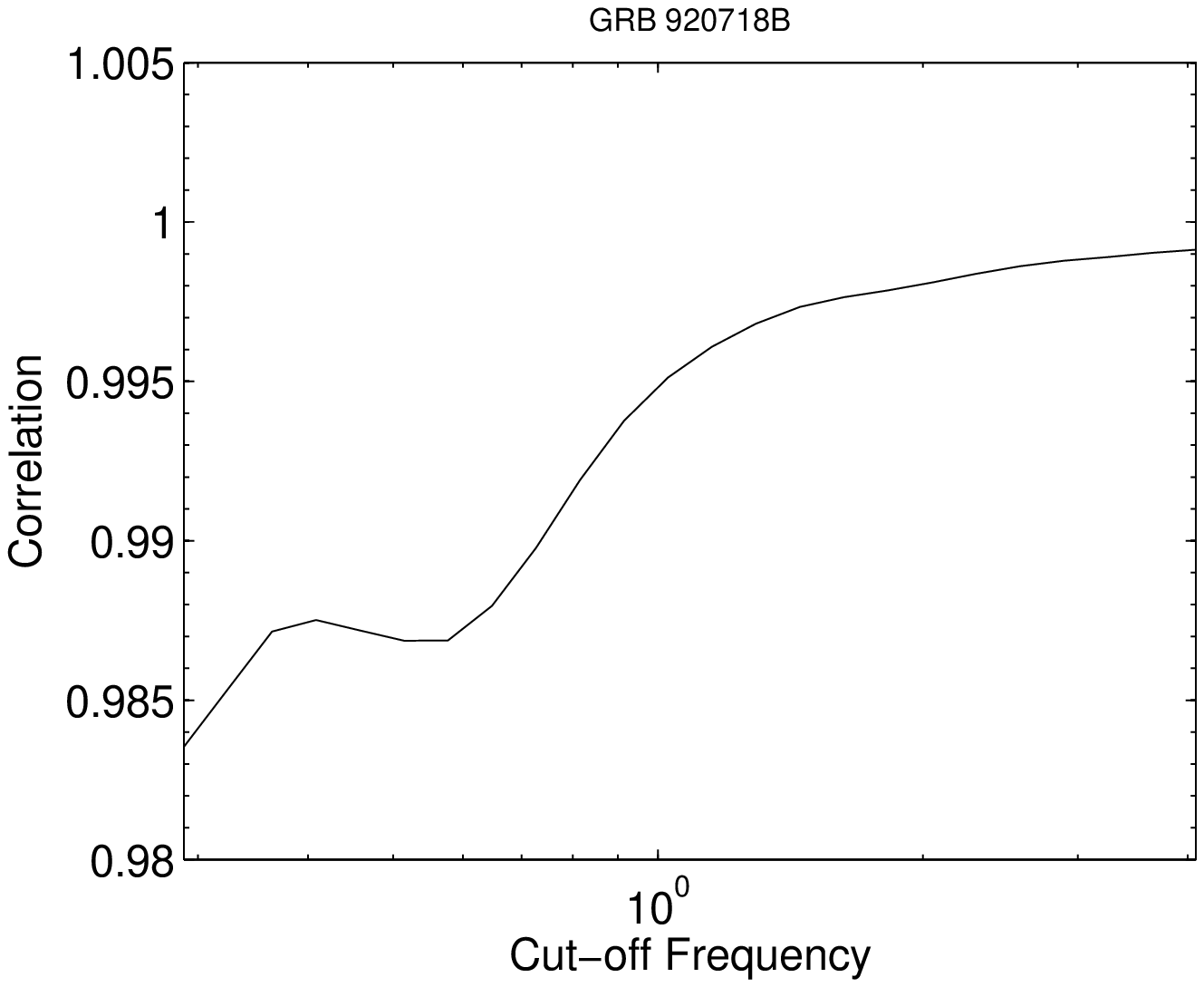}}
  \caption{The top and middle panel are original and synthetic lightcurve for GRB930120. The pulses that correspond to the identified frequencies are marked in different colors in the lightcurves. The time scales are rounded to the nearest 0.5.
  The bottom panel is lightcurve and SFC curve for GRB910522 and GRB920718B. }
  \label{fig:subfig} %% label for entire figure
\label{cases}
\end{figure}

We take GRB 930120 as an example in Group II (gaps/long tails). As
shown in Fig.\ref{cases}(a), the burst has two brief activities with
durations $\sim 10$ s, followed by the main emission episode. The
three episodes are separated by two gaps. The SFC curve
(Fig.\ref{cases}(b)) shows only one dip, which corresponds to the
main pulse with duration around $34.5$ s.

The puzzling fact is that the $\sim 10$ s feature and the high-frequency
spikes overlapping the main pulse are not captured. The missing
high-frequency spiky component is due to the low amplitude of this
component. The lack of the $\sim 10$ s component may be understood
in two ways.
% in the light curve (we call it the
%main pulse hereafter). It is also obvious that this pulse is
%superposed with even faster variabilities. So it's safe for us to
%add this burst into our sample revealing the superposition fact in
%GRB light curve.
%However, two pulses with about $\sim$10s timescale that are clearly
%identifiable in the light curve were lost by SFC $R_{\rm i}-f_{\rm
%c,i}$ curve. How could that happen? We propose two reasons for that:
(1) The amplitudes of those two pulses are too small compared with
the main pulse; (2) The existence of the gaps modified the durations
of those pulses from $\sim 10$ s to $\sim 30$ s, which is close to the
duration of the main pulse so that the two dips merge to one. To test
these possibilities, we perform several tests. First, we
manually remove the quiescent periods (the gaps) in the lightcurve.
% between the two
%$\sim$10s pulses from the light curve of GRB 930120, and then apply
%SFC to it. From Fig.7(g) $\sim$ Fig.7(h), we can see that the
The SFC curve still does not show the $\sim 10$ s component
(Fig.\ref{cases}(c,d)). Next, we manually increase the amplitudes
of the two pulses to be comparable to that of the main pulse,
the $\sim 10$ s component then shows up in the SFC curve
(Fig.\ref{cases}(e,f)).
%$\sim$10s component still not show up. However, if we additionally
%change the amplitude of those two pulses to the same order of the
%main pulse as shown in Fig.7(i) $\sim$ Fig.7(j), the expected
%$\sim$10s component eventually shows up in the SFC $R_{\rm i}-f_{\rm
%c,i}$ curve, indicating
%This proves that the small amplitude is the reason why
%SFC lose the component. For completeness, we did another test on
%this burst.
Finally, we increase the amplitude of the two pulses but do not
remove the gaps. The $\sim 10$ s dip in the SFC curve disappears
again (Fig.\ref{cases}(g,h)). This suggests that both reasons
(low amplitude and influence of gaps) play a role in missing
the $\sim 10$ s component in the original lightcurve.

%Without extracting the lags, we directly change the
%amplitude of those two pulses. Fig.7(k) $\sim$ Fig.7(l) show us that
%the $\sim$10s component is still missing in SFC $R_{\rm i}-f_{\rm
%c,i}$ curve. This should be due to the second reason we proposed
%above, and to some extent, this final test could prove the
%conclusions we obtained about SFC algorithm in Appendix 2.

In the irregular group (III) we chose GRB 910522 as an example.
The lightcurve is very noisy (Fig.\ref{cases}(i)), and the SFC
curve is irregular (Fig.\ref{cases}(j)).

%\textbf{We take GRB 910522 as an example for the second group, i.e.,
%bursts with noisy lightcurve. Fig.7(c) $\sim$ Fig.7(d) are the
%lightcurve and SFC curve for this case. Others are presented in the
%group website listed above.}

Finally, the group IV includes bursts with short durations or
poor temporal resolution. The SFC method is no longer applicable
to these bursts. An example (GRB 920718B) is presented in
Fig.\ref{cases}(k,l).

%\textbf{For the first group of GRBs, namely short bursts or long
%bursts whose lightcurves are with poor resolution, we just choose
%one burst GRB 920718B as an example to show in this paper. Fig.7(a)
%$\sim$ Fig.7(b) are the lightcurve and SFC curve for this case.}

\begin{thebibliography}

\bibitem[Barthelmy et al.(2005)]{Barthelmy05} Barthelmy, S.~D., et
al.\ 2005, \apjl, 635, L133
\bibitem[Beloborodov et al.(1998)]{Beloborodov98} Beloborodov, A.~M.,
Stern, B.~E., \& Svensson, R.\ 1998, \apjl, 508, L25
\bibitem[Beloborodov et al.(2000)]{Beloborodov00} Beloborodov, A.~M.,
Stern, B.~E., \& Svensson, R.\ 2000, \apj, 535, 158
%\bibitem[Cho et al.(2002)]{Cho02} Cho, J., Lazarian, A.,
%\& Vishniac, E.~T.\ 2002, \apj, 564, 291
\bibitem[Daigne \& Mochkovitch(2003)]{Daigne03} Daigne, F., \& Mochkovitch, R.\
2003, \mnras, 342, 587
\bibitem[Dermer \& Mitman(1999)]{Dermer99} Dermer, C.~D., \& Mitman, K.~E.\ 1999,
\apjl, 513, L5
\bibitem[Fishman \& Meegan(1995)]{Fishman95} Fishman, G.~J., \& Meegan, C.~A.\ 1995,
\araa, 33, 415
%\bibitem[Goldreich \& Sridhar(1995)]{Goldreich95} Goldreich, P., \& Sridhar, S.\ 1995,
%\apj, 438, 763
\bibitem[Golenetskii et al.(2009)]{Golenetskii09}
Golenetskii, S., Aptekar, R., Mazets, E., Pal'Shin, V., Frederiks,
D., Oleynik, P., Ulanov, M., \& Svinkin, D.\ 2009, GRB Coordinates
Network, 9647, 1
\bibitem[Kaneko et al.(2006)]{Kaneko06} Kaneko, Y., Preece,
R.~D., Briggs, M.~S., Paciesas, W.~S., Meegan, C.~A., \& Band,
D.~L.\ 2006, \apjs, 166, 298
\bibitem[Kobayashi et al.(1997)]{Kobayashi97} Kobayashi, S., Piran,
T., \& Sari, R.\ 1997, \apj, 490, 92
\bibitem[Kocevski et al.(2003)]{Kocevski03} Kocevski, D., Ryde, F.,
\& Liang, E.\ 2003, \apj, 596, 389
\bibitem[Kumar \& Narayan(2009)]{Kumar09} Kumar, P., \& Narayan, R.\ 2009, \mnras,
395, 472
\bibitem[Lazzati et al.(2009)]{Lazzati09} Lazzati, D., Morsony,
B.~J., \& Begelman, M.~C.\ 2009, \apjl, 700, L47
\bibitem[Lei et al.(2007)]{Lei07}Lei W.H. et al., 2007, A\&A, 468,
563
\bibitem[Li \& Fenimore(1996)]{Li96} Li, H., \& Fenimore, E.~E.\ 1996, \apjl,
469, L115
\bibitem[Lu et al.(2008)]{Lu08}Lu Y., Huang Y.F., and Zhang S.N.,
2008, \apj, 684, 1330
\bibitem[Lyutikov \& Blandford(2003)]{Lyutikov03} Lyutikov, M., \& Blandford,
R.\ 2003, arXiv:astro-ph/0312347
\bibitem[Maxham \& Zhang(2009)]{Maxham09} Maxham, A., \& Zhang, B.\ 2009, \apj, 707,
1623
\bibitem[McBreen et al.(2001)]{McBreen01} McBreen, S., Quilligan, F., McBreen, B.,
Hanlon, L., \& Watson, D.\ 2001, \aap, 380, L31
\bibitem[M\'esz\'aros \& Rees(1993)]{Meszaros93} M\'esz\'aros, P., \& Rees, M.~J.\ 1993, \apj,
405, 278
\bibitem[Morsony et al.(2010)]{Morsony10} Morsony, B.~J.,
Lazzati, D., \& Begelman, M.~C.\ 2010, \apj, 723, 267
\bibitem[Nakar\& Piran(2002)]{Nakar02} Nakar, E., \& Piran, T.\ 2002, \mnras, 331,
40
\bibitem[Narayan \& Kumar(2009)]{Narayan09} Narayan, R., \& Kumar, P.\ 2009, \mnras,
394, L117
\bibitem[Norris et al.(1996)]{Norris96} Norris, J.~P., Nemiroff,
R.~J., Bonnell, J.~T., Scargle, J.~D., Kouveliotou, C., Paciesas,
W.~S., Meegan, C.~A., \& Fishman, G.~J.\ 1996, \apj, 459, 393
\bibitem[Norris et al.(2000)]{Norris00} Norris, J.~P., Marani,
G.~F., \& Bonnell, J.~T.\ 2000, \apj, 534, 248
\bibitem[Oppenheim et al.(1998)]{opp98} Oppenheim, A.V., Schafer,
R.W., Buck, J.R., 1998, "Discrete-time signal processing" second
edit, Vol.2, 52
\bibitem[Portegies Zwart et al.(1999)]{Portegies Zwart99} Portegies
Zwart, S.~F., Lee, C.-H., \& Lee, H.~K.\ 1999, \apj, 520, 666
\bibitem[Piran(1999)]{Piran99} Piran, T.\ 1999, \physrep, 314,
575
\bibitem[Ramirez-Ruiz et al.(2001)]{RR01} Ramirez-Ruiz, E., Merloni, A., Rees, M. J.
\ 2001, \mnras, 324, 1147
\bibitem[Rees \& M\'esz\'aros(1994)]{Rees94} Rees, M.~J., \& M\'esz\'aros, P.\ 1994,
\apjl, 430, L93
\bibitem[Sari \& Piran(1997)]{Sari97} Sari, R., \& Piran, T.\ 1997, \apj, 485, 270
\bibitem[Shen \& Song(2003)]{Shen03} Shen, R.-F., \& Song, L.-M.\ 2003, \pasj, 55,
345
\bibitem[Tagliaferri et al.(2005)]{Tagliaferri05} Tagliaferri, G., et
al.\ 2005, \nat, 436, 985
\bibitem[Vetere et al.(2006)]{Vetere06} Vetere, L., Massaro, E., Costa, E., Soffitta,
P., \& Ventura, G.\ 2006, \aap, 447, 499
%\bibitem[Wu \& Zhang(2011)]{Wu11} Wu, X.-F., Zhang, B.\ 2011, \apj,
%Submitted
\bibitem[Zhang et al.(2006)]{Zhang06} Zhang, B., Fan, Y.~Z.,
Dyks, J., Kobayashi, S., M{\'e}sz{\'a}ros, P., Burrows, D.~N.,
Nousek, J.~A., \& Gehrels, N.\ 2006, \apj, 642, 354
\bibitem[Zhang \& Yan(2011)]{Zhang11} Zhang, B., \& Yan, H.\ 2011, \apj, 726, 90
\bibitem[Zhang \& Zhang (2011)]{ZZ11} Zhang, B., \& Zhang, B. 2011, \mnras, to be submitted
\bibitem[Zhang et al.(2003)]{Zhang03} Zhang, W., Woosley,
S.~E., \& MacFadyen, A.~I.\ 2003, \apj, 586, 356

\end{thebibliography}
\end{document}